\RequirePackage{lineno}
\documentclass[a4,nofootinbib,notitlepage,superscriptaddress]{revtex4-1}
\usepackage{bm}
\usepackage{lineno}
\usepackage[pdftex]{graphicx}

\newcommand{\I}{\mathrm{i}}

\usepackage{amsmath,amsfonts,amssymb,mathtools}
\usepackage{amsthm}

\usepackage{bbm,bm}
\usepackage{float}
\usepackage{mathrsfs}

\usepackage[normalem]{ulem}

\usepackage{physics}
\usepackage{xcolor}

\usepackage[utf8]{inputenc}
\usepackage{pgfplots}
\usepgfplotslibrary{groupplots}
\usepgfplotslibrary{fillbetween}
\usepackage{apptools}
\AtAppendix{\counterwithin{theorem}{section}}

\DeclarePairedDelimiterXPP{\sfTr}[1]{\mathsf{Tr}}{[}{]}{}{#1}
\DeclarePairedDelimiterXPP{\sfTrAbs}[1]{\mathsf{TrAbs}}{[}{]}{}{#1}
\DeclarePairedDelimiterXPP{\opTr}[1]{\mathrm{Tr}}{[}{]}{}{#1}
\DeclarePairedDelimiterXPP{\bbTr}[1]{\mathbb{T}\mathrm{r}}{[}{]}{}{#1}

\def\ANU{Centre for Quantum Computation and Communication Technology, Department of Quantum Science, Australian National University, Canberra, ACT 2601, Australia.}

  \def\Astar{Institute of Materials Research and Engineering, Agency for Science Technology and Research (A*STAR), 2 Fusionopolis Way, 08-03 Innovis 138634, Singapore}
\begin{document}
\title{Multiparameter estimation with two qubit probes in noisy channels}
\author{Lorc\'{a}n O. Conlon}
\email{lorcanconlon@gmail.com}
\affiliation{\ANU}
\affiliation{\Astar}
%\author{Jun Suzuki}
%\email{junsuzuki@uec.ac.jp}
%\affiliation{\UEC}
\author{Ping Koy Lam}
\affiliation{\ANU}
\affiliation{\Astar}
\author{Syed M. Assad}
\email{cqtsma@gmail.com}
\affiliation{\ANU}
\affiliation{\Astar}
%\affiliation{\NTU}
%\affiliation{Centre for Quantum Computation and Communication Technology, Department of Quantum Science,\\ Research School of Physics and Engineering, Australian National University, Canberra ACT 2601, Australia.}%
\date{\today}

\begin{abstract}
This work compares the performance of single and two qubit probes for estimating several phase rotations simultaneously under the action of different noisy channels. We compute the quantum limits for this simultaneous estimation using collective and individual measurements  by evaluating the Holevo and Nagaoka--Hayashi Cram\'{e}r-Rao bounds respectively. Several quantum noise channels are considered, namely the decohering channel, the amplitude damping channel and the phase damping channel. For each channel we find the optimal single and two qubit probes. Where possible we demonstrate an explicit measurement strategy which saturates the appropriate bound and we investigate how closely the Holevo bound can be approached through collective measurements on multiple copies of the same probe. We find that under the action of the considered channels, two qubit probes show enhanced parameter estimation capabilities over single qubit probes for almost all non-identity channels, i.e. the achievable precision with a single qubit probe degrades faster with increasing exposure to the noisy environment than that of the two qubit probe. However, in sufficiently noisy channels, we show that it is possible for single qubit probes to outperform maximally entangled two qubit probes. This work shows that, in order to reach the ultimate precision limits allowed by quantum mechanics, entanglement is required in both the state preparation and state measurement stages. It is hoped the tutorial-esque nature of this paper will make it easily accessible.
%Throughout this study we found persistent results which we show are true for a broad class of channels; a two qubit probe is able to estimate a second parameter with no decrease in the achievable precision, for a single qubit probe estimating a second parameter causes the achievable precision to at best decrease by a factor of 2 in the absence of any noise and for a single qubit probe it is impossible to construct an unbiased estimator for estimating a rotation about all three axes of the Bloch sphere. 
\end{abstract}
\maketitle
%\tableofcontents
\section{Introduction}

\begin{figure}[t]
\includegraphics[width=\textwidth]{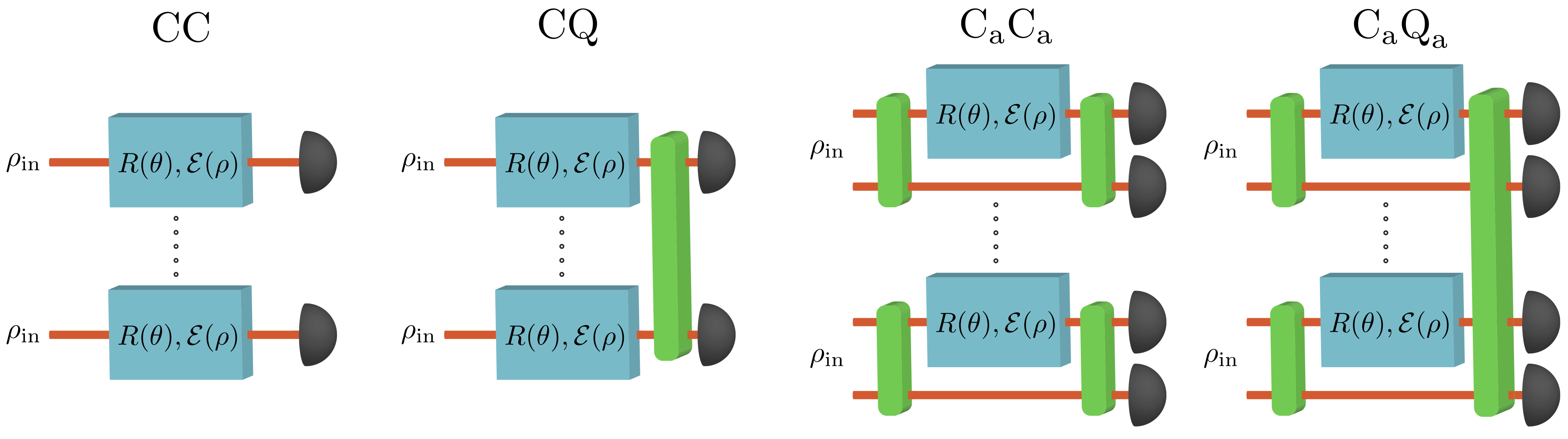}
\caption{\textbf{The four sensing schemes we consider; Classical-Classical (CC), Classical-Quantum(CQ), Classical with ancilla-Classical with ancilla($\text{C}_\text{a}$$\text{C}_\text{a}$) and Classical with ancilla-Quantum with ancilla($\text{C}_\text{a}$$\text{Q}_\text{a}$).} The probes $\rho_\text{in}$ are either unentangled single qubit, or entangled two qubit states, with the green box at the input symbolising entanglement generation. After the probes have passed through the channel, we can either perform a collective measurement or an individual measurement, symbolised by the green box at the output. The $\text{C}_\text{a}$$\text{C}_\text{a}$ scheme has a green box at the output to signify that any measurement can be performed on the two mode state, but collective measurements cannot be performed across multiple different states, as is the case with the CQ and $\text{C}_\text{a}$$\text{Q}_\text{a}$ schemes. The vertical dots in the schematic signify that many copies of the input probe state $\rho_\text{in}$ are used. The blue box $R(\theta)$, $\mathcal{E}(\rho)$ represents the quantum channels. $R(\theta)$ is a small rotation (or multiple rotations) by an angle $\theta$, the parameter we wish to measure and $\mathcal{E}(\rho)$ is a noisy quantum channel.}
\label{fig:scheme}
\end{figure}

Quantum parameter estimation involves preparing a quantum probe, allowing this
probe to interact with the system we wish to learn about and then
examining the probe at the output. The maximum precision with which certain
parameters can be estimated is dictated by the laws of quantum mechanics~\cite{robertson1929uncertainty, arthurs1965bstj,heisenberg1985quantentheoretische}. %By preparing $N$ identical, independent probes and averaging the results obtained it is possible to enhance the attainable precision. The minimum variance which can be attained using probes which are not entangled scales as ${1}/{N}$, however through the use of entangled probes this scaling can be improved to ${1}/{N^2}$~\cite{giovannetti2004quantum,giovannetti2006quantum, giovannetti2011advances}. 
Quantum resources have been proposed as a way to improve measurement sensitivity in optical interferometry~\cite{caves1981quantum,barnett2003ultimate,dorner2009optimal,demkowicz2009quantum,zhuang2018distributed,ge2018distributed,conlon2022enhancing}, quantum superresolution
\cite{tsang2016quantum, tsang2019resolving}, quantum
positioning~\cite{giovannetti2001quantum, lamine2008quantum}, and tests of fundamental physics~\cite{brady2022entangled,marchese2023optomechanics,shi2023ultimate}. Several experiments have demonstrated enhanced precision estimation through the use of quantum resources~\cite{higgins2007entanglement, kacprowicz2010experimental, yonezawa2012quantum,girolami2014quantum,strobel2014fisher,slussarenko2017unconditional, zhang2019quantum,mccormick2019quantum,wang2019heisenberg,aasi2013enhanced,guo2020distributed,liu2021distributed,backes2021quantum,casacio2021quantum,marciniak2022optimal,malia2022distributed,PhysRevLett.130.123603}. 

%Parameter estimation and quantum metrology are two of the areas of research that best showcase the advantage quantum mechanical resources can offer over their classical counterparts. 
 
Arguably, the full range of quantum mechanical effects is only revealed through multi-parameter metrology owing to the possible incompatibility of conjugate observables. There are many physically motivated reasons for studying multiparameter estimation. The simultaneous estimation of several parameters can enhance our ability to measure the different components of a magnetic field~\cite{baumgratz2016quantum,hou2020minimal,montenegro2022sequential,kaubruegger2023optimal}, multiple phase shifts~\cite{spagnolo2012quantum,humphreys2013quantum,yue2014quantum,gagatsos2016gaussian,ciampini2016quantum, pezze2017optimal,zhang2017quantum}, a phase shift and loss or phase diffusion~\cite{crowley2014tradeoff, szczykulska2017reaching}, and can improve the tracking of chemical processes~\cite{cimini2019quantum}. Additionally, quantum super-resolution for resolving two incoherent point sources of light~\cite{tsang2016quantum,chrostowski2017super,vrehavcek2017multiparameter} and many estimation problems concerning Gaussian quantum states~\cite{chiribella2006joint,monras2011measurement,Genoni2013,gao2014bounds,bradshaw2017tight,bradshaw2018ultimate,assad2020accessible,park2022optimal} can be cast as multiparameter estimation problems. Given this motivation, there has been significant experimental~\cite{steinlechner2013quantum, vidrighin2014joint,hou2016achieving,liu2018loss,li2023optimal} and theoretical~\cite{vaneph2013quantum,suzuki2015parameter,suzuki2016explicit,szczykulska2016multi,proctor2018multiparameter,gessner2018sensitivity,tsang2019quantum,carollo2019quantumness,tsang2020quantum,demkowicz2020multi,razavian2020quantumness,gessner2020multiparameter,lu2021incorporating,gebhart2021bayesian,albarelli2022probe,huang2021quantum,gianani2021kramers,hanamura2021estimation,di2022multiparameter,hosseiny2022estimating,fadel2022multiparameter,len2022multiparameter,xie2022quantum} interest in quantum multiparameter estimation. See Refs.~\cite{liu2019quantum,albarelli2020perspective,sidhu2020geometric,polino2020photonic} for recent reviews on the subject.

The physical process of quantum metrology can be described as a quantum channel with several different variations possible depending on the degree to which we wish to exploit quantum mechanical effects \cite{bennett1998quantum}. In this work, we consider four different schemes; Classical-Classical (CC), Classical-Quantum(CQ), Classical with ancilla-Classical with ancilla($\text{C}_\text{a}$$\text{C}_\text{a}$) and Classical with ancilla-Quantum with ancilla($\text{C}_\text{a}$$\text{Q}_\text{a}$). These four schemes are distinguished by how the states are prepared, either using entangled states or non-entangled states, and by how they are measured at the output, as shown in Fig.~\ref{fig:scheme}. Note that, on the state preparation side, we only consider either single-qubit or two-qubit states. This is distinct from previous studies that have considered the case where entanglement is generated across all input states before the channel~\cite{giovannetti2006quantum}. In our model, as the quantum states pass through the channel they experience a small rotation about all three axes by angles $\theta=(\theta_x,\theta_y,\theta_z)$ before experiencing some form of decoherence. The quantum states are then measured, either using collective measurements or individual measurements. Individual measurements here means that the probe states are measured one by one, in contrast to the most general measurement type, a collective measurement, which involves measuring multiple probe states simultaneously in an entangling basis. The quantum measurement strategies in Fig.~\ref{fig:scheme}, thus refer to performing a collective measurement on asymptotically many copies of the probe state. However, we can also consider performing a collective measurement on a finite number of copies of the probe state. It is known that collective measurements offer no advantage over individual measurements (i.e. the CC and CQ strategies are equivalent, as are $\text{C}_\text{a}$$\text{C}_\text{a}$ and $\text{C}_\text{a}$$\text{Q}_\text{a}$) for estimating only a single parameter or for estimating multiple parameters with pure states~\cite{Matsumoto2002}. However, for the most general metrology problem, multiparameter estimation with impure probe states, quantum resources can offer an advantage at both the state preparation and state measurement stages. The advantage of quantum resources at either of these stages can be thought of as quantum enhanced metrology. 

%Despite the myriad physical motivations behind quantum metrology many questions remain unanswered and specific results are known for only a few examples. 
%Another bound is the Gill-Massar bound~\cite{gill2005state}. 
There are several known methods to determine fundamental limits on how accurately the parameters of interest can be measured. One common approach is to use the quantum Fisher information (QFI). There are several variants of the QFI, including the QFI based on the symmetric logarithmic derivative (SLD bound), introduced by Helstrom~\cite{helstrom1967minimum,helstrom1968minimum} and the QFI based on the right logarithmic derivative (RLD bound)~\cite{yuen1973}. It is known that for estimating a single parameter, the SLD bound can always be saturated, making it a particularly important bound. However, for estimating multiple parameters the SLD bound may not be attainable if the optimal measurements for estimating each parameter individually do not commute. An attainable bound on the ultimate achievable precision in quantum parameter estimation theory was formulated by Holevo~\cite{holevo2011probabilistic,holevo1973statistical}, the Holevo Cram{\'{e}}r-Rao bound. Throughout this paper we shall simply refer to this as the Holevo bound. The Holevo bound is important as it known to be asymptotically attainable when one uses a collective measurement on infinitely many copies of the probe state~\cite{kahn2009local,yamagata2013quantum,yang2019attaining}. As such, the Holevo bound offers a way to investigate the precision attainable with the CQ and $\text{C}_\text{a}$$\text{Q}_\text{a}$ strategies in the asymptotic limit. In several different scenarios, utilising pure states and/or estimating a single parameter, the SLD bound and Holevo bound have been saturated experimentally~\cite{yu2022quantum,li2022geometric} or there exist theoretical proposals to saturate these bounds~\cite{bradshaw2018ultimate,helstrom1968minimum}. However, it has recently been proven that if the SLD bound or Holevo bound cannot be saturated with individual measurements, then they cannot be saturated with any physical measurement~\cite{conlon2022gap}. This property is known as gap persistence and is an important caveat to the statement that the Holevo bound is asymptotically attainable.

In light of the practical difficulties associated with the SLD bound and Holevo bound, another bound of particular interest is the Nagaoka Cram{\'{e}}r-Rao bound (Nagaoka bound)~\cite{nagaoka2005new} which applies when one is restricted to measuring the probe states individually, i.e. the CC and $\text{C}_\text{a}$$\text{C}_\text{a}$ strategies in Fig.~\ref{fig:scheme}. As it was originally introduced, the Nagaoka bound applies only to two parameter estimation. This bound was later generalised to the $n$-parameter case by Hayashi~\cite{hayashi1997linear,conlon2021efficient}, which we shall refer to as the Nagaoka--Hayashi bound (NHB)\footnote{For estimating one or two parameters the NHB reduces to the Nagaoka bound and so in these scenarios we shall use the two names interchangeably.}. The Nagaoka bound is known to be a tight bound for probes existing in a two-dimensional Hilbert space~\cite{nagaoka2005generalization}, i.e. there always exists a measurement which can achieve the same variance as the Nagaoka bound. However, it has since been shown that the NHB is not always a tight bound~\cite{hayashi2022tight}. Importantly, the NHB will always give a variance smaller than or equal to that of the Holevo bound as individual measurements are a subset of collective measurements. Experimentally, the collective measurements required to surpass the NHB and approach the Holevo bound are extremely challenging to implement, hence there have been a very limited number of demonstrations of such measurements~\cite{roccia2017entangling,hou2018deterministic,parniak2018beating,wu2019experimentally,wu2020minimizing,yuan2020direct,conlon2023approaching,conlon2023discriminating}. \footnote{Also note that the techniques demonstrated in Ref.~\cite{martinez2023certification} could in principle be used to implement collective measurements.} Recently, there has been a great deal of work developing new computational techniques for calculating the Holevo bound~\cite{bradshaw2018ultimate,albarelli2019evaluating,sidhu2021tight} and NHB~\cite{conlon2021efficient}.

%It has recently been shown that both the Holevo bound and NHB can be solved using a semidefinite program (SDP)~\cite{albarelli2019evaluating,conlon2021efficient}. The recasting of the Holevo bound as a SDP was originally performed to compute the ultimate limits for estimating a displacement in both quadratures of a Gaussian probe~\cite{bradshaw2018ultimate}.  Additionally, an analytic approach which provides upper and lower bounds to the Holevo bound for estimating two parameters has recently been developed~\cite{sidhu2021tight}. These techniques greatly simplify the task of calculating the ultimate bounds on the achievable precision.

This work aims to demonstrate the efficacy of these new techniques by finding the single and two qubit probes which optimise the Holevo and Nagaoka bounds for several different noise channels. There have been several previous experimental and
theoretical considerations as to how noisy channels affect the
achievable precision \cite{dorner2009optimal,kacprowicz2010experimental,genoni2011optical,datta2011quantum}. In this work we study specific examples of these noisy channels for
qubit probes. We consider the problem of simultaneous estimation of
three independent rotations around the $x$, $y$ and $z$ axes of the
Bloch sphere. We find a hierarchy between the four
different schemes in Fig.~\ref{fig:scheme} in terms of the
achievable precision subject to a noisy channel; $\text{C}_\text{a}$$\text{Q}_\text{a}$$\geq \{\text{C}_\text{a}\text{C}_\text{a}, \text{CQ}\}
\geq \text{CC}$, with no general ordering between $\text{C}_\text{a}$$\text{C}_\text{a}$ and $\text{CQ}$. Although one might expect that more entangled probes offer a quantum advantage over unentangled probes, this is not necessarily
true. In very noisy channels probes with more entanglement can offer a disadvantage. As expected, it was also found that, for a noisy channel, collective measurements
offer an advantage over individual measurements. Thus, we can consider
these quantum mechanical effects as offering an increased robustness
to noise. Typically, when estimating multiple parameters there is a trade-off between the number of parameters we wish to measure and the accuracy with which we can measure them~\cite{ragy2016compatibility,kull2020uncertainty}, however we show for multiple-qubit probes this is not necessarily true. Some of the other, more surprising, features of quantum metrology are evident in the examples considered, for example we find scenarios where states which experience decoherence outperform those which do not and we find discontinuities in the Holevo bound. We note that investigating these distinct schemes is different to many other works which have studied the scaling of the variance with the number of input probe states $N$, i.e. Heisenberg scaling (1/$N^2$) or scaling at the standard quantum limit (1/$N$)~\cite{ballester2005optimal,imai2007geometry,wang2019heisenberg,napolitano2011interaction,cimini2019quantum,hayashi2022global}.

%we find decohered states which are able to achieve a better measurement precision than states with less decoherence and

%
%We will begin by introducing the relevant bounds used in this paper in section \ref{Prelim}. In sections \ref{firstres} to \ref{lastres} we discuss the results obtained for specific channels. Finally in section \ref{generalisation} we generalise some of the results obtained to arbitrary channels and probe states. In section \ref{generalisation} we show that it is impossible, with a single qubit probe, to construct an unbiased estimator for estimating a rotation about all three axes. However,this does not eliminate the possibility of using a biased estimator to estimate these three parameters. It is known that biased estimators can outperform unbiased estimators in some situations~\cite{liu2016valid}. In section \ref{generalisation} we also consider how the estimation performance of different probes is affected by the estimation of a second and third parameter. Typically when estimating multiple parameters there is a trade-off between how many parameters we wish to measure and the accuracy with which we can measure them~\cite{ragy2016compatibility, kull2020uncertainty}., however we show for multiple-qubit probes this is not necessarily true
We begin by introducing the relevant bounds used in this paper in section \ref{Prelim}. In sections \ref{firstres} to \ref{lastres} we discuss the results obtained for specific channels. For each channel we consider single qubit probes, two qubit probes and the attainability of the Holevo bound. In the appendices we construct explicit measurement schemes which saturate the Nagaoka bound and NHB for many of the examples considered.
\section{Preliminaries}
\label{Prelim}

\subsection{Parameter estimation and quantum Fisher information}
%We consider a finite $d$-dimensional Hilbert space. 
The family of states being investigated,
$\rho_{\theta}$, are parameterised by
${\theta}=(\theta_{1},\dots, \theta_{n})$, where the $\theta_{j}$
are the unknown parameters we wish to estimate. In this paper the
$\theta_{j}$ are qubit rotations on the Bloch sphere. We can measure
the parameters we wish to estimate using a positive operator-valued
measure (POVM). A POVM is described by a set of positive linear
operators, $\Pi_{k}$, that sum up to the identity
\begin{equation}
\label{eq:POVMcondition}
\sum_{k}\Pi_{k}=\mathbbm{1}\;.
\end{equation}
The $k$-th outcome is realized with probability
$\text{Tr}\{\rho_{\theta}\Pi_{k}\}$. Based on these measurement outcomes we can
construct an unbiased estimator for the parameters of interest,
$\hat{\theta}$. Our estimated value is constructed from the estimator coefficients $\xi$
\begin{equation}
\hat{\theta}_j=\sum_{k}\xi_{j,k}p_{\theta}(k)\;,
\end{equation}
where $p_{\theta}(k)$ is the probability of obtaining the
measurement outcome denoted $k$ and the sum is over all possible
measurement outcomes. The aim of parameter estimation is to minimise the sum
of the mean squared error between our unbiased estimate and the actual
values we wish to measure, $\theta$. For estimating several parameters
simultaneously the mean-square error matrix,
$V_{\theta}$, has elements given by
\begin{equation}
\begin{split}
[V_{\theta}]_{jk}&=\sum_{x}\left(\xi_{j,x}-\theta_{j}\right)\left(\xi_{k,x}-\theta_{k}\right)p_{\theta}(x)\;.
%&=\left(\hat{\theta}_{j}-\theta_{j}\right)\left(\hat{\theta}_{k}-\theta_{k}\right)\;.
\end{split}
\end{equation}

If we have $N$ independent and identical copies of
the quantum state, $\rho_{\theta}$, the sum of the 
variance of our estimators is bounded by the quantum Cram{\'e}r--Rao bound
\begin{equation}
\text{Tr}\left\{ V_{\theta}\right\} \geq \frac{1}{N} \text{Tr}\left\{J(\rho_{\theta})^{-1}\right\}\;,
\label{QCRB}
\end{equation}
where $J(\rho_{\theta})$ is the QFI matrix with
elements,% \comment{The equation for the QCR bound is not
%true for a general QFI matrix. The LHS above corresponds to the usual
%variance only when we use the SLD QFI.}
\begin{equation}
  \label{eq:QFI}
  \left[J(\rho_{\theta})\right]_{jk}=\text{Tr}\left\{\frac{\partial\rho_{\theta}}{\partial
    \theta_j} \mathcal{L}_k 
\right\}\;,
\end{equation}
and $\mathcal{L}$ is the quantum analogue for the classical
logarithmic derivative. There is no
unique way to define the quantum logarithmic derivative and each
definition gives rise to a different QFI. %Although the different QFI's are not tight bounds in general, they have become popular owing to their ease of computation~\cite{pinel2013quantum}. 
Two of the most prominent QFI's are those based on the symmetric logarithmic derivative (SLD) and the right logarithmic derivative (RLD). The SLD and RLD operators combined with Eqs.~\eqref{QCRB} and \eqref{eq:QFI} give rise to the SLD bound, $C^\text{SLD}$, and the RLD bound, $C^\text{RLD}$ respectively. Although neither the SLD bound nor the RLD bound are attainable in general, both bounds are useful in certain scenarios. For this work, the SLD bound shall be used when we consider estimating a single parameter, as in this scenario it is known that the SLD bound is attainable. The SLD operators, $\mathcal{L}$, can be computed as
\begin{equation}
\label{eq:sld:comp}
\mathcal{L}_k =2\sum_{m,p}\ket{e_m}\frac{\bra{e_m}\frac{\partial\rho_{\theta}}{\partial
    \theta_k}\ket{e_p}}{\lambda_m+\lambda_p}\bra{e_p}\;,
\end{equation}
where $\ket{e_m}, \lambda_m$ are the eigenvectors and eigenvalues of the density matrix, $\rho=\sum_{m}\lambda_m\ket{e_m}\bra{e_m}$, and the sum is over all $\lambda_m+\lambda_p\neq0$~\cite{pinel2013quantum}. Thus, for any given problem the SLD QFI is generated in a completely deterministic manner meaning no optimisation is required. As we shall only use the SLD bound for single parameter estimation, the corresponding bound on the variance in estimating the parameter $\theta_k$ is given by
\begin{align}
\label{eq:SLDbound}
 v_k\geq C^\text{SLD}=\frac{1}{J(\rho_{\theta})_{kk}}\;,
\end{align}
with $J$ defined using the SLD operator, Eq.~\eqref{eq:sld:comp}.

\subsection{Holevo and Nagaoka--Hayashi bounds}
Holevo unified the SLD bound and the RLD bound through the Holevo bound, which we denote $\mathcal{H}$. The Holevo bound is achieved asymptotically and is assured to be at
least as informative as  $C^\text{SLD}$ or  $C^\text{RLD}$, i.e. $\mathcal{H}\geq C^\text{SLD},C^\text{RLD}$. The Holevo bound
involves a minimisation over $X=(X_1,X_2,\ldots,X_n)$, where $X_j$ are
Hermitian operators that satisfy the unbiased conditions
\begin{align}
\label{eq_xcon1}
\text{Tr}\left\{\rho_{\theta} X_j\right\}&=0 \;, \\
\label{eq_xcon2}
\text{Tr}\left\{\frac{\partial \rho_\theta}{\partial\theta_j} X_k\right\}&=\delta_{jk}\;.
\end{align}
The Holevo bound is
\begin{align}
\label{eq_hol2}
\mathcal{H} \coloneqq  \min_{X} \text{Tr}\left\{  Z_\theta[X]\right\} +\text{TrAbs} \left\{\Im Z_\theta[X]\right\}\;,
\end{align}
where
\begin{align}
\label{eq_zmat}
Z_\theta[X]_{jk} \coloneqq  \text{Tr} \left\{ \rho X_j X_k\right\}\;,
\end{align}
and  $\text{TrAbs}\{\text{Im}Z_{\theta}[X]\}$ is
the sum of the absolute values of the eigenvalues of the matrix $\text{Im}Z_{\theta}[X]$.
$Z$ takes the role of the inverse of the Fisher information matrix. The Holevo bound sets a limit to the sum of the variance of an unbiased estimate
\begin{align}
  \text{Tr}\{V_\theta\}\geq \mathcal{H}\;.
\end{align}
Holevo derived this bound in his original work~\cite{holevo1976noncommutative}, but the bound in the form shown above was introduced by
Nagaoka~\cite{nagaoka2005new}.  A major obstacle preventing the more
widespread use of the Holevo bound is that, unlike the RLD and SLD
bounds which can be calculated directly, the Holevo bound involves a
non-trivial optimisation problem. However, as mentioned in the introduction, these difficulties have been somewhat alleviated in recent years~\cite{bradshaw2018ultimate,albarelli2019evaluating,sidhu2021tight}.

%The Holevo bound is an upper bound on the accessible information obtained using collective measurements.

 % For two-qubits, we perform this
% optimisation numerically. For three-qubit probes in order to calculate
% the optimal X matrices we make certain simplifying assumptions about
% the symmetry of these matrices.

For estimating two parameters, Nagaoka derived the bound
\begin{align}
\label{eq_hol2N}
  \text{Tr}\{V_\theta\}\geq \mathcal{N} \coloneqq  \min_{X} \text{Tr}\left\{  Z_\theta[X]\right\}
  +\text{TrAbs} \left\{ \rho [X_1,X_2]\right\}\;,
\end{align}
valid if we are restricted to individual measurements. This bound is always
more informative than or equal to the Holevo bound. As mentioned in the introduction, for two-dimensional
systems, it is known to be attainable~\cite{nagaoka2005new} and was conjectured by Nagaoka to be
attainable in higher dimensional systems as well~\cite{nagaoka2005generalization}. For estimating more than two parameters with individual measurements we shall use the NHB~\cite{hayashi1997linear,conlon2021efficient}. The NHB can be computed from the following optimisation problem~\cite{conlon2021efficient}
 \begin{align}
 \text{Tr}\{V_\theta\}\geq\mathcal{N}\coloneqq   \min_{\mathbb{L},\,X}\left\{
                          \bbTr{\mathbb{S}_\theta
                          \mathbb{L}}\,\big|\, \mathbb{L}_{jk}=\mathbb{L}_{kj}\, \mathrm{
       Hermitian, }\, \mathbb{L}\geq {X} X^\intercal
    \right\} \;,\label{eq:NHB}
\end{align}
where $\mathbb{S}_\theta= {1}_n\otimes \rho_\theta$, ${1}_n$ is the $n\times n$ identity matrix\footnote{Note that a different notation is used for identity matrices which exist in the classical vector space, compared to those in the quantum Hilbert space as in Eq.~\eqref{eq:POVMcondition}.} and $X=(X_1,X_2,...,X_n)^\intercal$ is a vector of Hermitian estimator observables $X_j$ that satisfy the locally unbiased conditions, Eqs.~\eqref{eq_xcon1} and \eqref{eq_xcon2}. The matrix $\mathbb{S}_\theta$, exists on an extended quantum-classical Hilbert space. We use the symbol $\bbTr{\cdot}$ to denote trace over both classical and quantum systems. We shall use the symbol $\mathcal{N}$ to denote both the Nagaoka bound and NHB, and it will be obvious from the number of parameters being estimated which we are referring to. We can define the most informative bound as the precision achievable through individual measurements, $C^\text{MI}$, such that for 2 dimensional systems $C^\text{MI}=\mathcal{N}$~\cite{nagaoka2005new}. As mentioned in the introduction, the NHB is not always a tight bound~\cite{hayashi2022tight}. An alternative bound for individual measurements was introduced by Gill and Massar~\cite{gill2005state}. However, as this bound is, in general, less tight than the NHB we shall not consider it in this work.

There is a hierarchy between the bounds described above, $C^\text{MI}\geq\mathcal{N}\geq\mathcal{H}\geq\max(C^\text{RLD},C^\text{SLD})$. However, it is known that for
estimating a single parameter the SLD bound, the Holevo
bound, the NHB and the most informative bound coincide, $C^\text{MI}=\mathcal{N}=\mathcal{H}=C^\text{SLD}$~\cite{suzuki2016explicit,suzuki2019information}. This gives a simple method of
explicitly computing the achievable precision for estimating one
parameter.  Additionally, when estimating any number of parameters using pure states, the Holevo bound and the NHB are equal, $\mathcal{N}=\mathcal{H}$, i.e. the CC and CQ strategies are equivalent, as are the $\text{C}_\text{a}$$\text{C}_\text{a}$ and $\text{C}_\text{a}$$\text{Q}_\text{a}$ strategies. Finally, note that although~the Holevo bound is tighter than the SLD bound, it has recently been proven that the difference between the two is at most a factor of 2~\cite{carollo2019quantumness,tsang2020quantum}. 

\subsection{Quantum Channels}
\label{channel}
In this work, we consider estimating qubit rotations in a noisy channel. There are three rotation parameters we wish to estimate: $\theta_x$, $\theta_y$ and $\theta_z$. These three
parameters describe the rotation amplitudes about the three
axes of the Bloch sphere:
\begin{align}
  \label{eq:three_rotations}
  U(\theta_x,\theta_y,\theta_z) = \exp(\I \frac{\theta_z}{2}\sigma_z) \exp(\I \frac{\theta_y}{2}\sigma_y) \exp(\I \frac{\theta_x}{2}\sigma_x)\;,
\end{align}
where $\sigma_x$, $\sigma_y$ and $\sigma_z$ are the Pauli matrices. We insist that the rotation amplitudes are small so that the order of the rotations does not matter. Note that this assumption of small rotations is relevant provided it is possible to perform some prior characterisation of the quantity to be estimated, such as in Ref.~\cite{li2023optimal}. We are interested in finding the
optimal single-qubit and two-qubit probes for estimating
these parameters. We shall consider the three cases with: (i) a single
rotation when $\theta_y=\theta_z=0$, (ii) two rotations when
$\theta_z=0$ and (iii) all three rotations present. Depending on the number of parameters, the
optimal probe and measurement strategy will be different.
For certain channels, when estimating a single parameter $\theta_x$, as we shall show later, any pure state in the
$y{-}z$ plane will be optimal. 

The rotations transform an input quantum state $\rho$ to the state $ \rho_\theta=U(\theta_x,\theta_y,\theta_z)\rho U(\theta_x,\theta_y,\theta_z)^\dagger$. In our model, this rotated state is then subject to some noise. A noisy quantum channel can be described in the operator-sum representation as a linear map acting on the density matrix of the state subjected to the channel,
\begin{equation}
\label{krauss}
\mathcal{E}_\theta(\rho)=\sum_{a=1}^{K}M_{a}\rho_\theta M_{a}^{\dagger}\;,
\end{equation}
where the operators, $M_{a}$, obey the completeness relation, $\sum_{a=1}^{K}M_{a}^{\dagger}M_{a}=\mathbbm{1}$. In what follows, we shall drop the explicit dependence of the quantum channel on $\theta$. The operators are known as operational elements of the channel, or Kraus
operators~\cite{kraus1983states}. Note that, as is evident from Eq.~\eqref{krauss}, the overall quantum channel we consider involves rotating the quantum state before subjecting it to the noise. $K$ is known as the
Kraus number and satisfies $K\leq d^2$, where $d$ is the dimensions of
the system in question. For the map
$\mathcal{E}(\rho)$ to represent a deterministic, physical channel it
must be linear, trace-preserving and completely
positive~\cite{serafini2017quantum}. A deterministic channel here means
that we do not allow postselection on certain measurement
results. Quantum channels such as this are also called
`trace-preserving completely positive' maps. There are two distinct scenarios when considering estimation in the presence of noise. One is to perform the estimation using noisy probe states, $\mathcal{E}(\rho)$, i.e. the decoherence happens before the rotations. The other is to use pure probe states, where the decoherence occurs after the rotation. In this work we consider the second option. In all cases we assume that the noise parameter is known, i.e. we do not need to treat it as a nuisance parameter as has been considered before~\cite{suzuki2020nuisance,suzuki2020quantum}.
\begin{figure}[t]
\includegraphics[width=1\textwidth]{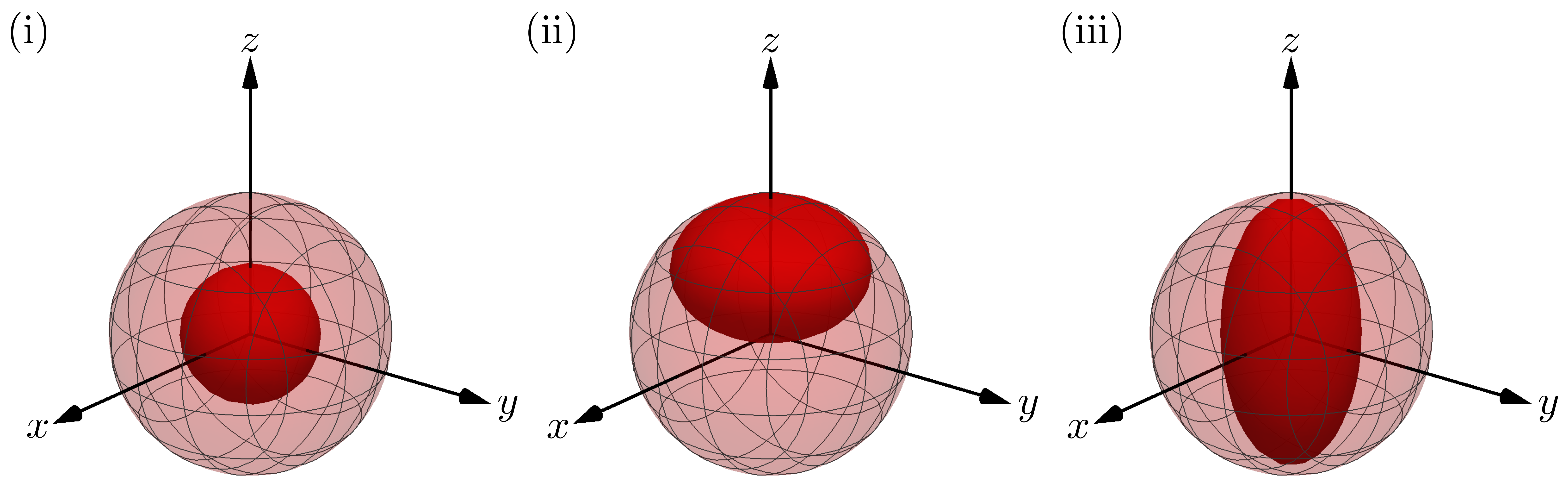}
\caption{\textbf{Effect of the different quantum channels on the Bloch sphere}. The light red spheres and dark red spheroids represent the Bloch sphere for pure states before and after, respectively, passing through a decoherence channel parameterised by a decoherence strength of 0.5. The decoherence channel, amplitude damping channel and phase damping channel are shown in figures i), ii) and iii) respectively.}
\label{fig:BS_effect}
\end{figure}

\section{Results}
We shall now present our results on the optimal estimation variances
for three channels: 1) decoherence channel, 2) amplitude damping channel and 3) phase damping
channel. Fig.~\ref{fig:BS_effect} shows the effect each channel has on the Bloch sphere. These three channels have been considered before in the context of the QFI~\cite{ozaydin2014phase, ma2011quantum}, and their physical motivation includes energy dissipation and decoherence in trapped ions~\cite{huelga1997improvement, myatt2000decoherence, turchette2000decoherence, nielsen2002quantum}. For each channel, we present
the optimal achievable precisions for estimating one and two qubit rotations with single qubit probes and one, two and three
qubit rotations with two qubit probes. For single parameter estimation with single qubit probes, the optimal probe state is determined from the SLD bound, Eq.~\eqref{eq:SLDbound}. In all other scenarios we find the optimal probe numerically. We compare the performance of using an entangled
two-qubit probe, where only one qubit passes through the channel, with that of having only a single qubit probe. Thus, in our work, computing the NHB for the single qubit and two-qubit probe states, corresponds to the CC and $\text{C}_\text{a}$$\text{C}_\text{a}$ strategies respectively. Computing the Holevo bound for the single qubit and two-qubit probe states, corresponds to the CQ and $\text{C}_\text{a}$$\text{Q}_\text{a}$ strategies respectively. When we wish to compute the precision attainable with collective measurements on a finite number of copies, $M$, of the probe state, we shall evaluate the NHB for the state $\rho^{\otimes M}$, scaled by a factor of $M$ to ensure a fair comparison in terms of resources used. Note that the Holevo bound for the state $\rho^{\otimes M}$, scaled by a factor of $M$, also provides a bound on the precision attainable with collective measurements on $\rho^{\otimes M}$. However, this bound is only guaranteed to be tight as $M\to\infty$. In many of the problems we consider we are able to obtain analytic solutions for the matrices which optimise the Holevo and Nagaoka bounds based on the symmetries of the system. However, an analytic solution was not always possible and in this case a combination of numerics and guesswork was used to obtain results. %All of the results presented in the main text 
%We present numerical simulations which support our analytic results.% {\color{red}The probe space is convex}, and both the NHB and Holevo-bound are convex functions, so we are guaranteed that there exist no local minima, so finding the optimal probe is easy numerically.

\subsection{Decoherence channel}
\label{firstres}
We first explicitly compute the achievable precisions for individual and
collective measurements with one and two-qubit probes under
the action of a decoherence channel.  This channel is represented
by the following Kraus operators
\begin{align}
\begin{array}{rclcrcl}
    M_{0}&=&\sqrt{1-\frac{3\epsilon}{4}}\mathbbm{1}&,&M_{1}&=&\sqrt{\frac{\epsilon}{4}}\sigma_{x}\;,\\
M_{2}&=&\sqrt{\frac{\epsilon}{4}}\sigma_{y}&, &M_{3}&=&\sqrt{\frac{\epsilon}{4}}\sigma_{z}\;,
\end{array}
\end{align}
where $0\leq \epsilon \leq 1$ parametrises the decoherence strength.
These Kraus operators act only on the probe qubit as we wish to consider the situation where only a single
qubit passes through the channel \cite{preskillnotes}, as shown in Fig.~\ref{fig:scheme}.

\subsubsection{Single qubit probe}
\label{singqub}
\begin{figure}
\includegraphics[width=1\textwidth]{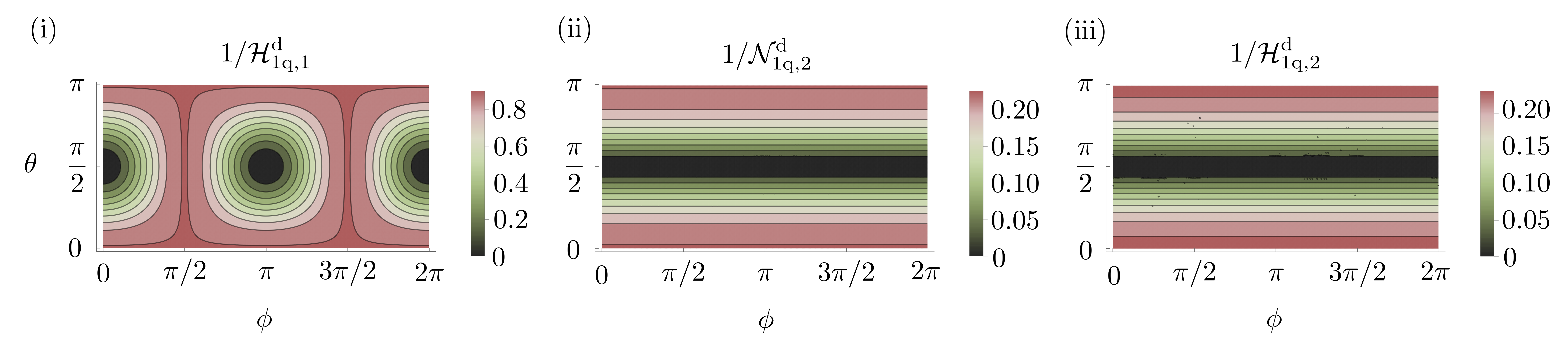}
    \caption{{\bf Parameter estimation in a decoherence channel with single qubit probes on the Bloch sphere, $\boldsymbol{\ket{\psi}=\text{cos}(\theta/2)\ket{0}+e^{\I\phi}\text{sin}(\theta/2)\ket{1}}$}.
  i) Optimal achievable precision for estimating a single parameter obtained using the SLD bound, Eq.~\eqref{eq:SLDappendix}. The optimal probes lie in the Y-Z plane of the Bloch sphere. ii), iii) Optimal achievable precision for estimating two parameters simultaneously when using individual and collective measurements respectively obtained using the semidefinite programmes put forward in Refs.~\cite{conlon2021efficient} and \cite{albarelli2019evaluating}. These plots show that the state $\ket{\psi}=\ket{0}$, which is one of many optimal states for estimating a single parameter, remains optimal for estimating two parameters. Contours are shown for $\epsilon=0.05$. }%
    \label{BS_DC}%
\end{figure}

With a single qubit, the optimal probe for sensing a rotation about
the $x$-axis is any pure state in the Y-Z plane of the Bloch sphere. This is easily verified as, for estimating a single parameter, the Holevo bound coincides with the SLD bound, Eq.~\eqref{eq:SLDbound}. For the probe $\ket{\psi}=\text{cos}(\theta/2)\ket{0}+e^{\I\phi}\text{sin}(\theta/2)\ket{1}$ and a particular choice of the decoherence parameter, $\epsilon=0.05$, we show the Holevo bound as a function of the Bloch sphere angles, $\theta$ and $\phi$, in Fig.~\ref{BS_DC} (i). This figure verifies our claim that any probe in the Y-Z plane ($\phi=\pi/2,3\pi/2$) is optimal. The computation of the SLD bound is described in appendix~\ref{apennewSLD}. We now consider the state $\ket{0}$, one of the many possible optimal states. After the decoherence channel this probe is left in the the state
\begin{equation}
\rho=\ket{0}\bra{0}(1-\epsilon)+\frac{\epsilon}{2}\;.
\end{equation}
%When estimating a single parameter we use the SLD bound to compute the optimal precision. 
For a single qubit the optimal $X_{x}$ matrix for estimating a rotation about the
$x$-axis is
\begin{equation}
\label{xone}
X_{x}=\frac{\I}{1-\epsilon}\left(\ket{0}\bra{1}-\ket{1}\bra{0}\right)\;.
\end{equation}
We show how $X_{x}$ is computed in appendix \ref{apenxmat}. Using this matrix, the Holevo bound is given by
\begin{equation}
\label{Hol1}
v_x\geq \mathcal{H}^\text{d}_{1 \text q,1} =\frac{1}{(1-\epsilon)^{2}}\;.
\end{equation}
We use the superscript `d' to denote the decoherence channel, the first subscript `1q'
to denote a one-qubit probe and the second subscript `1' to denote single parameter estimation.
%In this case the SLD bound is tight. %but the RLD bound is not,
%$v_{1,\text{RLD}}(\theta_{x})=\frac{2\epsilon
%  (1-\frac{\epsilon}{2})}{(1-\epsilon)^{2}}$. 
For single parameter estimation, an individual measurement
  provides as much information as a collective measurement~\cite{ragy2016compatibility}.  Hence
  \begin{align}
    \label{NagD1}
    v_x \geq \mathcal{N}^\text{d}_{1\text q,1} = \frac{1}{(1-\epsilon)^2}\;,
  \end{align}
where we use $\mathcal{N}$ to indicate the Nagaoka bound. In appendix~\ref{apenDC1q} we show an explicit measurement reaching this bound. 
We now consider estimating two parameters, rotations about the $x$ and $y$ axes. The $\ket{0}$ probe is
still optimal as shown in Fig.~\ref{BS_DC} (ii). The
optimal precision with individual measurements can be obtained using
the Nagaoka bound and is given by
\begin{equation}
\label{Nag12}
v_x+v_y \geq \mathcal{N}^\text{d}_{1\text q,2}=\frac{4}{(1-\epsilon)^{2}}\;,
\end{equation} 
which is exactly four times the single parameter Nagaoka bound. This
implies that an optimal strategy for estimating both parameters is to
estimate each parameter separately: use $N/2$ probes to estimate $\theta_x$
and the remaining $N/2$ probes to estimate $\theta_y$. We show a measurement strategy reaching this bound in appendix~\ref{apenDC1q}. The  $X$-matrix which is required to achieve this Nagaoka bound is given by
\begin{equation}
X_{y}=\frac{1}{1-\epsilon}\left(\ket{0}\bra{1}+\ket{1}\bra{0}\right)\;.
\end{equation}
The same $X$-matrix also optimises the Holevo bound, giving
\begin{equation}
\label{Hol12}
v_x+v_y \geq \mathcal{H}^\text{d}_{1\text q,2}=\frac{4-2\epsilon}{(1-\epsilon)^{2}}\;,
\end{equation}
which is slightly smaller than the corresponding Nagaoka bound when $\epsilon$ is
not equal to 0 or 1. Thus, in this case, we can achieve better
precision by performing a collective measurement. The optimal strategy
gives
\begin{align}
v^*_x=v^*_y = \frac{1}{2}\mathcal{H}^\text{d}_{1\text q,2}=\frac{2-\epsilon}{(1-\epsilon)^2}\;,
\end{align}
which is always greater than $\mathcal{H}^\text{d}_{1\text q,1}$. This
indicates that by estimating the second parameter we lose some precision in our
estimate of the first parameter as can be expected.% As the Holevo bound does not equal the Nagaoka bound for any $0<\epsilon<1$, the gap persistence theorem shows that the Holevo bound cannot be saturated by any physical measurement in this scenario~\cite{conlon2022gap}.

%
%If we try to estimate a third parameter using the single qubit probe
%the performance deteriorates further, We can show that for a general
%one qubit probe $\ket{\psi}=\ket{0}a+\ket{1}b$, it is impossible to
%construct an unbiased estimator for estimating a rotation about all
%three axis, if $a$ and $b$ are both real numbers. This is
%because $\partial \rho /\partial \theta_z$ is a multiple of
%$\partial \rho /\partial \theta_x$
%\begin{equation}
%\frac{\partial \rho}{\partial \theta_x}= \frac{\partial \rho}{\partial \theta_z}\left(\frac{b^{2}-a^{2}}{2ab}\right).
%\end{equation}
%Thus,it is impossible to construct an unbiased estimator to
%simultaneously estimate $\theta_{x}$ and $\theta_{z}$ with a single
%qubit with real coefficients---every estimator is biased. We extend this result in the appendix. %~\comment{(What about
%  %a probe with complex coefficients, pointing to (1,1,1)?)} \change{I think even with complex coefficients we still can't construct an unbiased estimator}

\subsubsection{Two-qubit probe}

\begin{figure}%
   \includegraphics[width=1\textwidth]{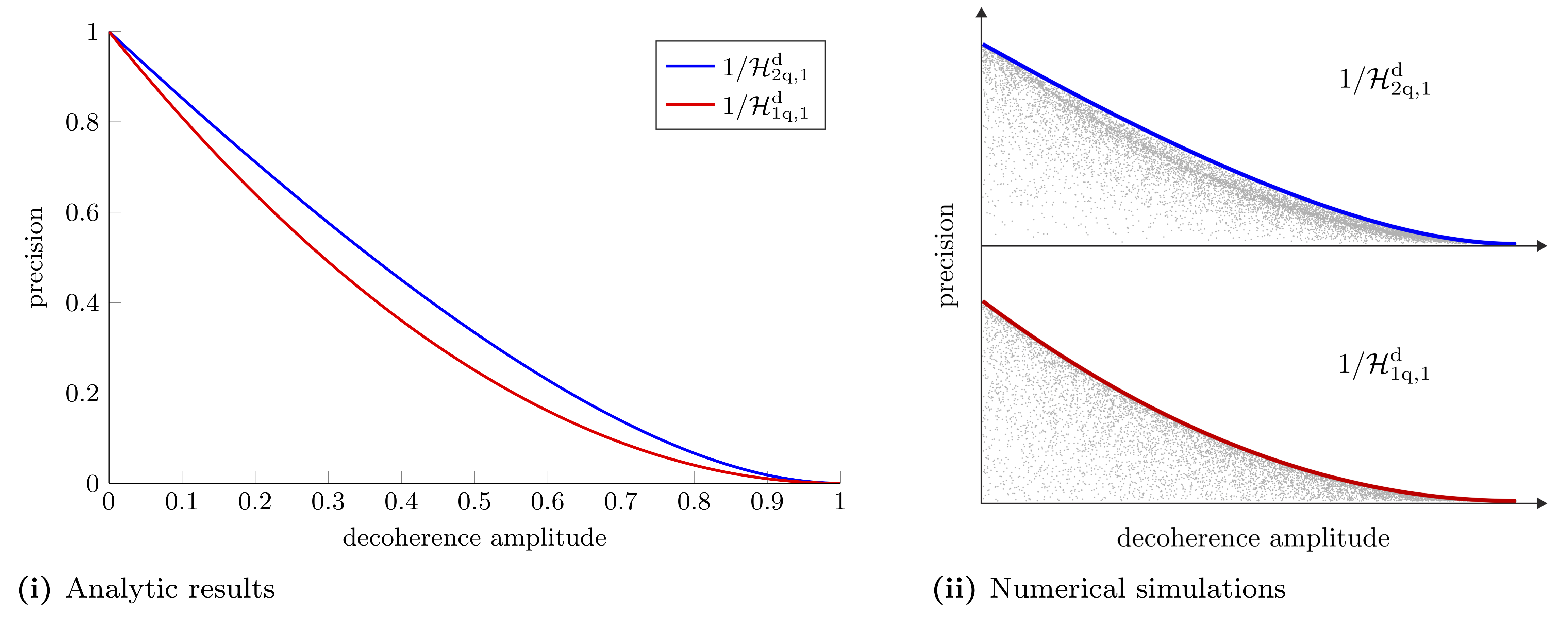}
    \caption{{\bf Single parameter estimation in a decoherence channel.}
  (i) Optimal achievable precision for estimating a single parameter using one
  and two-qubit probes. These are given by the Holevo bounds Eqs.~\eqref{Hol1} and \eqref{Hol2qDC}
  respectively. (ii) Numerical simulations supporting the analytic results. For each plot in (ii) 10,000 random pure state probes are tested as the decoherence amplitude is varied from 0 to 1. The grey dots represent the precision from the random probes and the coloured lines show the precision from the optimal probes.}%
    \label{Indv_singl}%
\end{figure}

%\begin{figure}[t]
%  \input{figures/Indv_singl.tikz}
%\caption{{\bf Single parameter estimation in a decoherence channel.}
%  Optimal achievable precision for estimating a single parameter using one
%  and two-qubit probes. These are given by the Holevo bounds~(\ref{Hol1}) and (\ref{Hol2qDC})
%  respectively. The Holevo bounds can be attained by an individual
%  measurement which means that collective measurements do not offer
%  any advantage.}
%\label{Indv_singl}
%\end{figure}
For estimating one, two or three rotations with a two-qubit probe, we numerically verify that any maximally
entangled two-qubit probe is equally optimal. Note that, in a slightly different setting, the maximally entangled state was proven to be optimal in the noiseless case~\cite{fujiwara2001estimation}. We consider the following probe
\begin{align}
\ket{\psi_{0}} &=\frac{1}{\sqrt{2}} \left(\ket{01}+\ket{10}\right)\;.
\label{2qstart}
\end{align}
After passing through the channel the probe becomes
\begin{equation}
\rho=(1-\epsilon)\ket{\psi_{0}}\bra{\psi_{0}}+\frac{\epsilon}{4}\mathbbm{1}\;.
\end{equation}
For estimating one parameter with this probe the optimal $X$ matrix is
\begin{equation}
\label{xtwo}
X_{x}=\frac{\I}{(1-\epsilon)}\left(\ket{\psi_{0}}\bra{\psi_{2}}-\ket{\psi_{2}}\bra{\psi_{0}}\right)\;,
\end{equation}
where
\begin{align}
\ket{\psi_{2}} &=\frac{1}{\sqrt{2}} \left(\ket{00}+\ket{11}\right)\;.
\end{align}
The bounds for both variances, using individual and collective measurements, are given by 
\begin{equation}
\label{Hol2qDC}
v_{x} \geq \mathcal{H}^\text{d}_{2\text q,1}=\mathcal{N}^\text{d}_{2\text q,1}=\frac{2-\epsilon}{2(1-\epsilon)^{2}}\;,
\end{equation}
which coincides with the SLD bound. In appendix~\ref{app2q2cDC} we show that by performing a measurement on both qubits we can achieve this precision. We note that $\mathcal{H}^\text{d}_{2\text
  q,1}\leq\mathcal{H}^\text{d}_{1\text q,1}$ which indicates that using a
two-qubit probe gives a better estimation precision even though the
second probe does not go through the channel. We quantify the
estimation precision as
\begin{align}
  \label{eq:4}
  \text{precision} = \frac{1}{v^*_x}\;,
\end{align}
where $v^*_x$ is the variance obtained from the optimal strategy. We compare single-qubit and two-qubit probes for estimating a single parameter in Fig.~\ref{Indv_singl}.

We now proceed to estimating two parameters using an entangled
two-qubit probe. Similar to the single parameter case, we can write
%\begin{align}
%\frac{\partial \rho}{\partial \theta_y}&=\frac{(1-\epsilon)}{2}[\ket{\psi_{0}}\bra{\psi_{3}}+\ket{\psi_{3}}\bra{\psi_{0}}]\\
%\frac{\partial \rho}{\partial \theta_z}&=\frac{i(1-\epsilon)}{2}[\ket{\psi_{1}}\bra{\psi_{0}}-\ket{\psi_{0}}\bra{\psi_{1}}]\;,
%\end{align}
\begin{align}
X_{y}=\frac{-1}{(1-\epsilon)}\left(\ket{\psi_{0}}\bra{\psi_{3}}+\ket{\psi_{3}}\bra{\psi_{0}}\right)\;,\\
X_{z}=\frac{-\I}{(1-\epsilon)}\left(\ket{\psi_{1}}\bra{\psi_{0}}-\ket{\psi_{0}}\bra{\psi_{1}}\right)\;,
\label{xzone}
\end{align}
where
\begin{align}
  \ket{\psi_{1}} &=\frac{1}{\sqrt{2}} \left(\ket{01}-\ket{10}\right) \;, \\
\ket{\psi_{3}} &=\frac{1}{\sqrt{2}} \left(\ket{00}-\ket{11}\right)\;,
\label{2qend}
\end{align}
to arrive at
\begin{align}
\label{Hol22DCjoint}
v_x+v_y \geq  \mathcal{H}^\text{d}_{2\text{q},2} =  \frac{2-\epsilon}{(1-\epsilon)^2}\;.
\end{align}
The optimal variance of each parameter is
\begin{align}
\label{Hol22DCindividual}
  v^*_x=v^*_y= \frac{1}{2}   \mathcal{H}^\text{d}_{2\text{q},2} =  \frac{2-\epsilon}{2(1-\epsilon)^2}\;,
\end{align}
which is exactly equal to the single parameter result
$\mathcal{H}^\text{d}_{2\text q,1}$ in Eq.~\eqref{Hol2qDC}. Hence, we
find that with a two-qubit probe, we can estimate a second parameter
without any degradation in the precision of the first. For the two-qubit probe we are also able to compute the Nagaoka bound,
\begin{equation}
\label{Nag22DC}
v_x+v_y \geq \mathcal{N}^\text{d}_{2\text{q},2}=\frac{4-\epsilon}{2(1-\epsilon)^{2}}\;.
\end{equation}
For a general $\epsilon$, this Nagaoka bound is larger than the corresponding Holevo
bound indicating that an individual measurement is inferior to a
collective measurement. In appendix \ref{app2q2cDC} we construct a measurement scheme which saturates this bound.

\begin{figure}%
\includegraphics[width=1\textwidth]{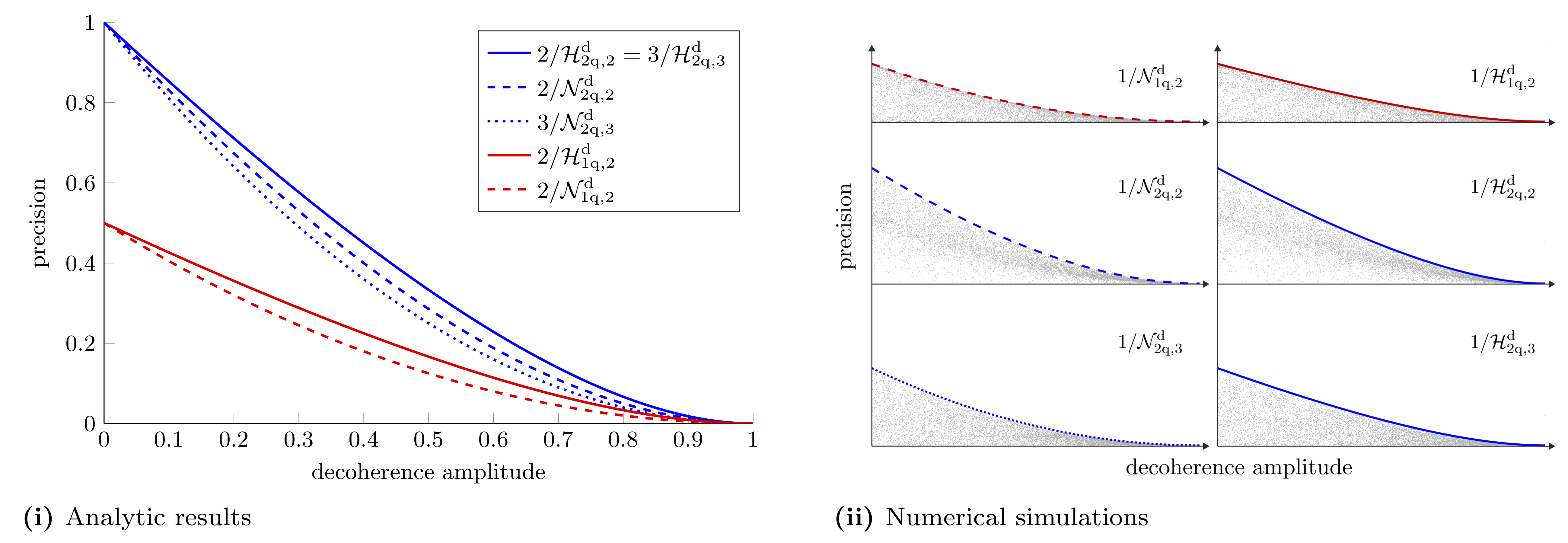}
  \caption{{\bf Multiple parameter estimation in a decoherence channel.}
   (i) Holevo bounds and NHBs with one and two
    qubit probes for simultaneously estimating multiple parameters
    in a decoherence channel. With the single qubit probe, for estimating two parameters, the
   Nagaoka bound and Holevo bound are given by Eqs.~\eqref{Nag12} and
    \eqref{Hol12} respectively. For the two qubit probe the Holevo and
    Nagaoka bounds for estimating two parameters are given by Eqs.~\eqref{Hol22DCjoint} and \eqref{Nag22DC}
    respectively. For estimating three parameters the Holevo bound and NHB are given by Eqs~\eqref{Hgen3DC} and \eqref{Ngen3DC} respectively. (ii) Numerical simulations to support analytic results. As before, each plot shows the precision attained with 10,000 random probes, which always lies below the precision attained with the optimal probe.}
\label{Indv_mult}
\end{figure}

 Finally, for estimating three parameters, with the optimal $X_{z}$ matrix given in Eq.~\eqref{xzone}, we find that
\begin{align}
\label{Hgen3DC}
v_x+v_y+v_z \geq  \mathcal{H}^\text{d}_{2\text{q},3} =  \frac{6-3\epsilon}{2(1-\epsilon)^2}\;,
\end{align}
and the optimal variance of each parameter is
\begin{align}
  v^*_x=v^*_y=v^*_z =\frac{1}{3}   \mathcal{H}^\text{d}_{2\text{q},3} =  \frac{2-\epsilon}{2(1-\epsilon)^2}\;.
\end{align}
This is once again equal to the single parameter result
$\mathcal{H}^\text{d}_{2\text q,1}$ in Eq.~(\ref{Hol2qDC}). Hence, we
find that with a two-qubit probe, we can estimate all three parameters
simultaneously just as well as we can estimate just one parameter. The ability of entangled probe states to avoid trade-offs in multiparameter estimation has been observed before~\cite{baumgratz2016quantum,bradshaw2017tight,bradshaw2018ultimate}.
However, when restricted to individual measurements the NHB is given by
\begin{align}
v_x+v_y+v_z \geq  \mathcal{N}^\text{d}_{2\text{q},3} =  \frac{3}{(1-\epsilon)^2}\;,
\label{Ngen3DC}
\end{align}
and the optimal variance of each parameter is
\begin{align}
  v^*_x=v^*_y=v^*_z =\frac{1}{3}   \mathcal{N}^\text{d}_{2\text{q},3} =  \frac{1}{(1-\epsilon)^2}\;.
\end{align}
We see that when we are restricted to individual measurements the estimation of another parameter further degrades measurement precision. The same $X$ matrices which optimise the Holevo bound, optimise the NHB. In appendix~\ref{decohapen3} we show that there exists a measurement strategy which saturates the NHB in this case. The differences between individual and collective measurement precisions
are highlighted in Fig.~\ref{Indv_mult}. This figure shows the hierarchy
between the different schemes mentioned earlier in the text:
$\text{C}_\text{a}$$\text{Q}_\text{a}$$\geq \{\text{C}_\text{a}\text{C}_\text{a}\text{,CQ}\} \geq$ CC. For this particular example we find that $\text{C}_\text{a}$$\text{C}_\text{a}$$\geq$CQ.

\begin{figure}[t]
\includegraphics[width=0.9\textwidth]{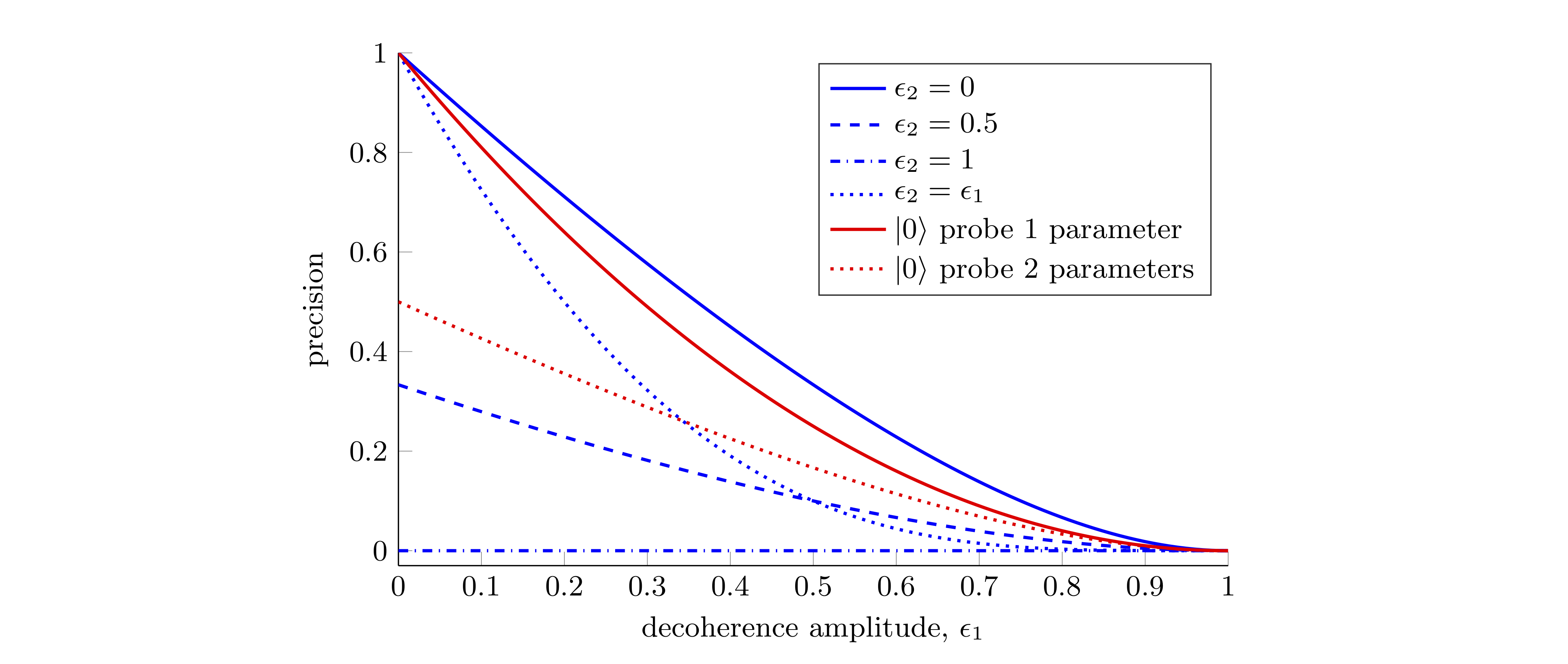}
  \caption{{\bf Two parameter estimation in a decoherence channel affecting both qubits.}
   Precision that can be achieved with one and two
    qubit probes for simultaneously estimating one or two parameters
    in a channel where both qubits experience some decoherence. For the single qubit probe the
    Holevo bounds for estimating one and two parameters are given by Eqs.~(\ref{Hol1}) and
    (\ref{Hol12}) respectively. For the two qubit probe the Holevo bound is given by Eq.~(\ref{2qdecoh}).}
\label{Indv_decoh_DC}
\end{figure}

\subsubsection{Decoherence of both qubits}
Thus far we have considered the case where the second qubit experiences no decoherence. This lack of decoherence is equivalent to storing the second qubit in a perfect quantum memory, something which is not feasible with current technology.
Thus, we now consider the channel where both qubits in the two qubit probe experience some decoherence. We expose the two qubit probe to the channel where the first and second qubit experience decoherence amplitudes of $\epsilon_{1}$ and $\epsilon_{2}$ respectively. Under the action of this channel for estimating either one or two parameters the maximally entangled two qubit probe achieves a Holevo bound of 
\begin{equation}
\label{2qdecoh}
\mathcal{H}^\text{2d}_{2\text q,1}=\frac{1}{2}\mathcal{H}^\text{2d}_{2\text q,2}=\frac{1-\frac{1}{2}(\epsilon_{1}+\epsilon_{2})+\frac{1}{2}\epsilon_{1}\epsilon_{2}}{(1-\epsilon_{1})^{2}(1-\epsilon_{2})^{2}}\;.
\end{equation}
Although we have shown the advantages offered by two qubit probes over single qubit probes, this expression highlights the dangers associated with using highly entangled probes for estimation. Eq~(\ref{2qdecoh}) is symmetric in $\epsilon_{1}$ and $\epsilon_{2}$, which is somewhat surprising given that the rotation we are trying to estimate acts on the first qubit only. In spite of this, decoherence of the second qubit is equally damaging to our estimation ability. We see that when the second qubit is fully decohered we are unable to estimate with any precision at all, regardless of the decoherence of the first qubit. This is shown in Fig.~\ref{Indv_decoh_DC}

Perhaps the most physically relevant scenario is the one in which both qubits experience the same decoherence, i.e. $\epsilon_{1}=\epsilon_{2}$. In this situation for estimating a single parameter the single qubit probe described in section \ref{singqub} always outperforms the two qubit probe considered. For estimating two parameters with the two qubit probe the Holevo bound, Eq~(\ref{2qdecoh}), remains unchanged. Even though, when using the maximally entangled two qubit probe, we can estimate a second parameter for free, when the noise in the system is sufficiently high the single qubit probe can still outperform the two qubit probe for estimating two parameters. Thus, we can see that the optimal probe to use in this instance depends on the decoherence amplitudes experienced by both qubits. It is worth noting that in this high noise regime a different two qubit probe will be optimal and the optimised two qubit probe will always perform better than or equal to the optimised single qubit probe. The fact that the maximally entangled two qubit probe is no longer optimal is a reflection of the fact that highly entangled states are very susceptible to loss and noise, see Refs.~\cite{dorner2009optimal,demkowicz2012elusive}.

\begin{figure}[t]
\includegraphics[width=0.7\textwidth]{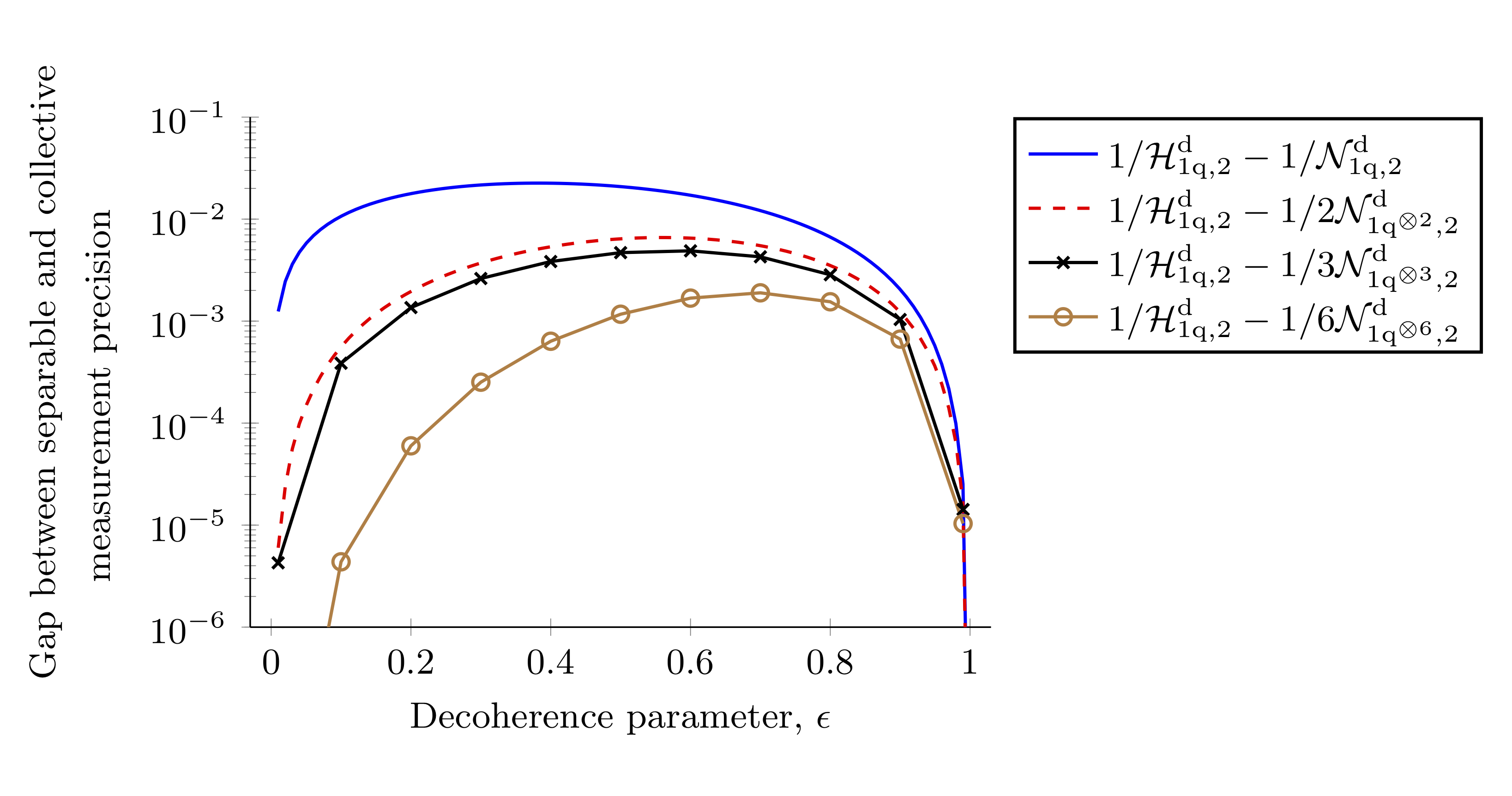}
  \caption{\textbf{Achievable precision using collective measurements versus individual measurements in the decohering channel.} Difference between the inverse of the Holevo bound for a single qubit probe and the inverse of the Nagaoka bound for \textit{M} copies of the same probe. As \textit{M} increases the Nagaoka bound tends towards the Holevo bound as expected. For one and two copies of the probe the Nagaoka bound is obtained analytically, for three and six copies of the probe we numerically obtain the Nagaoka bound for $\epsilon$ in the range 0.01$\rightarrow$0.99. \label{decohmcop}}
\end{figure}

\subsubsection{Collective measurements on multiple copies of the state}
Before concluding discussions on the decoherence channel, we consider what happens when we can perform collective measurements on \textit{M} copies of the single qubit probe, i.e. we have the state $\rho^{\otimes \textit{M}}$ available to measure. As $\textit{M}\rightarrow\infty$, we expect to find $\mathcal{N}_{1q^{\otimes \textit{M}}, 2}^{d}\rightarrow\mathcal{H}_{1q^{\otimes \textit{M}},2}^{d}=\frac{1}{\textit{M}}\mathcal{H}_{1q,2}^{d}$, where we have used the additivity of the Holevo bound~\cite{hayashi2008asymptotic}. Let us first consider $\textit{M}=2$. With two copies the probe becomes $\rho^{\otimes 2}=\text{diag}\{\left(1-\frac{\epsilon}{2}\right)^2, \frac{\epsilon}{2}\left(1-\frac{\epsilon}{2}\right), \frac{\epsilon}{2}\left(1-\frac{\epsilon}{2}\right), \frac{\epsilon^2}{4}\}$. The derivatives of this matrix with respect to the parameters of interest are given by:
\begin{equation}
  \begin{gathered}
  \frac{\partial \rho_\theta}{\partial \theta_x}=
  \begin{pmatrix}0&\I A&\I A&0\\
   -\I A&0&0&\I B\\
    -\I A&0&0&\I B\\
    0&-\I B&-\I B&0
  \end{pmatrix}\;\text{and}\qquad
  \frac{\partial \rho_\theta}{\partial \theta_y}=\frac{1}{4}
  \begin{pmatrix}0&A&A&0\\
   A&0&0& B\\
   A&0&0& B\\
    0& B& B&0
  \end{pmatrix}\;,
  \end{gathered}
\end{equation}
where $A=\frac{1}{2}(1-\epsilon)\left(1-\frac{\epsilon}{2}\right)$ and $B=\frac{1}{4}(1-\epsilon)\epsilon$. For this state the optimal $X_x$ and $X_y$ matrices are given by
\begin{equation}
  \begin{gathered}
  X_x=
  \frac{\I}{2(1-\epsilon)}
  \begin{pmatrix}0&1& 1&0\\
   -1&0&0&1\\
    -1&0&0&1\\
    0&-1&-1&0
  \end{pmatrix}\;\text{and}\qquad
  X_y=\frac{1}{2(1-\epsilon)}
  \begin{pmatrix}0&1&1&0\\
   1&0&0&1\\
   1&0&0&1\\
    0&1&1&0
  \end{pmatrix}\;,
  \end{gathered}
\end{equation}
which give a Nagaoka bound of $\mathcal{N}_{1q^{\otimes 2}, 2}^{d}=(2-\epsilon+\frac{\epsilon^2}{2})/(1-\epsilon)^2$. So it is clear that with just two copies of the probe the Nagaoka bound is close to the Holevo bound. A similar result has been observed recently for optical magnetometry systems~\cite{friel2020attainability} and a measurement saturating the two-copy Nagaoka bound has been implemented experimentally~\cite{conlon2023approaching}. In appendix~\ref{app:1q2cDC} we show a measurement scheme which attains this bound. With the development of recent techniques, it is now possible to compute the Nagaoka bound efficiently~\cite{conlon2021efficient}. This allows us to compute the Nagaoka bound for many copies of the single qubit probe. In Fig.~\ref{decohmcop} we compare the Holevo bound to the Nagaoka bound for an increasing number of copies of the single qubit probe. We plot the difference in achievable precisions as a measure of how close the two bounds are. As expected, with an increasing number of copies of the probe state, the Nagaoka bound tends to the Holevo bound. However, these results are purely theoretical and in an experimental implementation with current capabilities such precisions can only be reached for measurements on a limited number of copies of the probe state~\cite{conlon2023approaching}. Note that if we are restricted to performing separable measurements (non-entangling POVMs) on the state $\rho^{\otimes M}$, it is known that there is no advantage compared to individual measurements~\cite{hayashi2016quantum}.

\newpage
\subsection{Amplitude Damping Channel}
We now consider an amplitude damping channel. This channel models
the decay of a two level atom from the excited state to the ground
state. The Kraus operators for this channel are:
\begin{equation}
\label{ampeq}
M_{0}=
\begin{pmatrix}
1&0\\
0&\sqrt{1-p}\\
\end{pmatrix}\;,\quad
M_{1}=
\begin{pmatrix}
0&\sqrt{p}\\
0&0\\
\end{pmatrix}\;.
\end{equation}
These operators model an atom which if it is in the excited state will
decay to the ground state with probability $p$, and if it is in
the ground state will remain unaffected. Thus, when we consider this
channel we get significantly different variances depending on the
probe we chose for the single qubit case. For example, the
$\ket{0}$ and $\ket{1}$ states, corresponding to the ground and
excited states respectively, will behave differently
depending on the decoherence amplitude~\cite{preskillnotes}. If we
wish to consider the time evolution of an atomic system we can
imagine applying these operators to our quantum state once per time
interval. In each time interval the atomic system has a certain
probability of decaying and as $t\rightarrow\infty$ all of the atoms
end up in the ground state.

\subsubsection{Single qubit probe}
\begin{figure}
\includegraphics[width=1\textwidth]{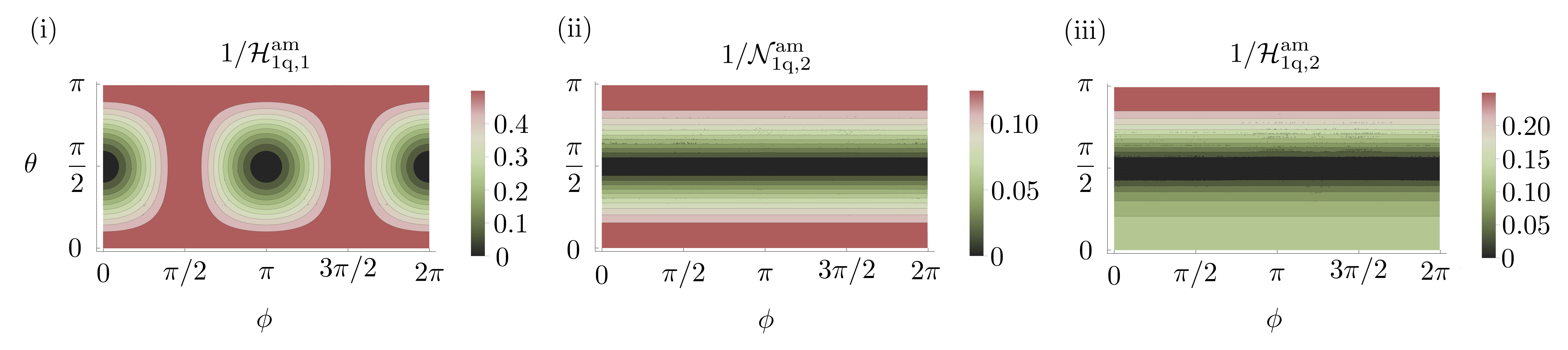}
\caption{\textbf{Parameter estimation with a single qubit probe in the amplitude damping channel with the probe $\boldsymbol{\ket{\psi}=\text{cos}(\theta/2)\ket{0}+e^{\I\phi}\text{sin}(\theta/2)\ket{1}}$.} i) The precision achievable when estimating a rotation about the $x$-axis. ii) and iii) show the precision achievable for simultaneously estimating rotations about the $x$ and $y$-axis when using individual measurements on each copy of the probe state and collective measurements on multiple copies of the probe state respectively. For these figures $p=0.5$ and we can see that when estimating two parameters the $\ket{1}$ state ($\theta=\pi$) outperforms the $\ket{0}$ state ($\theta=0$) only if we allow for collective measurements. This is in spite of the fact that the $\ket{1}$ state experiences the decoherence whereas the $\ket{0}$ state does not.}
\label{fig:AD_xy_cont}
\end{figure}

Similar to the decoherence channel in the single qubit case, the optimal probe for sensing a rotation about
the $x$-axis is any pure state which lies in the Y-Z plane of the Bloch sphere, shown in Fig.~\ref{fig:AD_xy_cont} (i).
These probes are able to estimate a
single parameter with a Holevo bound of
\begin{equation}
v_x\geq \mathcal{H}^\text{am}_{1\text{q},1}=\frac{1}{1-p}\;.
\label{adfirst}
\end{equation}
At first sight, it might seem strange that the state $\ket{0}$ which
is unaffected by the channel will perform just as well as $\ket{1}$
which decays through the channel. The reason for this is that the
probe undergoes the rotation before the decoherence. After the information
has already been encoded onto the probe, the amplitude damping channel
destroys the information at the same rate for both probes. This can be
seen from the derivatives
\begin{align}
  \frac{\partial }{\partial \theta_x}\mathcal{E}\left(\ket{0}\bra{0}\right)=
  -\frac{\partial }{\partial \theta_x}\mathcal{E}\left(\ket{1}\bra{1}\right)=
   -\frac{\sqrt{1-p}}{2} \sigma_y\;,
\end{align}
which only differ in sign for the two probes. Here $\mathcal{E}$
represents the amplitude damping channel. The $X$ matrix which saturates this Holevo bound is 
\begin{equation}
\label{xxampd}
X_{x}=\pm \frac{1}{\sqrt{1-p}}\sigma_y\;,
\end{equation}
where the $+$ is for the $\ket{1}$ state and the $-$ is for the $\ket{0}$ state. For estimating a single parameter using these probes the Holevo bound is equal to the SLD bound.

For two parameter estimation using individual measurements, the two probes $\ket{0}$ and $\ket{1}$
remain optimal, as shown in Fig~\ref{fig:AD_xy_cont} (ii). We find that for either probe, the Nagaoka bound is given by
\begin{equation}
v_x+v_y \geq \mathcal{N}^\text{am}_{1\text q,2}=\frac{4}{1-p}\;,
\label{amp1nagjoint}
\end{equation}
which is exactly four times the optimal single parameter estimate. In appendix~\ref{apenAD1q} we describe the measurement strategy required to reach this bound. The optimal
strategy is to use half of the probes to estimate $\theta_x$ and the
remaining half to estimate $\theta_y$. The variance in each estimate is
\begin{equation}
v_x^{*}=v_y^{*} =\frac{1}{2} \mathcal{N}^\text{am}_{1\text q,2}=\frac{2}{1-p}\;.
\label{amp1nag}
\end{equation}

Interestingly, when allowing for collective measurements, the optimal probe for
estimating a rotation about the $x$ and $y$ axes is the state
$\ket{1}$, which now outperforms the state $\ket{0}$. This is
very surprising since the $\ket{1}$ probe is affected by the
channel, while the $\ket{0}$ probe isn't. This can be viewed as \textit{decoherence assisted metrology}. We can understand this phenomenon by observing that although the state $\ket{0}$ does not experience any decoherence, the rotated state $\ket{0}+(\theta_y-\mathrm{i}\theta_x)/2\ket{1}$ does experience decoherence. Fig.~\ref{fig:AD_xy_cont} (iii) depicts the difference between the two probes $\ket{0}$ and $\ket{1}$.The difference in the probe
performance can be attributed to the difference in the partial
derivatives of the probe after the channel. The Holevo bounds for these two probes apply in the asymptotic limit, but this limiting behaviour is already present if we consider a collective measurement on two probes. We can see that
\begin{align}
  \label{eq:3}
  \frac{\partial}{\partial \theta_x}( \rho \otimes \rho) =
  \frac{\partial \rho}{\partial \theta_x} \otimes \rho+
  \rho \otimes \frac{\partial \rho}{\partial \theta_x}\;,
\end{align}
will be different for the probes $\ket{0}$ and $\ket{1}$ even if they
have the same $\partial \rho/\partial \theta_x$. We will return to this in section 4. The probe $\ket{1}$ achieves a Holevo bound for estimating two parameters of\footnote{The Holevo bound obtained from the suboptimal probe $\ket{0}$ is
$4/(1-p)$ which coincides with the result for an individual
measurement. Thus, collective measurements do not provide any advantage
when using the probe $\ket{0}$.}
\begin{equation}
  v_x+v_y \geq \mathcal{H}^\text{am}_{1\text q,2}=
  \bigg\{\begin{array}{lll}
           4&\text{for}&p \leq 1/2\;,\\
           \frac{4p}{1-p}&\text{for}&p > 1/2\;.
                                      \end{array}
                                     \;.
\label{amphol2joint}
\end{equation}
For every $0<p<1$, a collective measurement will give greater
precision compared to an individual measurement.
The minimum variance attained by the optimal probe $\ket{1}$ is given by
\begin{equation}
  v_x^{*}=v_y ^{*} = \frac{1}{2} \mathcal{H}^\text{am}_{1\text q,2}=
    \bigg\{\begin{array}{lll}
           2&\text{for}&p \leq 1/2\;,\\
           \frac{2p}{1-p}&\text{for}&p > 1/2\;.
                                      \end{array}
                                      \;.
\label{amphol2}
\end{equation}
This is plotted in Fig.~\ref{Ampdamp}. The $X_y$ matrix required to saturate this bound is
\begin{equation}
\label{ytwo}
X_{y}=\frac{-1}{\sqrt{1-p}}\sigma_x\;.
\end{equation}

\subsubsection{Two-qubit probe}
\begin{figure}%
\includegraphics[width=1\textwidth]{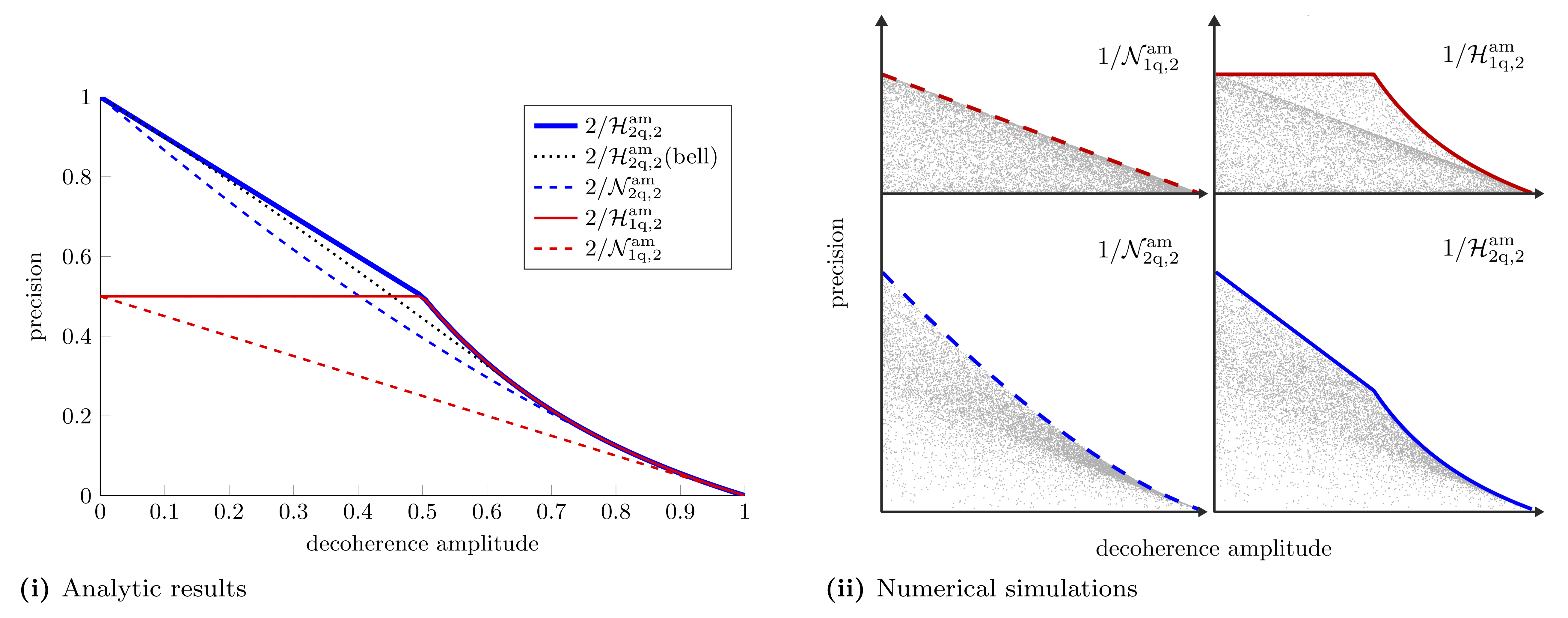}
\caption{{\bf Two parameter estimation in an amplitude damping
    channel.} (i) Achievable precisions using single and two qubit probes to
  estimate two parameters under the amplitude damping channel. The Nagaoka bound and Holevo bounds for the single qubit probe are given by
  Eqs. \eqref{amp1nagjoint} and \eqref{amphol2joint} respectively. They can be
  achieved when the probe is $\ket{1}$.
  For the two qubit probe the Nagaoka bound is given by Eq.
  \eqref{ampnag2} and is achieved with a maximally entangled probe. The
   Holevo bound is given by Eq.~\eqref{opholad}. The optimal probe
   depends on $p$. The Holevo bound for the maximally entangled
   probe Eq.~\eqref{amphol2ME} is plotted in black for comparison. (ii) Numerical simulations which support the analytic results.}
  \label{Ampdamp}
\end{figure}

We now consider the two qubit case where the first qubit is exposed to
the amplitude damping channel and the second qubit is passed through
the identity channel. We consider the probe $\frac{1}{\sqrt{2}}(\ket{0, 1}+\ket{1, 0})$ which exhibits some of the advantages offered by two qubit probes. Using this probe we are able to estimate a rotation around the $x$-axis with a Holevo bound given by
\begin{equation}
v_x \geq \mathcal{H}^\text{am}_{2\text{q},1}=\mathcal{N}^\text{am}_{2\text{q},1}=\frac{1}{1-p}\;.
\label{adhol2}
\end{equation}
Again this coincides with the SLD bound. Unlike the decoherence
channel, for estimating a single parameter under the action of this
channel the entangled two qubit probe doesn't actually offer any
advantage. For estimating two parameters using individual measurements, the Bell state is optimal and the minimum variance that can be obtained is
\begin{equation}
  v_x+v_y \geq \mathcal{N}^\text{am}_{2\text q,2}=
  \left\{\begin{array}{lll}
           \frac{16\left(3-\sqrt{1-p}\right)\left(1-\sqrt{1-p}\right)}{p(8+p)(1-p)}&\text{for}&0<p \leq 2\sqrt{2}-2\;,\\
           \frac{4p}{1-p}&\text{for}& 2\sqrt{2}-2 < p <1\;.
                                      \end{array}
                                      \right.
\label{ampnag2}
\end{equation}
For all $0<p<1$, this probe is always better than the optimal single qubit probe. The $X$ matrices required to saturate this bound when $p<2\sqrt{2}-2$ are
\begin{equation}
  \begin{gathered}
  X_x=
  \begin{pmatrix}0&-a\I& -b\I&(-1+\I)w\\
   a\I&0&0&a\I\\
    b\I&0&0&b\I\\
    (-1-\I)w&-a\I&-b\I&0
  \end{pmatrix}\;\text{and}\qquad
  X_y=
  \begin{pmatrix}0&-a& -b&(-1-\I)w\\
   -a&0&0&a\\
    -b&0&0&b\\
    (-1+\I)w&a&b&0
  \end{pmatrix}\,,
  \end{gathered}
\end{equation}
where $a=2/(4(1-p)+(4-p)\sqrt{1-p})$, $b=2(3+p)/((4+p)(1-p)+(4+2p)\sqrt{1-p})$ and $w=\sqrt{(a^2+b^2)/2}$. When $p>2\sqrt{2}-2$ the required matrices are
\begin{equation}
  \begin{gathered}
  X_x=
  \begin{pmatrix}0&-\I& \frac{-\I}{\sqrt{1-p}}&c_{14+}\\
   \I&c_{22}&-c_{23}\I&\frac{\I p}{2\sqrt{1-p}}-\sqrt{1-p}c_{34}\\
    \frac{\I}{\sqrt{1-p}}&c_{23}\I&\frac{-c_{22}}{1-p}&\frac{\I p}{2(1-p)}+c_{34}\\
    c_{14-}&\frac{-\I p}{2\sqrt{1-p}}-\sqrt{1-p}c_{34}&\frac{-\I p}{2(1-p)}+c_{34}&-c_{44}
  \end{pmatrix}\;,
    \end{gathered}
\end{equation}
and
\begin{equation}
  \begin{gathered}
  X_y=
  \begin{pmatrix}0&-1& \frac{-1}{\sqrt{1-p}}&c_{14-}\\
   -1&-c_{22}&-\I c_{23}&\frac{p}{2\sqrt{1-p}}-\I\sqrt{1-p}c_{34}\\
    \frac{-1}{\sqrt{1-p}}&\I c_{23}&\frac{c_{22}}{1-p}&\frac{p}{2(1-p)}+\I c_{34}\\
    c_{14+}&\frac{p}{2\sqrt{1-p}}+\I\sqrt{1-p}c_{34}&\frac{p}{2(1-p)}-\I c_{34}&c_{44}
  \end{pmatrix}\,,
    \end{gathered}
  \end{equation}
where $c_{14\pm}=\frac{-1\pm\I}{\sqrt{2}}\sqrt{\frac{2-p}{1-p}}$, $c_{22}=\sqrt{(p^2+4p-4)/(2(2-p))}$, $c_{23}=\sqrt{(p^2+4p-4)/(2(1-p)(2-p))}$, \newline $c_{34}=\sqrt{p^2+4p-4}/(2(1-p))$ and $c_{44}=p\sqrt{p^2+4p-4}/(\sqrt{2(2-p)}(1-p))$. These matrices are constructed such that $X_x$ and $X_y$ commute. A POVM can then be formed from the simultaneous eigenvectors of the two matrices which saturates the Nagaoka bound. 
% Although we cannot find an analytic expression for the optimal POVM required to reach the NHB, as we can with the other channels, we can numerically verify that there always exists a 4-outcome projective POVM that can saturate this bound for all p.

When allowing for collective measurements, the maximally entangled
probe gives
\begin{equation}
  v_x+v_y \geq \mathcal{H}^\text{am}_{2\text q,2}(\text{bell})=
  \left\{\begin{array}{lll}
           \frac{(2-p)^2}{2(1-p)^2}&\text{for}& p \leq 2/3\;,\\
           \frac{4p}{1-p}&\text{for}& 2/3 < p <1\;.
                                      \end{array}
                                      \right.
\label{amphol2ME}
\end{equation}
Plotting both $\mathcal{H}^\text{am}_{2\text q,2}$ and
$\mathcal{H}^\text{am}_{1\text q,2}(\text{bell})$ in Fig.~\ref{Ampdamp}, we see that at some
values of $p$, the single qubit probe $\ket{1}$ outperforms the
maximally entangled probe. This means that the maximally entangled
probe is not always the optimal two qubit probe for sensing the
channel. The optimal two qubit probe depends on the noise in the channel. The optimal two qubit probe can be written as
  $\ket{\psi}=a\ket{0,1}+b\ket{1,0}$, where $a$ and $b$ depend
  on $p$. With an optimised probe, we get
\begin{equation}
  v_x+v_y \geq \mathcal{H}^\text{am}_{2\text q,2}=
  \bigg\{\begin{array}{lll}
           \frac{2}{1-p}&\text{for}& p \leq 1/2\;,\\
           \frac{4p}{1-p}&\text{for}& 1/2 < p <1\;.
                                      \end{array}
                                      %\right
                                      \label{opholad}
\end{equation}
The maximally entangled probe is only optimal when $p=0$ and when
$p \geq 2/3$. For $p \geq 1/2$ the probe $\ket{1,0}$, which performs
identically to the single qubit probe $\ket{1}$, is optimal.  When
$p<1/2$, the optimal two-qubit probe is
\begin{align}
  \label{eq:3}
  \ket{\psi} = \sqrt{\frac{1-2p}{2-2p}}\ket{0,1}+\frac{1}{\sqrt{2-2p}}\ket{1,0}\;.
\end{align}
The optimal $X_x$ and $X_y$ matrices corresponding to this probe are
\begin{align}
  \label{eq:8}
  X_x&=\frac{-\I}{\sqrt{1-p}}\left(\ket{0}\bra{1}-\ket{1}\bra{0} \right)
      \otimes \ket{0}\bra{0}-
      \frac{\I \sqrt{1-2p}}{1-p} \ket{1}\bra{1} \otimes \left(\ket{1}\bra{0}-\ket{0}\bra{1} \right) \;,\\
  X_y&=-\frac{1}{\sqrt{1-p}}\left(\ket{0}\bra{1}+\ket{1}\bra{0} \right)
      \otimes \ket{0}\bra{0}+
      \frac{\sqrt{1-2p}}{1-p} \ket{1}\bra{1} \otimes \left(\ket{1}\bra{0}+\ket{0}\bra{1} \right)\;.
\end{align}
By direct substitution, one can check that these matrices satisfy
the unbiased conditions. By direct substitution, we get the $Z$ matrix
as
\begin{align}
  Z=\begin{pmatrix}
 \frac{1}{1-p}&0\\
 0&\frac{1}{1-p}
 \end{pmatrix}\;,
\end{align}
which gives $\mathcal{H}=2/(1-p)$. The attainable precisions for estimating two parameters with both single and two qubit probes are plotted in Fig.~\ref{Ampdamp}. Notably this example shows that in different regimes, it is possible to have either $\text{C}_\text{a}$$\text{C}_\text{a}$$>$CQ $\left(1/\mathcal{N}^{\text{am}}_{2q,2}>1/\mathcal{H}^{\text{am}}_{1q,2}\right)$ or CQ$>$$\text{C}_\text{a}$$\text{C}_\text{a}$ $\left(1/\mathcal{H}^{\text{am}}_{1q,2}>1/\mathcal{N}^{\text{am}}_{2q,2}\right)$.

%For measuring a rotation about the $x$ and $y$ axis using this
%probe the Holevo bound coincides with the Nagaoka bound and both are
%given by\comment{The $X$ and $Y$ matrices didn't give this result?}
%\begin{equation}
%v_x+v_y \geq \mathcal{H}^\text{am}_{2\text q,2}= \mathcal{N}^\text{am}_{2\text q,2} =\frac{2}{1-p}.
%\end{equation}
%Thus,in this instance collective measurements offer no advantage over individual measurements. The optimal strategy is to have the same estimation ability in both directions
%\begin{equation}
%v_x^{*}=v_y ^{*} = \frac{1}{2} \mathcal{H}^\text{am}_{2\text q,2}=\frac{1}{1-p}\;.
%\label{amphol2}
%\end{equation}
%Thus,using this probe we are able to estimate the second parameter for
%free. We see that the two qubit probe offers an advantage over the
%single qubit probes for estimating multiple parameters.

\begin{figure}[t]
\includegraphics[width=1\textwidth]{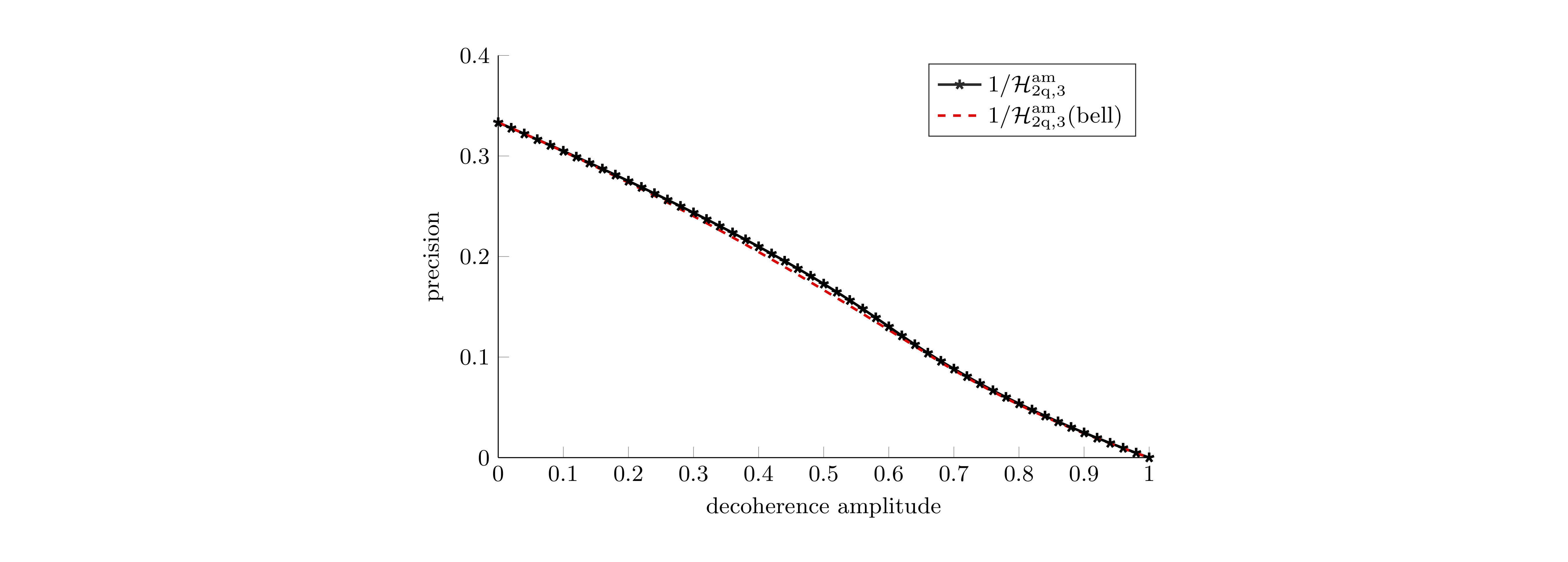}
\caption{{\bf Collective measurements for three parameter estimation in an amplitude damping channel.} Note that now the optimal $v_{x}=v_{y}\neq v_{z}$ and so instead we plot $1/\mathcal{H}_{3}$. The optimal probe was found numerically. The Holevo bound for the Bell state, (Eq.~\ref{adall3mark2}) is almost optimal.}
\label{ad3opt}
\end{figure}

Using the maximally entangled state as our probe, we find that the Holevo bound for estimating three parameters is given by
\begin{equation}
v_{x}+v_{y}+v_{z}\geq\mathcal{H}_{2q,3}^\text{am}= 
 \bigg\{\begin{array}{lll}
           \frac{(3-2p)(2-p)}{2(1-p)^{2}}&\text{for}& p \leq 2/3\;,\\
           \frac{(2+7p)}{2-2p}&\text{for}& 2/3 < p <1\;.
                                      \end{array}
                                      \label{adall3mark2}
\end{equation}
The optimal $X_x$, $X_y$ and $X_z$ matrices for this probe are 
%\begin{align}
%X_x&=i\,c_1(\ket{0,0}\bra{0,1}-\ket{0,1}\bra{0,0})+i\,c_2(\ket{0,0}\bra{1,0}-\ket{1,0}\bra{0,0})+ic_3(\ket{1,1}\bra{1,0}-\ket{1,0}\bra{1,1}),\\
%X_y&=c_1(\ket{0,0}\bra{0,1}+\ket{0,1}\bra{0,0})+c_2(\ket{0,0}\bra{1,0}+\ket{1,0}\bra{0,0})-c_3(\ket{1,1}\bra{1,0}+\ket{1,0}\bra{11}),\\
%X_z&=i\,c_2(-\ket{0,1}\bra{1,0}+\ket{1,0}\bra{0,1}),
%\end{align}
\begin{equation}
  \begin{gathered}
  X_x=-
\I
  \begin{pmatrix}0&c_1& c_2&0\\
   -c_1&0&0&0\\
    -c_2&0&0&-c_3\\
    0&0&c_3&0
  \end{pmatrix}\;\text{,}\qquad
  X_y=-
  \begin{pmatrix}0&c_1& c_2&0\\
   c_1&0&0&0\\
    c_2&0&0&-c_3\\
    0&0&-c_3&0
  \end{pmatrix}\,\text{and}\qquad
  X_z=\I 
  \begin{pmatrix}0&0& 0&0\\
   0&0&-c_2&0\\
    0&c_2&0&0\\
    0&0&0&0
  \end{pmatrix}\,,
  \end{gathered}
\end{equation}
where 
\begin{align}
c_1&= \bigg\{\begin{array}{lll}
           \frac{p}{2-2p}&\text{for}& 0\leq p \leq 2/3 \;,\\
           1&\text{for}&  2/3< p <1\;,
                                      \end{array}
                                     \;,\\
c_2&=\frac{1}{\sqrt{1-p}}\;,\\
c_3&=\frac{1}{1-p}-c_1\;.\\
\end{align}
We note that the maximally entangled probe is not the optimal
probe for estimating three parameters under the action of this
channel. In Fig.~\ref{ad3opt} however, it can be seen that the maximally entangled
probe achieves a variance which is very close to the numerically
optimised Holevo bound.

\subsubsection{Decoherence of both qubits}
\begin{figure}[t]
\includegraphics[width=0.5\textwidth]{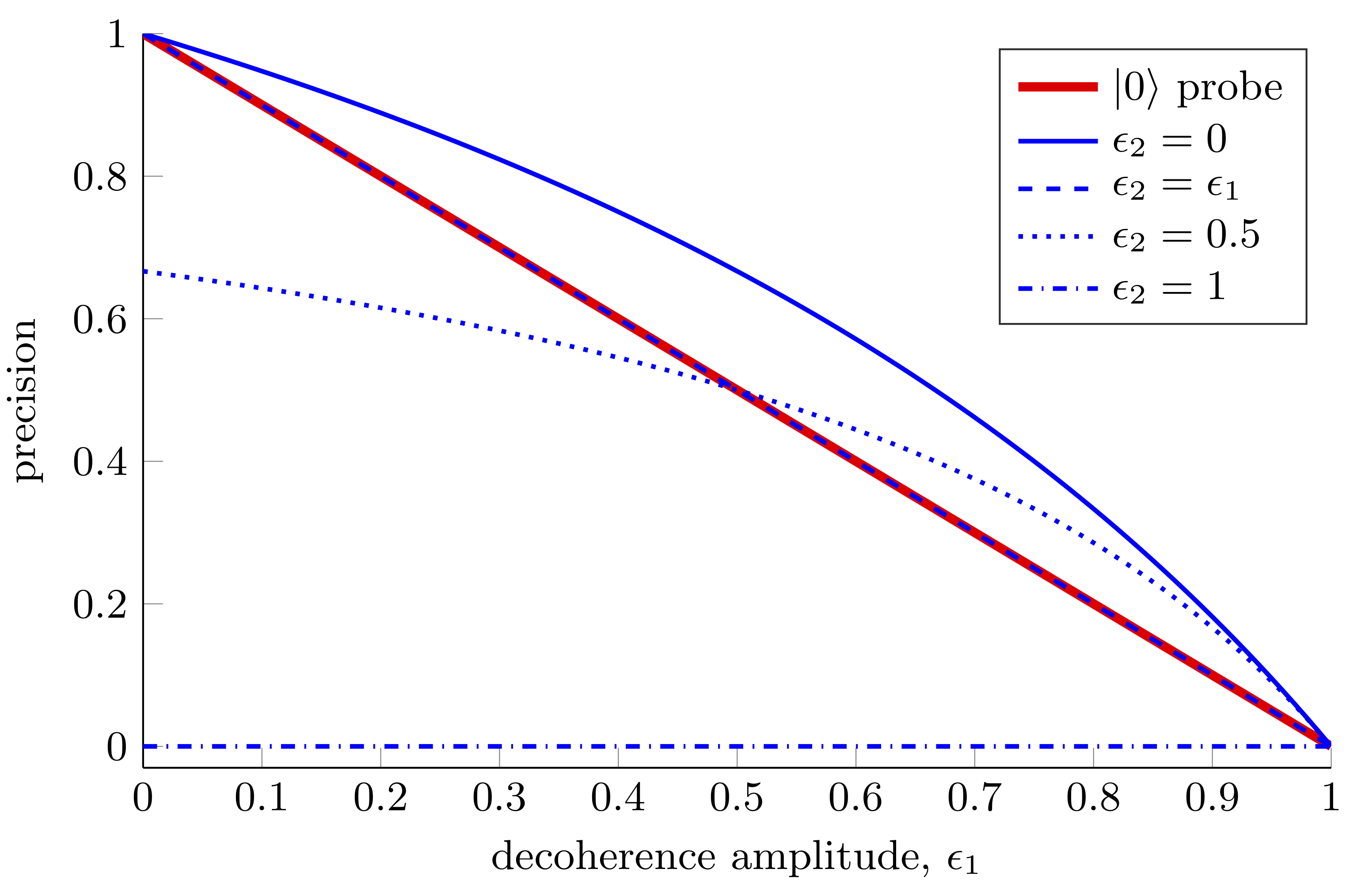}
\caption{{\bf Single parameter estimation in an amplitude damping
    channel where both qubits are exposed to the noise.} We plot the Holevo bounds achievable using single and two qubit probes to estimate a single parameter, $\theta_{z}$ with both qubits exposed to the amplitude damping channel. The
  Holevo bound for the single qubit probe is given by
  Eq. (\ref{adfirst}). For the two qubit probe the Holevo bound is given by Eq (\ref{addual}).
  }
  \label{ad_both_decoh}
\end{figure}

If we consider the more realistic channel where each qubit is individually subject to the amplitude damping channel then the estimation performance can only at best match the performance of the channel where the second qubit is not affected by the channel. For estimating $\theta_{z}$ using a $\frac{1}{\sqrt{2}}\left(\ket{0,1}+\ket{1,0}\right)$ probe with both qubits exposed to separate amplitude damping channels, with probabilities $p_{1}$ and $p_{2}$ respectively, the Holevo bound is given by
\begin{equation}
\label{addual}
v_z \geq \mathcal{H}^\text{2am}_{2\text q,1}=\frac{1}{2-2p_{1}}+\frac{1}{2-2p_{2}}\;.
\end{equation}
%When $p_{2}=0$ this reduces to the Holevo bound for estimating only $\theta_{z}$ using the two qubit probe considered. 
We can compare this to the Holevo bound for estimating $\theta_{z}$  using a single qubit probe, which is given by Eq.~\eqref{adfirst}, if we orient the probe such that it is optimal for sensing a rotation about the $z$-axis. We again see that, although we are trying to estimate a rotation on the first qubit only, the Holevo bound is symmetric in $p_{1}$ and $p_{2}$. If the second qubit is maximally exposed to the amplitude damping channel the first qubit can no longer be used for parameter estimation. When both qubits are exposed to the same decoherence amplitude, $p_{1}=p_{2}$, the Holevo bound for the two qubit probe is equal to the Holevo bound for the single qubit probe. Thus, although more entangled probes offer some advantages under certain conditions, they also have weaknesses that single qubit probes do not have, shown in Fig.~\ref{ad_both_decoh}.  

\subsubsection{Collective measurements on multiple copies of the state}
As with the decoherence channel, we now consider performing collective measurements on \textit{M} copies of the single qubit state. The
probes $\ket{0}$ and $\ket{1}$ perform equally well for simultaneous
estimation of two rotations about the $x$ and $y$ axes when we are restricted
to individual measurements. Both probes give a Nagaoka bound of $\mathcal{N}^{\text{am}}_{1\text{q},2}\left(\ket{0}\right)=\mathcal{N}^{\text{am}}_{1\text{q},2}\left(\ket{1}\right) = 4/(1-p)$. The Nagaoka bound is known to be attainable for a two-dimensional state. An optimal measurement is to perform a measurement of $\sigma_x$ half of
the time and a measurement of $\sigma_y$ the other half of the time, as shown in appendix \ref{apenAD1q}. However, we know from the Holevo bound, that when allowing for collective measurements on infinitely many probes, the probe $\ket{1}$ actually outperforms
$\ket{0}$. For
the probe $\ket{0}$, collective measurements do not provide any
advantage over individual measurements, but for the probe $\ket{1}$, collective measurements
improve the precision. We show how this is possible by constructing an explicit
estimator to approach the Holevo bound. We first consider doing a
collective measurement on the probe $\ket{1,1}$.

After the amplitude damping channel, the probe $\ket{1}$ transforms to
the mixed state $p \ket{0}\bra{0}+(1-p)\ket{1}\bra{1}$. Two copies of this
will be in the state $\rho_1=p^2 \ket{0,0}\bra{0,0}+p(1-p)\ket{0,1}\bra{0,1}+p(1-p) \ket{1,0}\bra{1,0}+(1-p)^2\ket{1,1}\bra{1,1}$. As this is a mixed state, collective measurements may offer an advantage. By performing the optimisation over the matrices $X$ and $Y$, we find that the
Nagaoka bound for the state $\ket{0,0}$ and $\ket{1,1}$ is
\begin{align}
  \label{nagad1q2c}
  \mathcal{N}^{\text{am}}_{1\text{q}^{\otimes2},2}\left(\ket{0,0}\right) &= \frac{2}{1-p}\;,\\
  \mathcal{N}^{\text{am}}_{1\text{q}^{\otimes2},2}\left(\ket{1,1}\right) &= \frac{2}{1-p}-2p\;.
    \label{nagad1q2c22}
\end{align}
The result for $\ket{0,0}$ is not surprising. As $\ket{0,0}$ is unaffected by this channel, it remains as a pure state and hence collective measurements cannot offer any benefit. Therefore, the minimum variance is exactly half of
the individual measurement case, meaning that the optimal
estimation is to measure each probe individually.  But the result for
$\ket{1,1}$ is smaller than half of the individual measurement
case. This is now a four dimensional state and it is not certain that
the Nagaoka bound can still be attained\footnote{Nagaoka conjectured
  it is always attainable, but only proved for certain specific
  cases.}.  We construct an explicit measurement which
saturates the bound in appendix \ref{11est}, showing that the bound is attainable. 

\begin{figure}[t]
\includegraphics[width=1\textwidth]{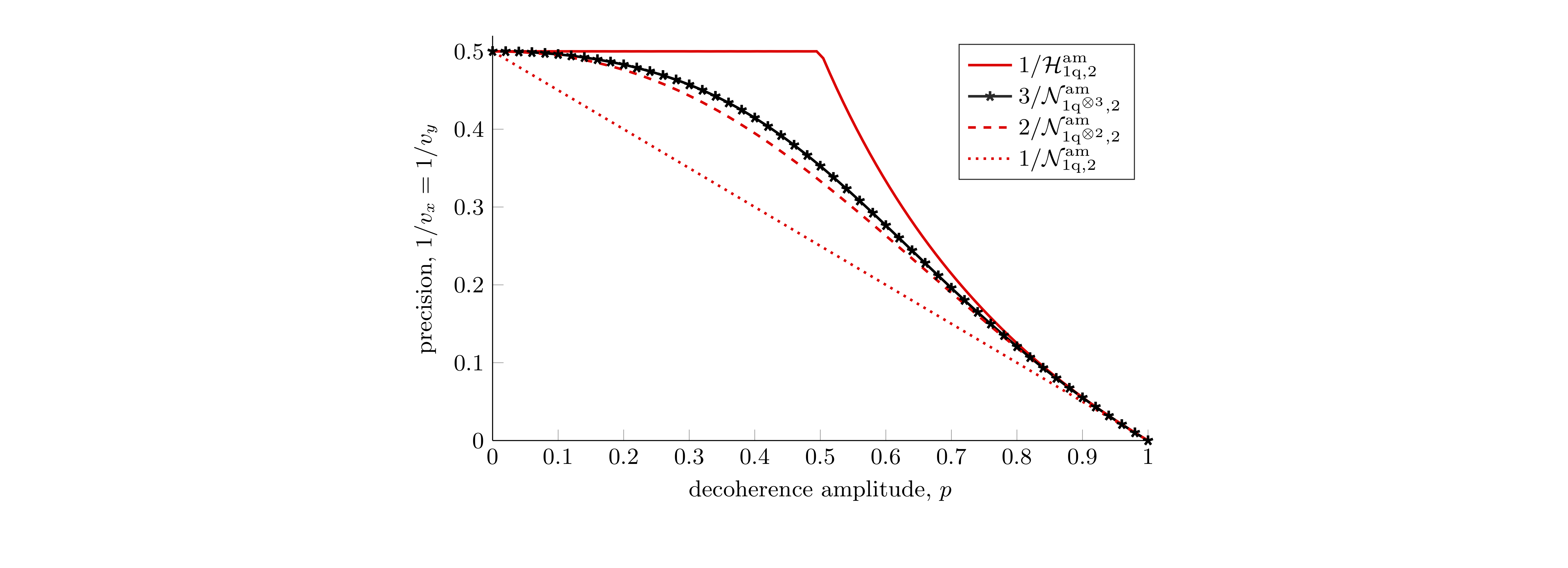}
\caption{{\bf Collective measurements for two parameter estimation in
    an amplitude damping channel.}
  Optimal achievable precision for estimating two parameters using the
  probe $\ket{1}$ when doing individual measurements $\mathcal{N}^{\text{am}}_{1\text{q},2}$, Eq.~\eqref{ampnag2}, and
  when measuring the probes in pairs, $\mathcal{N}^{\text{am}}_{1\text{q}^{\otimes2},2}$, Eq.~\eqref{nagad1q2c22}. The Nagaoka bound for measuring three qubits collectively is plotted as $\mathcal{N}^{\text{am}}_{1\text{q}^{\otimes3},2}$. We did not check that $\mathcal{N}^{\text{am}}_{1\text{q}^{\otimes3},2}$ is
  attainable, but it is still a valid bound. Finally, Holevo's bound applies when we are able to
  measure infinitely many copies simultaneously. \label{AD_collective}}
\end{figure}

The optimal $X$ and $Y$ matrices that give the Nagaoka bound for
$\ket{1,1}$ probe are\footnote{For the probe $\ket{0,0}$, the optimal
  matrices have the same form except for a sign change.}
%\begin{align}
%  \label{eq:1}
%  X&=\frac{1}{\sqrt{1-p}} \big(\ket{y+,y+}\bra{y+,y+}-\ket{y-,y-}\bra{y-,y-} \big)\;,\\
%  Y&=-\frac{1}{\sqrt{1-p}} \big(\ket{x+,x+}\bra{x+,x+}-\ket{x-,x-}\bra{x-,x-} \big)\;.
%\end{align}
\begin{align}
  \label{eq:1}
 X_x&=\frac{1}{\sqrt{1-p}} \big(\ket{y+,y+}\bra{y+,y+}-\ket{y-,y-}\bra{y-,y-} \big)\;,\\
  X_y&=-\frac{1}{\sqrt{1-p}} \big(\ket{x+,x+}\bra{x+,x+}-\ket{x-,x-}\bra{x-,x-} \big)\;,
\end{align}
where $\ket{y\pm}=\left(\ket{0}\pm \mathrm{i}\ket{1}\right)/\sqrt{2}$ and $\ket{x\pm}=\left(\ket{0}\pm \ket{1}\right)/\sqrt{2}$ are the eigenvectors of $\sigma_y$ and $\sigma_x$ respectively. In Fig.~\ref{AD_collective} we show how the Nagaoka bound tends towards the Holevo bound for the probe $\ket{1}$, with an increasing number of copies of the probe. This shows the same effect as Fig.~\ref{decohmcop}, but in a slightly different representation.

\newpage
\subsection{Phase Damping Channel}
\label{lastres}
Finally, we consider the phase damping channel. This channel describes the loss of information without the loss of energy and is unique to quantum systems. It is represented by the Kraus operators
\begin{equation}
M_{0}=\sqrt{1-\frac{\epsilon}{2}}\mathbbm{1}\;,\quad M_{1}=\sqrt{\frac{\epsilon}{2}}\sigma_{z}\;.
\end{equation}
This channel can be thought of as applying $\sigma_{z}$ with probability $\frac{\epsilon}{2}$. We can see why this channel is uniquely quantum mechanical by considering repeated applications of the channel to an arbitrary density matrix. If the initial quantum state is
\begin{equation}
\rho_{0}=\begin{pmatrix}
a&b\\
b^{*}&c\\
\end{pmatrix}\;,
\end{equation}
then after the action of this channel the state becomes
\begin{equation}
\rho_{\epsilon}=\begin{pmatrix}
a&(1-\epsilon)b\\
(1-\epsilon)b^{*}&c\\
\end{pmatrix}\;.
\end{equation}
Repeated application of this channel is equivalent to allowing $\epsilon\rightarrow 1$. The density matrix loses its off-diagonal elements and ends up as a completely diagonal matrix, i.e. the quantum state has become a completely classical mixture of $\ket{0}$'s and $\ket{1}$'s. In this way the quantum mechanical properties of the system are lost. Thus, this channel does not have a direct classical analogue.

\subsubsection{Single Qubit Probe}
\begin{figure}
\includegraphics[width=1\textwidth]{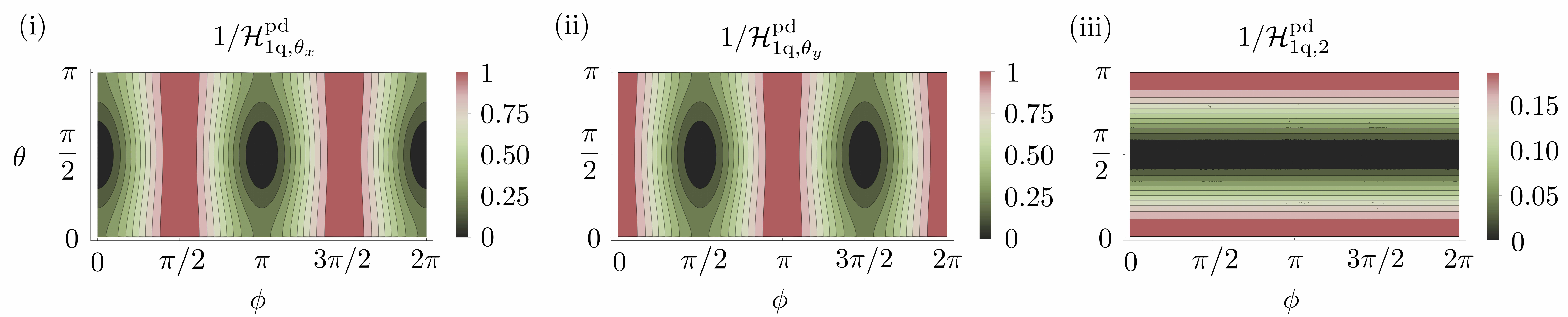}
\caption{\textbf{Parameter estimation in the phase damping channel with the probe $\boldsymbol{\ket{\psi}=\text{cos}(\theta/2)\ket{0}+e^{\I\phi}\text{sin}(\theta/2)\ket{1}}$.} (i) and (ii) show the precision achievable when estimating a rotation about the $x$-axis and $y$-axis respectively. We see that probes which are optimal for estimating $\theta_x$ are not optimal for estimating $\theta_y$. This explains why, although it is possible to estimate a single parameter without a degradation in precision as the decoherence increases, the same is not possible when estimating two parameters. (iii) Holevo bound for estimating both rotations simultaneously with multiple copies of the same probe. There is a discontinuity in all figures at $\theta=0$ or $\pi$. For all figures $\epsilon=0.5$.}
\label{fig:PD_xy_comp}
\end{figure}

Geometrically this channel represents a loss of information about
the $x$ and $y$ components of the Bloch vector. When restricted to real probes, for estimating a rotation about the $x$-axis the probes
$\ket{0}$ and $\ket{1}$ perform identically under the action of this
channel. Both probes are able to estimate a rotation about either the
$x$ or the $y$-axis individually with a Holevo bound given by Eq.
\eqref{Hol1}. This coincides with the SLD bound for these probes as is expected. The $X_{x}$ matrix required to saturate the Holevo bound is $\sigma_y/(1-\epsilon)$. However, by allowing complex probes the Holevo bound can be improved. The reason for this is that the phase damping channel affects the Bloch sphere in an asymmetric way. One optimal probe is $\ket{\psi}=\frac{1}{2}\left((1+\I)\ket{0}+(1-\I)\ket{1}\right)$ which can achieve a Holevo bound of 1, no matter what the decoherence amplitude. The $X_x$ matrix required to saturate this is $X_x=-\ket{0}\bra{0}+\ket{1}\bra{1}$. However, as is evident from Fig.~\ref{fig:PD_xy_comp} (i), almost any probe in the Y-Z plane of the Bloch sphere is optimal. For probes of the form $\ket{\psi}=\text{cos}(\theta/2)\ket{0}+e^{\I\phi}\text{sin}(\theta/2)$, any state with $\phi=\pi/2$ or $3\pi/2$ will be optimal, except for states where $\theta$ is exactly equal to $0$ or $\pi$. Similarly if we want to estimate a rotation about the $y$-axis only any state with $\phi=0$ or $\pi$ will be optimal, except when $\theta$ is exactly equal to $0$ or $\pi$. The discontinuity in the Holevo bound exactly at these extreme points has been observed before and corresponds to a point where the rank of the state changes~\cite{vsafranek2017discontinuities,vsafranek2018simple,seveso2019discontinuity,rezakhani2019continuity,goldberg2021taming,ye2022quantum}. The small rotation which we are trying to estimate changes the states $\ket{0}$ or $\ket{1}$ which are are rank 1 states and are not decohered to states which have rank 2 and are decohered.

For estimating a rotation about the $x$ and $y$ axes simultaneously, consider the probe state $\ket{\psi}=a\ket{0}+\sqrt{1-a^2}\ket{1}$, with $a=1-\delta$, where $\delta$ is small. With this probe as $\delta\rightarrow 0$, the Holevo bound is given by
\begin{equation}
\label{hol12pd}
v_x+v_y \geq \mathcal{H}^\text{pd}_{1\text q,2}=    \bigg\{\begin{array}{lll}
           4&\text{for}&\epsilon \leq 1-1/\sqrt{2}\;,\\
           \frac{1}{(2-\epsilon)(1-\epsilon)^2\epsilon}&\text{for}&\epsilon > 1-1/\sqrt{2}\;.
                                      \end{array}
\end{equation}
It might be expected that one could always estimate $\theta_x$ and $\theta_y$ with a minimum variance of 4, given that it is possible to estimate both individually with a variance of 1, simply by using half the probes to estimate $\theta_x$ and the other half to estimate $\theta_y$. However,  Fig.~\ref{fig:PD_xy_comp} shows why this is not possible. The optimal probes for estimating each angle are different and so the multiparameter estimation is degraded. The $X_{y}$ matrix required to saturate this Holevo bound is given by
%\begin{equation}
%X_{y}=\begin{bmatrix}
%0&\frac{-1}{1-\epsilon}\\
%\frac{-1}{1-\epsilon}&\frac{2}{a\sqrt{1-a^2}} 
%\end{bmatrix},
%\end{equation}
%when $\epsilon\leq 1-\frac{1}{\sqrt{2}}$ and 
%\begin{equation}
%X_{y}=\frac{1}{\epsilon(2-\epsilon)}\begin{bmatrix}
%0&\frac{-1}{1-\epsilon}\\
%\frac{-1}{1-\epsilon}&\frac{1}{a\sqrt{1-a^2}} 
%\end{bmatrix},
%\end{equation}
%
\begin{equation}
X_{y}= \left\{\begin{array}{lll}
           \begin{pmatrix}
0&\frac{-1}{1-\epsilon}\\
\frac{-1}{1-\epsilon}&\frac{2}{a\sqrt{1-a^2}} 
\end{pmatrix}&\text{for}&\epsilon \leq 1-1/\sqrt{2}\;,\\
           \frac{1}{\epsilon(2-\epsilon)}\begin{pmatrix}
0&-1+\epsilon\\
-1+\epsilon&\frac{1}{a\sqrt{1-a^2}} 
\end{pmatrix}&\text{for}&\epsilon > 1-1/\sqrt{2}\;.
                                      \end{array}\right.
\end{equation}

We see when $a=0$ or $a=1$ these matrices contain infinities, but this is not reflected in the QFI which is optimised as $a\rightarrow1$. In fact the matrix $X_y$ only satisfies the unbiased conditions in Eqs.~(\ref{eq_xcon1}) and (\ref{eq_xcon2}), when $\delta\rightarrow0$. Interestingly exactly at $\delta=0$, so that $a=1$, the Holevo bound is only $\frac{4}{(1-\epsilon)^2}$. This discontinuity is again explained by the changing rank of the state at this point and is shown in Fig~\ref{fig:PD_xy_comp} (iii). 

For estimating the same two parameters with individual measurements we find that the Nagaoka bound is given by
\begin{equation}
\label{nag12pd}
v_x+v_y \geq \mathcal{N}^\text{pd}_{1\text q,2}= \frac{(2-\epsilon)^{2}}{(1-\epsilon)^{2}}\;.
\end{equation}
Thus, collective
measurements offer an advantage over individual measurements in this
particular instance. In appendix \ref{nagexp} we present a measurement saturating the Nagaoka bound, and show that it requires an unequal number of measurements of the two parameters. This is due to the fact that when estimating these two parameters individually we do not obtain the same minimum variance. To saturate the Nagaoka bound we require the following $X_y$ matrix
\begin{equation}
X_y=\frac{1}{\sqrt{1-a^2}}\left(\left(a-\frac{1}{a}\right)\ket{0}\bra{0}+a\ket{1}\bra{1}\right)\;.
\end{equation}
Again there is a discontinuity exactly at $a=1$; in this case the Nagaoka bound is $\frac{4}{(1-\epsilon)^2}$. 

\subsubsection{Two-qubit probe}
We note that some of the results presented for two-qubit probes under the action of this channel have already been presented in Ref.~\cite{conlon2021efficient}, however we include them here for completeness. We consider the scenario where the first qubit is subject to the phase damping channel and the second auxiliary qubit experiences only the identity channel. The probes $\ket{0, 0}$ and $\ket{1, 1}$ perform identically to their single qubit counterparts. Thus, these probes are still unable to estimate a rotation about all three axes. This is to be expected since the probes are separable, hence the idler qubit has no effect.

\begin{figure}%
\includegraphics[width=1\textwidth]{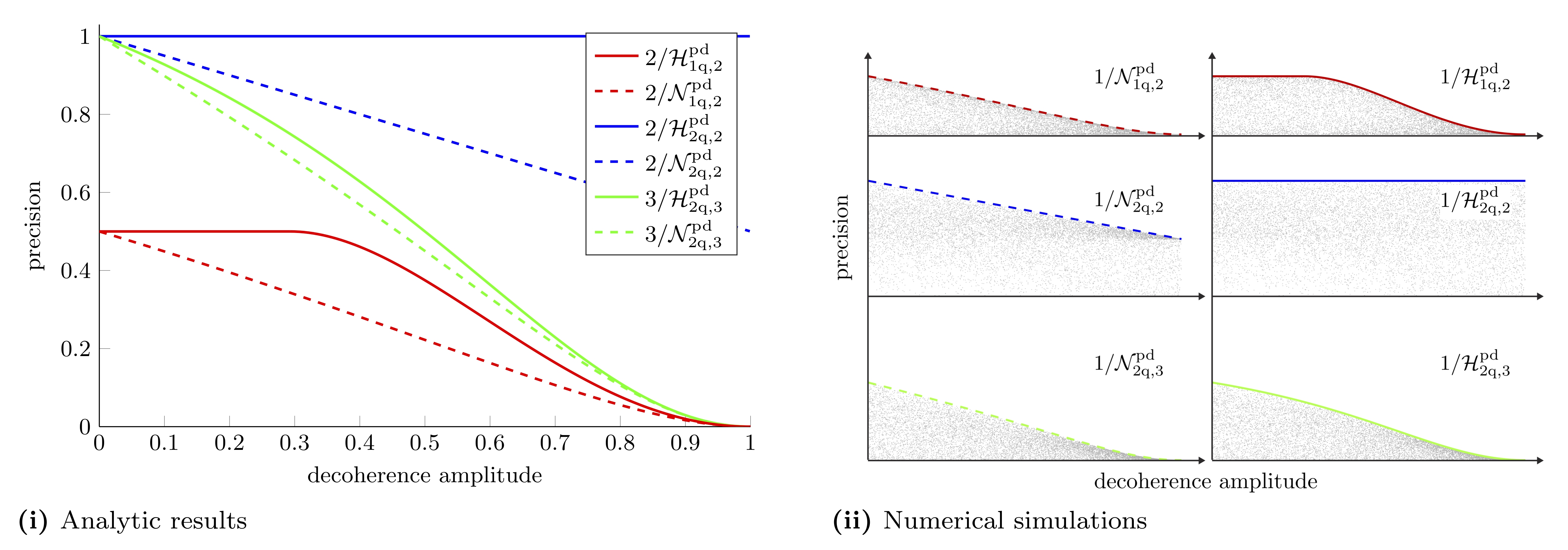}
    \caption{{\bf Multiple parameter estimation in a phase damping channel.} (i) Achievable precision for various probes under the action of the phase damping channel. For estimating $\theta_{x}$ and $\theta_{y}$ simultaneously with the probe $\frac{1}{\sqrt{2}}[\ket{0, 1}+\ket{1, 0}]$ the Holevo bound is equal to 1 regardless of how noisy the channel is. The corresponding Nagaoka bound is given by Eq (\ref{nagpd}). For estimating a rotation about all three axes simultaneously the Holevo and NHBs are given by Eqs (\ref{holpd32q}) and (\ref{nagpd3}) respectively. For the $\ket{0}$ probe estimating $\theta_{x}$ and $\theta_{y}$ simultaneously the Holevo bound is given by Eq (\ref{hol12pd}) and the Nagaoka bound is given by Eq (\ref{nag12pd}). (ii) Numerical simulations supporting the analytic results. }%
    \label{fig:PDexample}%
\end{figure}

For estimating a rotation about the $x$-axis the probe $\frac{1}{\sqrt{2}}\left(\ket{0, 1}+\ket{1, 0}\right)$ is optimal. When we estimate a rotation about either the $x$-axis or the $y$-axis individually using this probe the Holevo bound is independent of the channel parameter $\epsilon$. The Holevo bound remains unaffected when we estimate a rotation about both the $x$ and $y$ axes simultaneously. 
\begin{equation}
\label{hol22pd}
v_x+v_y \geq \mathcal{H}^\text{pd}_{2\text q,2}=2\;.
\end{equation}
This is optimised when $v_{x}^{*}=v_{y}^{*}=1$. Thus, the entangled probe is able to estimate a second parameter at no additional cost. The $X_{x}$ and $X_{y}$ matrices which saturate the Holevo bound are given by
\begin{align}
X_{x}&=\I\left(\ket{\psi_{0}}\bra{\psi_{2}}-\ket{\psi_{2}}\bra{\psi_{0}}+\ket{\psi_{1}}\bra{\psi_{3}}-\ket{\psi_{3}}\bra{\psi_{1}}\right)\;,\\
X_{y}&=-\left(\ket{\psi_{2}}\bra{\psi_{1}}+\ket{\psi_{1}}\bra{\psi_{2}}+\ket{\psi_{0}}\bra{\psi_{3}}+\ket{\psi_{3}}\bra{\psi_{0}}\right)\;.
\end{align}
For estimating these two parameters simultaneously the Bell state is still optimal and the Nagaoka bound is given by:
\begin{equation}
\label{nagpd}
v_x+v_y \geq \mathcal{N}^\text{pd}_{2\text q,2}=\frac{4}{2-\epsilon}\;.
\end{equation}
The $X$ matrices required to achieve the minimum possible variance using individual measurements are given by
\begin{align}
X_{x}=\frac{2\I}{2-\epsilon}\left(\ket{\psi_{0}}\bra{\psi_{2}}-\ket{\psi_{2}}\bra{\psi_{0}}\right) \;,\\
X_{y}=\frac{-2}{2-\epsilon}\left(\ket{\psi_{0}}\bra{\psi_{3}}+\ket{\psi_{3}}\bra{\psi_{0}}\right) \;.
\end{align}
Again owing to the symmetry in the channel the total variance is minimised when $v_{x}^{*}=v_{y}^{*}=\frac{1}{2}\mathcal{N}^\text{pd}_{2\text q,2}$. Thus, for estimating these two parameters individual measurements are not as powerful as collective measurements. For estimating a rotation about all three axes simultaneously this probe is able to achieve a Holevo bound of
\begin{equation}
\label{holpd32q}
v_x+v_y+v_z \geq \mathcal{H}^\text{pd}_{2\text q,3}=2+\frac{1}{(1-\epsilon)^{2}}\;.
\end{equation}
The $X_{z}$ matrix required to saturate this Holevo bound is
\begin{equation}
X_{z}=\frac{\I}{(1-\epsilon)}\left(\ket{\psi_{0}}\bra{\psi_{1}}-\ket{\psi_{1}}\bra{\psi_{0}}\right)\;.
\label{pd2q3}
\end{equation}
The same $X_z$ matrix optimises the NHB for estimating all three rotations simultaneously and we have
\begin{equation}
v_x+v_y+v_z \geq \mathcal{N}^\text{pd}_{2\text q,3}=\frac{4}{2-\epsilon}+\frac{1}{(1-\epsilon)^{2}}\;.
\label{nagpd3}
\end{equation}

In appendix \ref{phasedampapen} we construct a measurement scheme which saturates this bound. This channel showcases many of the benefits of using quantum resources in parameter estimation as summarised in Fig.~\ref{fig:PDexample}.\\
\begin{figure}[t]
\includegraphics[width=1\textwidth]{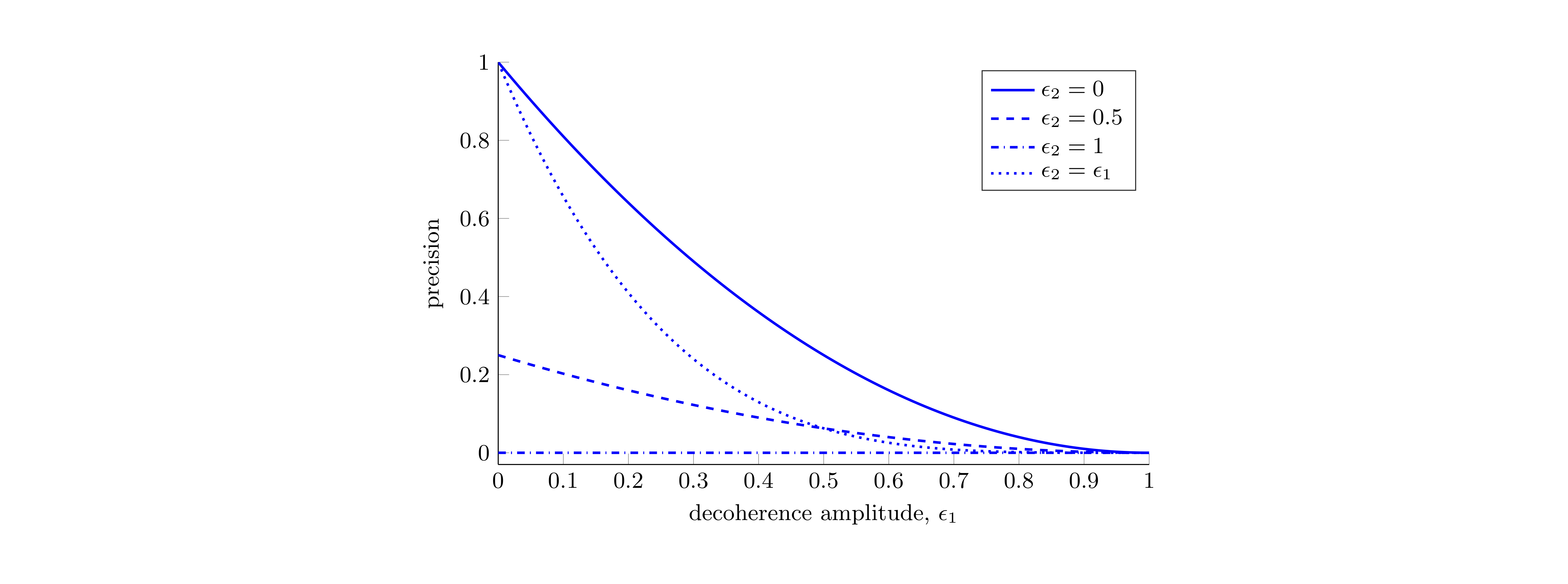}
\caption{{\bf Single parameter estimation in a phase damping
      channel.} Achievable precision for estimating $\theta_{z}$ using the $\frac{1}{\sqrt{2}}(\ket{0,1}+\ket{1,0})$ probe with both qubits exposed to separate phase damping channels. The Holevo bound is given by Eq (\ref{pddualdecoh}). When $\epsilon_{2}=0$ we obtain the achievable precision using the best possible single qubit probe.
\label{PhaseDamp2chvv}}
\end{figure}
\subsubsection{Decoherence of both qubits}
As with previous channels, if we consider the channel where both qubits are exposed to the phase damping channel individually, the performance can only worsen compared to the original channel. However, as mentioned above when using the probe $\frac{1}{\sqrt{2}}\left(\ket{0,1}+\ket{1,0}\right)$ to estimate $\theta_{x}$ or $\theta_{y}$ we can achieve a Holevo variance of 1 regardless of the noise in the channel. This is still true when we consider decohering both qubits. Nevertheless, the effects of exposing the second qubit to the phase damping channel can be examined by considering a rotation about the $z$-axis. This is shown in Fig.~\ref{PhaseDamp2chvv} where we plot the Holevo bound for the $\frac{1}{\sqrt{2}}\left(\ket{0,1}+\ket{1,0}\right)$ probe with both qubits exposed to separate phase damping channels. Using this probe to estimate a rotation about the $z$-axis the Holevo bound is given by
\begin{equation}
v_z \geq \mathcal{H}^\text{2pd}_{2\text q,1}=\frac{1}{(1-\epsilon_{1})^{2}(1-\epsilon_{2})^{2}}\;.
\label{pddualdecoh}
\end{equation}
When $\epsilon_{2}=0$ we recover the case where the second qubit is exposed to the identity channel, but for all other values of $\epsilon_{2}$ the performance of this probe worsens. For this particular example, when the auxillarly qubit experiences no decoherence the single qubit and two qubit probes perform equally well. Thus, when the secondary qubit experiences any decoherence the two qubit probe performs worse than the single qubit probe. In this situation, which is more realistic, using an entangled two qubit probe can bring a disadvantage as the additional ancillary qubit introduces extra noise into the system which lowers the achievable precision.

\subsubsection{Collective measurements on many copies of the state}
\begin{figure}[t]
\includegraphics[width=0.45\textwidth]{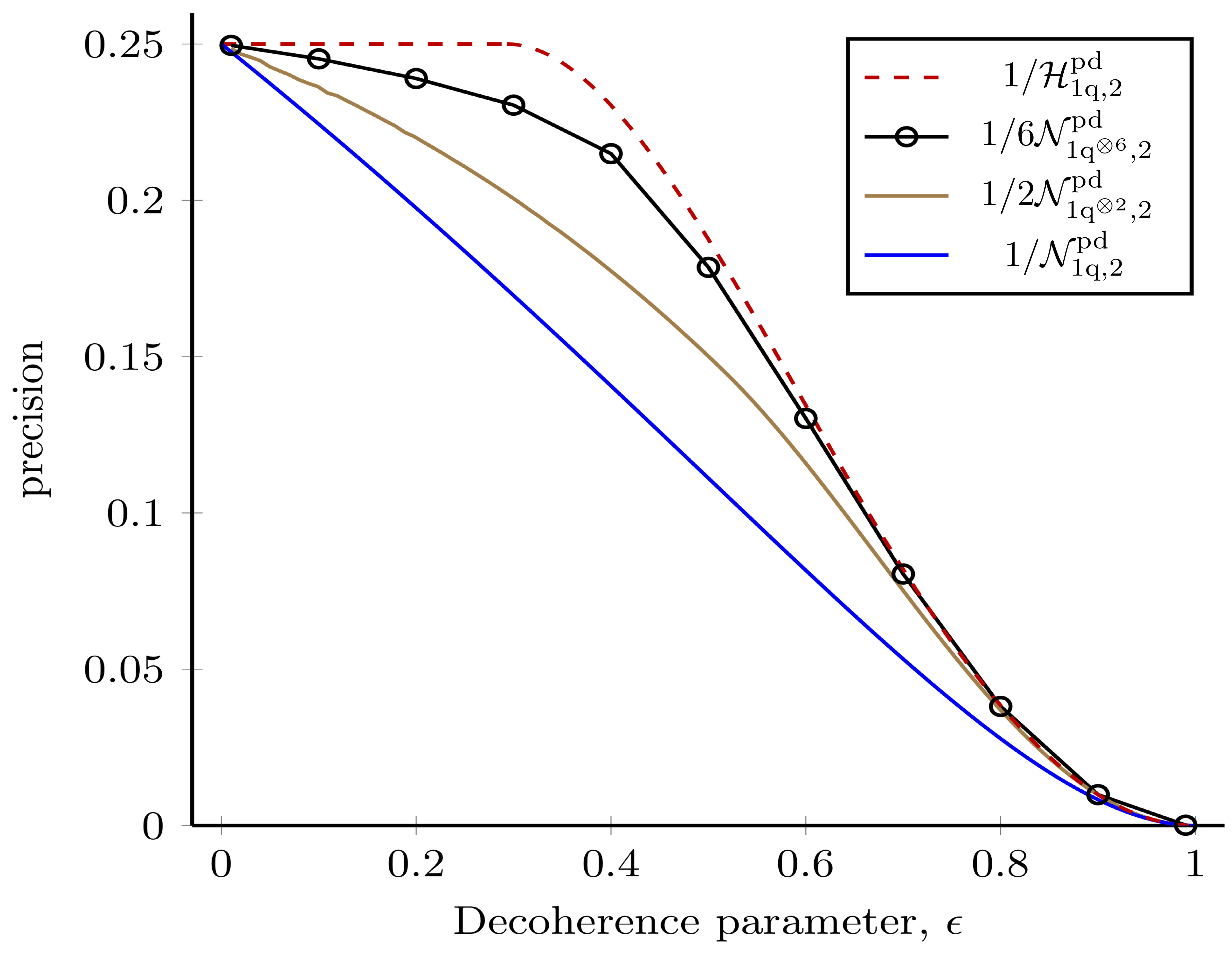}
\caption{\textbf{Achievable precision using collective measurements versus individual measurements in the phase damping channel.} Achievable precision when performing a collective measurement on \textit{M} copies of the same probe. As the number of copies of the probe increases the Nagaoka bound tends towards the Holevo bound as expected. For a single copy of the probe the Nagaoka bound is obtained analytically, for two and six copies of the probe we numerically obtain the Nagaoka bound for $\epsilon$ in the range 0.01$\rightarrow$0.99. \label{PDnagtendhol}}
\end{figure}

Finally, we consider what happens when we can perform measurements on more than one copy of the single qubit probe. Owing to the singularity when the single qubit probe, $\ket{\psi}=\text{a}\ket{0}+\sqrt{1-\text{a}^2}\ket{1}$, has $\text{a}=1$, it is difficult to find analytic results for many copies of this probe. However, it is easy to numerically verify that with an increasing number of copies of the state the Nagaoka bound tends to the Holevo bound as expected. This is plotted in Fig.\ref{PDnagtendhol}.

\newpage

\section{Conclusion}
This case study has examined the performance of single and two qubit probes for multiparameter estimation under the action of several quantum noise channels. The two qubit probes considered offer an increased robustness to environmental noise compared to single qubit probes. It is shown that entanglement is a necessary resource to reach the ultimate limits in quantum metrology, required in both the state preparation and state measurement stages. Having the option of employing entanglement at both of these stages always results in a better or equivalent measurement precision than using entanglement at only one of these stages. However, there is no general hierarchy between using entanglement in the state preparation stage only and using entanglement in the measurement stage only. Indeed, different scenarios have been presented where each of these settings outperforms the other. We have also found that while certain channels allow two qubit probes to outperform single qubit probes, these situations are somewhat unrealistic. In particular, under realistic channels, for a two qubit probe to obtain the maximum advantage over single qubit probes we require that the second qubit is stored in a perfect quantum memory. If the quantum memory we store the second qubit in adds too much noise, while the first qubit passes through the channel, then the single qubit probe can outperform certain two qubit probes. Thus, we have shown that although entanglement in principle improves parameter estimation in qubit states, it also offers potential drawbacks under certain conditions. For each channel we considered the attainability of the Holevo bound with a finite number of copies of the probe state. It was found that a collective measurement on a limited number of copies of the probe state was sufficient to attain a precision close to the Holevo bound. This suggests that the ultimate limits in quantum metrology may be approximately attainable in the near future. 
%It should be noted however, that in any scenario where the Holevo bound does not equal the Nagaoka bound, it is known that no physical measurement exists which saturates the Holevo bound~\cite{conlon2022gap}.

There are several possible ways to extend the results in this paper. We have focused on estimating small fixed rotations, but in a more realistic scenario, the quantum channel may be dynamic, meaning that the rotations we wish to sense are continually changing. In this situation, quantum resources can still offer an advantage~\cite{yonezawa2012quantum}. However, the assumption made in this paper that the angles are well known, i.e. we are performing local estimation, may no longer hold. Our results could be extended to this case through Bayesian estimation~\cite{suzuki2023bayesian,rubio2020bayesian}. Finally, it would be interesting to investigate whether these results can be applied to quantum key distribution, where it is important to estimate the parameters of noisy channels~\cite{thearle2016estimation,wang2017measurement,wang2022twin}.

\section*{Acknowledgements}

This research was funded by the Australian Research Council Centre of Excellence CE170100012, Laureate Fellowship FL150100019 and the Australian Government Research Training Program Scholarship.

%{\color{red}%
%
%
%Bayesian Thermometry :
%
%1) Global Quantum Thermometry
%
%2) Fundamental Limits in Bayesian Thermometry and Attainability via Adaptive Strategies
%}

%Finally we have generalised some of the results obtained. We have shown that it is impossible to construct an unbiased estimator for estimating a rotation about all three axis of the Bloch sphere simultaneously using a single qubit probe. We have also shown that for a broad class of channels, namely those whose Krauss operators can be written in terms of Pauli matrices, a two qubit probe will always be able to estimate a second and third parameter with no loss of precision in the first estimation. Finally we have shown that for a single qubit probe estimating a second parameter causes the achievable precision to degrade by a factor 2. 

\section{Appendix}
\appendix
%\section{Symmetry}
%\label{apena}
%Some of the channels discussed are symmetric about all three axis of the Bloch sphere. In these situations, 
%due to the symmetry of the problem, the optimal probe and measurement must have
%$v_x=v_y$ or $v_x=v_y=v_z$. Let's consider two parameters and assume
%there exists an optimal scheme with $v_x  \neq v_y$. We will show that this leads
%to a contradiction by constructing a scheme which gives a lower
%combined variance $\tilde{v}_x+\tilde{v}_{y} < v_{x}+v_{y}$. The scheme is as follows. If scheme A provides a variance of $\text{Var}\hat\theta_{xA}=v_{1}$ and
%$\text{Var}\hat \theta_{yA}=v_2$, then we can construct scheme B with  $\text{Var}\hat\theta_{xB}=v_{2}$ and
%$\text{Var}\hat \theta_{yB}=v_1$. Measuring the qubits with scheme A half of the
%time and scheme B half of the time will give us two estimates for
%$\theta_x$ and two estimates for $\theta_y$. Consider the estimator
%\begin{align}
%  \label{eq:15}
%\hat \theta_x&=\frac{v_2}{v_1+v_2}\hat\theta_{xA} +
%\frac{v_1}{v_1+v_2}\hat \theta_{xB}\;,\\
%\hat \theta_y&=\frac{v_1}{v_1+v_2}\hat\theta_{yA} +
%\frac{v_2}{v_1+v_2}\hat \theta_{yB}\;.
%\end{align}
%This scheme will have a combined variance of
%\begin{equation}
%  \label{eq:17}
%\text{Var}\hat\theta_{x}  + \text{Var}{\hat\theta_{y}} = \frac{4 v_{1} v_{2}}{v_{1}+v_{2}}
%  < v_{1}+v_{2}\;,
%\end{equation}
%when $v_{1} \neq v_{2}$. Therefore the optimal scheme must have $v_{1}=v_{2}$.
%
%

\section{Computing the SLD bound for single parameter estimation in the decoherence channel}
\label{apennewSLD}
In Fig.~\ref{BS_DC} (i) we present results for single parameter estimation using single qubit probes, obtained through the SLD bound. In this appendix we provide all the necessary information to compute the SLD bound. We consider an input probe state of the form $\ket{\psi}=\text{cos}(\theta/2)\ket{0}+e^{\I\phi}\text{sin}(\theta/2)\ket{1}$. Assuming $\theta_x$ is small, the derivative of the probe state with respect to $\theta_x$, after passing through the decoherence channel is given by
\begin{equation}
\frac{\partial\rho}{\partial\theta_x}=\frac{1-\epsilon}{2}\begin{pmatrix}
\text{sin}(\theta)\text{sin}(\phi)&\mathrm{i}\text{cos}(\theta)\\
-\mathrm{i}\text{cos}(\theta)&\text{sin}(\theta)\text{sin}(\phi)
\end{pmatrix}\;.
\end{equation}
The unnormalised eigenvectors of the probe state after the channel are given by
\begin{equation}
\ket{e_1}=\begin{pmatrix}
\text{e}^{-\mathrm{i}\phi}\left(\frac{1}{\text{tan}(\theta)}-\frac{1}{\text{sin}(\theta)}\right)\\
1
\end{pmatrix}\qquad\text{and}\qquad\ket{e_2}=\begin{pmatrix}
\text{e}^{-\mathrm{i}\phi}\left(\frac{1}{\text{tan}(\theta)}+\frac{1}{\text{sin}(\theta)}\right)\\
1
\end{pmatrix}\;,
\end{equation}
with corresponding eigenvalues of $\lambda_1=\epsilon/2$ and $\lambda_2=1-\epsilon/2$. This allows the SLD operators to be computed from Eq.~\eqref{eq:sld:comp} as
\begin{equation}
\mathcal{L}_x=(1-\epsilon)\begin{pmatrix}
\text{sin}(\theta)\text{sin}(\phi)&\mathrm{i}\text{cos}(\theta)\\
-\mathrm{i}\text{cos}(\theta)&-\text{sin}(\theta)\text{sin}(\phi)
\end{pmatrix}\;.
\end{equation} 
Using this operator and Eq.~\eqref{eq:SLDbound}, the SLD bound can be computed as 
\begin{equation}
\label{eq:SLDappendix}
v_x\geq C^\text{SLD}=\frac{1}{(1-\epsilon)^2(\text{cos}(\theta)^2+\text{sin}(\theta)^2\text{sin}(\phi)^2)}\;.
\end{equation}
It is clear that the variance is minimised when $\phi=\pi/2,3\pi/2$ or when $\theta=0,\pi$. For these values of $\theta$ and $\phi$ it is easily verified that the SLD bound coincides with the Holevo bound, Eq.~\eqref{Hol1}. For the amplitude damping and phase damping channels the SLD bound can be computed in a similar manner.

\section{Explicit construction of $X_{x}$ matrix for single parameter estimation in the decoherence channel}
\label{apenxmat}
Here we demonstrate how the optimal $X_{x}$ matrix for estimating a single parameter using a single qubit probe subject to the decoherence channel is constructed. For this we consider the real single qubit probe, $\ket{\psi}=\ket{0}$. After subjecting this probe to a small rotation about the $x$-axis and some decoherence in the channel we can write
\begin{gather}
\frac{\partial \rho}{\partial \theta_x}=\frac{\mathrm{i}}{2}(1-\epsilon)\left(\ket{0}\bra{1}-\ket{1}\bra{0}\right)\;,\\
\rho=\left(1-\frac{\epsilon}{2}\right)\ket{0}\bra{0}+\frac{\epsilon}{2}\ket{1}\bra{1}\;,
\end{gather}
where $\rho$ is the probe after it has passed through the channel and we have assumed the rotation $\theta_{x}$ is small. We now construct a matrix $X_{x}$
\begin{equation}
X_{x}=A\ket{0}\bra{0}+B\ket{1}\bra{1}+C\left(\ket{0}\bra{1}-\ket{1}\bra{0}\right)\;,
\end{equation}
which must be Hermitian and satisfy the conditions given by Eqs.~\eqref{eq_xcon1} and \eqref{eq_xcon2}. As $\frac{\partial \rho}{\partial \theta_x}$ is purely imaginary this constrains the form of $X_{x}$. From Eqs.~\eqref{eq_xcon1} and~\eqref{eq_xcon2} we get the following conditions
\begin{equation}
C=\frac{\mathrm{i}}{(1-\epsilon)} 
\qquad\text{and}\qquad
A=\frac{-B\epsilon}{2-\epsilon}\;.
\end{equation}
We can then evaluate the Holevo bound using Eq.~\eqref{eq_hol2}.
\begin{equation}
\text{Tr}\{\rho.X_{x}.X_{x}\}=\frac{1}{(1-\epsilon)^2}+\frac{B^2\epsilon}{2-\epsilon}\;.
\end{equation}
This is minimised by taking $A=B=0$, which leaves
\begin{equation}
\text{Tr}\{\rho.X_{x}.X_{x}\}=\frac{1}{(1-\epsilon)^{2}}\;.
\end{equation}
This gives the Holevo bound in Eq.~\eqref{Hol1} and the $X_x$ matrix given by Eq.~\eqref{xone}.

\section{Measurement scheme attaining the Nagaoka bound for single qubit probes subject to the decoherence channel}
\label{apenDC1q}
In this appendix we show a measurement scheme which saturates the Nagaoka bound for single qubit probes estimating one and two parameters in the decoherence channel. From the eigenvectors of the $X_{x}$ and $X_{y}$ matrices for this probe we can construct the optimal POVM for saturating the Nagaoka bound. The POVMs necessary are
\begin{align}
\label{POVM1qDC}
\Pi_{1}&=\frac{1}{4}\left(-\I\ket{0}+\ket{1}\right)\left(\I\bra{0}+\bra{1}\right)\;,\\
\Pi_{2}&=\frac{1}{4}\left(\I\ket{0}+\ket{1}\right)\left(-\I\bra{0}+\bra{1}\right)\;,\\
\Pi_{3}&=\frac{1}{4}\left(\ket{0}+\ket{1}\right)\left(\bra{0}+\bra{1}\right)\;,\\
\Pi_{4}&=\frac{1}{4}\left(-\ket{0}+\ket{1}\right)\left(-\bra{0}+\bra{1}\right)\;.
\label{POVM1qDCend}
\end{align}
The probability of each of the four outcomes is 1/4. We can construct unbiased estimators from these POVMs, using the estimator coefficients $\xi_{x,2}=\xi_{y,3}=-\xi_{x,1}=-\xi_{y,4}=\frac{2}{1-\epsilon}$ and $\xi_{i,j}=0$ for all other $i$ and $j$. This measurement strategy gives a variance of 
\begin{align}
v_x=\frac{1}{4}\left(\xi_{x,1}^2+\xi_{x,2}^2\right) \;,\\
v_y=\frac{1}{4}\left(\xi_{y,3}^2+\xi_{y,4}^2\right) \;,
\end{align}
which gives a total variance of $v_x+v_y=\frac{4}{(1-\epsilon)^2}$ which coincides with Eq.~\eqref{Nag12}. We can also see that by ignoring the POVM's corresponding to estimating $\theta_y$, $\Pi_3$ and $\Pi_4$, the remaining POVMs multiplied by 2 attain the Nagaoka bound for estimating a single parameter from Eq~\eqref{NagD1}.

\section{Measurement saturating the Nagaoka bound for estimating two parameters with a two qubit probe subject to the decoherence channel}
\label{app2q2cDC}
\begin{figure}[t]
\includegraphics[width=1\textwidth]{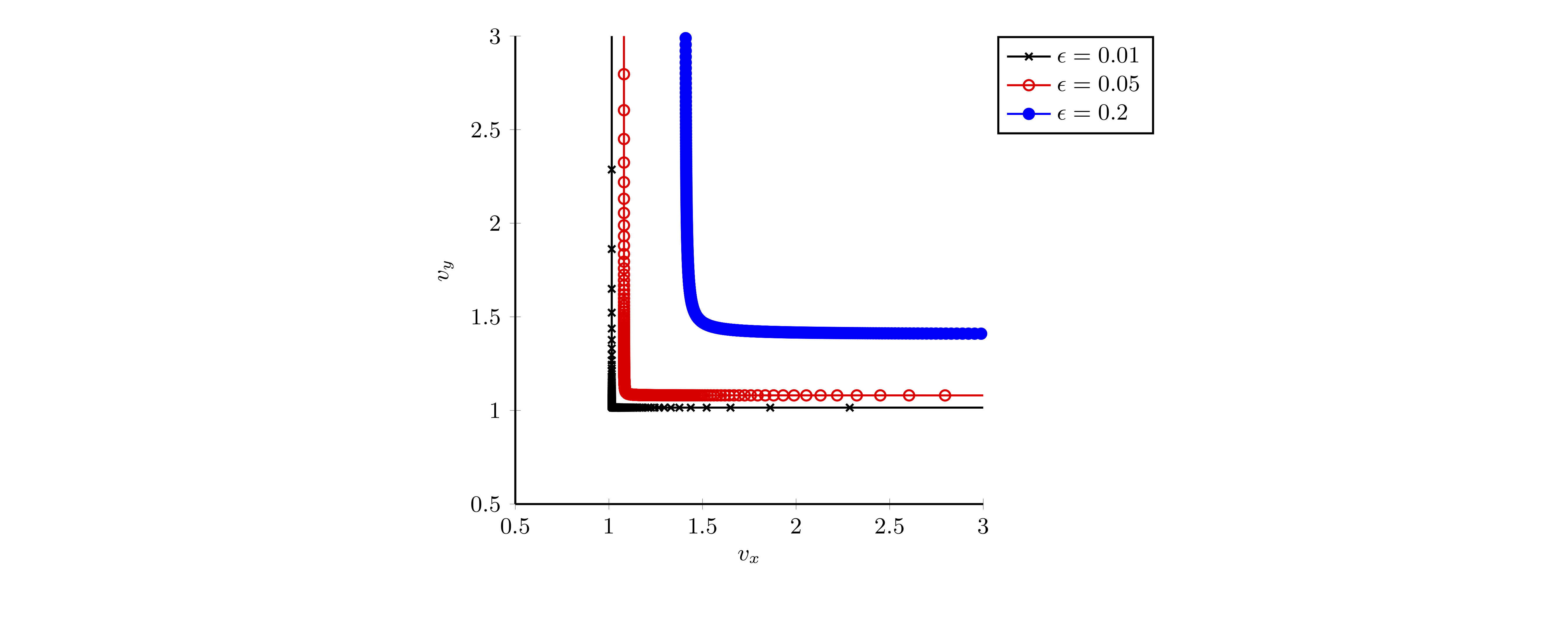}
\caption{\textbf{Tradeoff curve between measuring $\theta_x$ and $\theta_y$ with a two qubit probe in the decoherence channel.} Variance in each estimate given by the measurement strategy in appendix~\ref{app2q2cDC}, for estimating $\theta_x$ and $\theta_y$ simultaneously. We see that we can learn more about $\theta_x$ at the expense of knowing less about $\theta_y$. \label{DCH2N}}
\end{figure}

With a two qubit probe, we claim that individual measurements can achieve a variance of $\frac{4-\epsilon}{2(1-\epsilon)^{2}}$, Eq.\eqref{Nag22DC}. We define the four sub-normalised projectors
\begin{align}
  \begin{drcases}\ket{\phi_1}\\
    \ket{\phi_2}\end{drcases}=\frac{1}{2}
  \begin{pmatrix}1\\\pm a\I\\\pm a\I\\1\end{pmatrix}\qquad \text{and}\qquad
  \begin{drcases}\ket{\phi_3}\\
    \ket{\phi_4}\end{drcases}=\frac{1}{2}
  \begin{pmatrix}1\\\mp b\\\mp b\\-1\end{pmatrix}\;,\;
\end{align}
where $a$ and $b$ are two non-zero real parameters satisfying
$a^{2}+b^{2}\leq 1$. An optimal strategy that saturates the Nagaoka bound for estimating $\theta_x$ and $\theta_y$ consists of measuring
the five-outcome POVM with $\Pi_j = \ket{\phi_j}\bra{\phi_j}$ for $j=1,2,3,4$
and $\Pi_{5}=1-(\Pi_1 + \Pi_2 + \Pi_3 +\Pi_4)$.  The probability for
each POVM outcome is
\begin{equation}
\begin{aligned}
  \begin{drcases}p_1\\
    p_2\end{drcases}&=a^2\left(\frac{1}{2}-\frac{3\epsilon}{8}\right)+\frac{\epsilon}{8}\;,\\
  \begin{drcases}p_3\\
    p_4\end{drcases}&=b^2\left(\frac{1}{2}-\frac{3\epsilon}{8}\right)+\frac{\epsilon}{8}\;,\\
  p_5&=1-0.5\epsilon-\left(a^2+b^2\right)\left(1-\frac{3\epsilon}{4}\right)\;.
\end{aligned}
\end{equation}
From this POVM, we can construct unbiased estimators for $\theta_{x}$ and $\theta_{y}$ with
\begin{equation}
  \begin{gathered}
\xi_{x,1} = -\xi_{x,2} =\frac{1}{(1-\epsilon)a} \;,\qquad\xi_{x,3}=\xi_{x,4}=\xi_{x,5}=0\;,\\
\xi_{y,3}=-\xi_{y,4}=\frac{1}{(1-\epsilon)b}\;,\qquad \xi_{y,1}=\xi_{y,2}=\xi_{y,5}=0\;.
\end{gathered}
\end{equation}
In the asymptotic limit, the variances in our estimate of $\theta_{x}$ and
$\theta_{y}$ are
\begin{equation}
  \begin{aligned}
    v_{x}&=\xi_{x,1}^2 \,p_1 + \xi_{x,2}^2\, p_2
    =\frac{a^2\left(1-\frac{3\epsilon}{4}\right)+\frac{\epsilon}{4}}{a^2(1-\epsilon)^2}\;,\\
    v_{y}&=\xi_{y,3}^2 \,p_3 + \xi_{y,4}^2\,
    p_4=\frac{b^2\left(1-\frac{3\epsilon}{4}\right)+\frac{\epsilon}{4}}{b^2(1-\epsilon)^2} \;,
  \end{aligned}
\end{equation}
which are optimised when $a=b=\frac{1}{\sqrt{2}}$. The sum $v_x+v_y=\frac{4-\epsilon}{2(1-\epsilon)^2}$
saturates the Nagaoka bound as claimed. We can also use the same POVM's to saturate Eq.~\eqref{Hol2qDC}, the Holevo (and Nagaoka) bound for estimating a single parameter in the decoherence channel. For this we require $a=1, b=0$ and we see $v_x=\frac{2-\epsilon}{2(1-\epsilon)^2}$. The cost of this is however, that we learn nothing about $v_y$. This measurement strategy gives rise to a tradeoff curve, between how much we can learn about $\theta_x$ and how much we can learn about $\theta_y$ as shown in Fig.\ref{DCH2N}. 

\section{Measurement scheme required to saturate the NHB for estimating three parameters with the optimal two qubit probe in the decoherence channel}
\label{decohapen3}
We now show that the NHB can also be saturated when we estimate all three parameters simultaneously using a two qubit probe. However, we now require a new set of POVMs:

\begin{align}
  \begin{drcases}\ket{\psi_{x,1}}\\
    \ket{\psi_{x,2}}\end{drcases}=
  \frac{1}{2}\begin{pmatrix}1\\\pm \frac{\I}{\sqrt{3}}\\\pm \frac{\I}{\sqrt{3}} \\1\end{pmatrix}\qquad \text{and}\qquad
  \begin{drcases}\ket{\psi_{y,1}}\\
    \ket{\psi_{y,2}}\end{drcases}=
 \frac{1}{2} \begin{pmatrix}1\\\pm \frac{1}{\sqrt{3}}\\\pm \frac{1}{\sqrt{3}} \\-1\end{pmatrix}\;.
\end{align}
The POVM's which we will use to estimate $\theta_x$ and $\theta_y$ are then given by 
\begin{align}
\Pi_{x,j}&=\ket{\psi_{x,j}}\bra{\psi_{x,j}}\;,\\
\Pi_{y,j}&=\ket{\psi_{y,j}}\bra{\psi_{y,j}}\;.
\end{align}
This gives the first four POVMs. The POVMs required to estimate $\theta_z$ are given by 
\begin{equation}
\begin{drcases}\Pi_{z,1}\\
    \Pi_{z,2}\end{drcases}=\begin{pmatrix}
0&0&0&0\\
0&\frac{1}{3}&-\frac{1}{6}\mp\frac{\I}{\sqrt{12}}&0\\
0&-\frac{1}{6}\pm\frac{\I}{\sqrt{12}}&\frac{1}{3}&0\\
0&0&0&0\\
\end{pmatrix}\;.
\end{equation}
These POVM's already sum to the identity and so no extra POVMs are required. We see that from these we can construct unbiased estimators for $\theta_x$, $\theta_y$ and $\theta_z$ as
\begin{align}
\hat{\theta}_x&=\frac{\sqrt{3}}{1-\epsilon}\left(\Pi_{x,1}-\Pi_{x,2}\right)\;,\\
\hat{\theta}_y&=\frac{\sqrt{3}}{1-\epsilon}\left(\Pi_{y,2}-\Pi_{y,1}\right)\;,\\
\hat{\theta}_z&=\frac{\sqrt{3}}{1-\epsilon}\left(\Pi_{z,1}-\Pi_{z,2}\right)\;,\\
\end{align}
which give variances of
\begin{align}
v_x=\left(\frac{\sqrt{3}}{1-\epsilon}\right)^2\left(\text{Tr}\{\rho\Pi_{x,1}\}+\text{Tr}\{\rho\Pi_{x,2}\}\right)\;,\\
v_y=\left(\frac{\sqrt{3}}{1-\epsilon}\right)^2\left(\text{Tr}\{\rho\Pi_{y,1}\}+\text{Tr}\{\rho\Pi_{y,2}\}\right)\;,\\
v_z=\left(\frac{\sqrt{3}}{1-\epsilon}\right)^2\left(\text{Tr}\{\rho\Pi_{z,1}\}+\text{Tr}\{\rho\Pi_{z,2}\}\right)\;.\\
\end{align}
This gives a total variance of 
\begin{equation}
v_{\text{tot}}=v_x+v_y+v_z=\frac{3}{(1-\epsilon)^2}\;,
\end{equation}
which coincides with the variance from the NHB given in Eq.~\eqref{Ngen3DC}. 

\section{Measurement saturating the Nagaoka bound for estimating two parameters with two copies of a single qubit probe subject to the decoherence channel}
\label{app:1q2cDC}
Using single qubit probes in the decoherence channel the Holevo bound is given by Eq.~\eqref{Hol12}, $\mathcal{H}^\text{d}_{1\text q,2}=\frac{4-2\epsilon}{(1-\epsilon)^{2}}$, and the Nagaoka bound is given by Eq.~\eqref{Nag12}, $\mathcal{N}^\text{d}_{1\text q,2}=\frac{4}{(1-\epsilon)^{2}}$. We also claim that with two copies of the single qubit probe the Nagaoka bound becomes $\mathcal{N}^\text{d}_{1\text q^{\otimes 2},2}=(2-\epsilon+\frac{\epsilon^2}{2})/(1-\epsilon)^2$. The measurement strategy which saturates this bound is very similar to that in appendix~\ref{app2q2cDC}. We use the following four sub-normalised projectors
\begin{align}
  \begin{drcases}\ket{\phi_1}\\
    \ket{\phi_2}\end{drcases}=\frac{1}{2}
  \begin{pmatrix}1\\\pm a\I\\\pm a\I\\-1\end{pmatrix}\qquad \text{and}\qquad
  \begin{drcases}\ket{\phi_3}\\
    \ket{\phi_4}\end{drcases}=\frac{1}{2}
  \begin{pmatrix}1\\\mp b\\\mp b\\1\end{pmatrix}\;,\;
\end{align}
where $a$ and $b$ are two non-zero real parameters satisfying
$a^{2}+b^{2}\leq 1$. An optimal strategy that saturates the Nagaoka bound for estimating $\theta_x$ and $\theta_y$ consists of measuring
the five-outcome POVM with $\Pi_j = \ketbra{\phi_j}$ for $j=1,2,3,4$
and $\Pi_{5}=1-(\Pi_1 + \Pi_2 + \Pi_3 +\Pi_4)$.  The probability for
each POVM outcome is
\begin{equation}
\begin{aligned}
p_1=p_2&=\frac{1}{4}\left(1+(a^2-1)\epsilon+\frac{1}{2}(1-a^2)\epsilon^2\right)\;,\\
p_3=p_4&=\frac{1}{4}\left(1+(b^2-1)\epsilon+\frac{1}{2}(1-b^2)\epsilon^2\right)\;,\\
  p_5&=\frac{1}{4}(2-a^2-b^2)(2-\epsilon)\epsilon\;.
\end{aligned}
\end{equation}
From this POVM, we can construct unbiased estimators for $\theta_{x}$ and $\theta_{y}$ with
\begin{equation}
  \begin{gathered}
\xi_{x,2} = -\xi_{x,1} =\frac{1}{(1-\epsilon)a} \;,\qquad\xi_{x,3}=\xi_{x,4}=\xi_{x,5}=0\;,\\
\xi_{y,3}=-\xi_{y,4}=\frac{1}{(1-\epsilon)b}\;,\qquad \xi_{y,1}=\xi_{y,2}=\xi_{y,5}=0\;.
\end{gathered}
\end{equation}
In the asymptotic limit, the variances in our estimate of $\theta_{x}$ and
$\theta_{y}$ are
\begin{equation}
  \begin{aligned}
    v_{x}&=\xi_{x,1}^2 \,p_1 + \xi_{x,2}^2\, p_2
    =\frac{1+(a^2-1)\epsilon+\frac{1}{2}(1-a^2)\epsilon^2}{2a^2(1-\epsilon)^2}\;,\\
    v_{y}&=\xi_{y,3}^2 \,p_3 + \xi_{y,4}^2\,
    p_4=\frac{1+(b^2-1)\epsilon+\frac{1}{2}(1-b^2)\epsilon^2}{2b^2(1-\epsilon)^2}\;,
  \end{aligned}
\end{equation}
which are optimised when $a=b=\frac{1}{\sqrt{2}}$. The sum $v_x+v_y=(2-\epsilon+\frac{\epsilon^2}{2})/(1-\epsilon)^2$
saturates the Nagaoka bound as claimed.

\section{POVM's attaining the Nagaoka bound for single qubit probes subject to the amplitude damping channel}
\label{apenAD1q}
Owing to the similarities between the optimal $X$ matrices required for the amplitude damping channel and the decoherence channel, the POVMs required to saturate the Nagaoka bound in the amplitude damping channel are the same as those given in Eqs.~\eqref{POVM1qDC} to \eqref{POVM1qDCend}. In order to construct unbiased estimators we require the estimator coefficients $\xi_{x,1}=\xi_{y,3}=-\xi_{x,2}=-\xi_{y,4}=\frac{2}{\sqrt{1-p}}$ and $\xi_{i,j}=0$ for all other $i$ and $j$. This measurement gives $v_x+v_y=\frac{4}{1-p}$, in agreement with Eq.~\eqref{amp1nag}. As before we can consider only the first two POVMs, scaled by a factor of two and we arrive at the Holevo bound (and Nagaoka bound) for estimating a single parameter, $v_x=\frac{1}{1-p}$.

\section{Optimal estimator for estimating two parameters
  in the amplitude damping channel with two copies of the single-qubit probe}
  \label{11est}
We use the same POVMs as described in appendix \ref{app:1q2cDC} to saturate the Nagaoka bound in Eq.~\eqref{nagad1q2c22}. The probability for
each POVM outcome is now
\begin{equation}
\begin{aligned}
p_1=p_2&=\frac{1}{4}\left(1+2(a^2-1)p+2(1-a^2)p^2\right)\;,\\
p_3=p_4&=\frac{1}{4}\left(1+2(b^2-1)p+2(1-b^2)p^2\right)\;,\\
  p_5&=\left(2-a^2-b^2\right)(1-p)p\;.
\end{aligned}
\end{equation}
We require the following estimator coefficients
\begin{equation}
  \begin{gathered}
\xi_{x,1} = -\xi_{x,2} =\frac{1}{a\sqrt{1-p}} \;,\qquad\xi_{x,3}=\xi_{x,4}=\xi_{x,5}=0\;,\\
\xi_{y,3}=-\xi_{y,4}=\frac{1}{b\sqrt{1-p}}\;,\qquad \xi_{y,1}=\xi_{y,2}=\xi_{y,5}=0\;.
\end{gathered}
\end{equation}
In the asymptotic limit, the variances in our estimate of $\theta_{x}$ and
$\theta_{y}$ become
\begin{equation}
  \begin{aligned}
    v_{x}&=\xi_{x,1}^2 \,p_1 + \xi_{x,2}^2\, p_2
    =\frac{0.5+(a^2-1)p+(1-a^2)p^2}{a^2(1-p)^2}\;,\\
    v_{y}&=\xi_{y,3}^2 \,p_3 + \xi_{y,4}^2\,
    p_4=\frac{0.5+(b^2-1)p+(1-b^2)p^2}{b^2(1-p)^2}\;,
  \end{aligned}
\end{equation}
which are optimised when $a=b=\frac{1}{\sqrt{2}}$. The sum $v_x+v_y=\frac{2-2p+2p^2}{1-p}$
saturates the Nagaoka bound in Eq.~\eqref{nagad1q2c22} as claimed.
  
 \section{Explicit construction of the optimal estimator for estimating a single parameter
  in the phase damping channel with a single qubit probe}
   From Fig.~\ref{fig:PD_xy_comp} we know that for estimating a small rotation about the $x$-axis, the probe $\ket{\psi}=\text{cos}(\theta/2)\ket{0}+\I\text{sin}(\theta/2)$ is optimal for all $\theta$ except $\theta=0$ or 1. By constructing the optimal estimator for this channel we show why this is true. After the channel the probe and its derivative become
   \begin{equation}
   \rho=\begin{pmatrix}
   \text{cos}(\frac{\theta}{2})^2&-\frac{\I}{2}(1-\epsilon)\text{sin}(\theta)\\
   \frac{\I}{2}(1-\epsilon)\text{sin}(\theta)&\text{sin}(\frac{\theta}{2})^2
   \end{pmatrix}\qquad\text{and}\qquad
   \frac{\partial\rho}{\partial\theta_x}=\begin{pmatrix}
   \frac{\text{sin}(\theta)}{2}&\frac{\I}{2}(1-\epsilon)\text{cos}(\theta)\\
  -\frac{\I}{2}(1-\epsilon)\text{cos}(\theta)&-\frac{\text{sin}(\theta)}{2}
   \end{pmatrix}\;,
   \end{equation}
  where we have taken the limit $\theta_x\rightarrow0$. For this derivative the optimal $X_x$ matrix is
  \begin{equation}
  X_x=\begin{pmatrix}
  \text{tan}(\frac{\theta}{2})&0\\
  0&-\text{cot}(\frac{\theta}{2})
  \end{pmatrix}\;,
  \end{equation}
  which gives a Holevo bound equal to 1 irrespective of the value of $\epsilon$. However, the optimal matrix $X_x$ contains infinities for either $\ket{0}$ and $\ket{1}$ and so a different sub-optimal solution is found exactly at these points.

\section{Explicit construction of the optimal estimator for estimating two parameters
  in the phase damping channel with a single qubit probe}
  \label{nagexp}
Consider the probe $\ket{\psi}=a\ket{0}+b\ket{1}$ where $a=\cos\theta$ and
$b=\sin \theta$ are real numbers. This probe passes through a channel
rotating it by $\theta_x$ around the $x$-axis and $\theta_y$
around the $y$-axis. It then goes through a phase damping channel
parametrised by $\epsilon$. The probe after the channel becomes

\begin{align}
  \label{eq:1}
  \rho(\theta_x,\theta_y) =
  \begin{pmatrix}
    a^2 + a b \theta_y & (1-\epsilon)\left(a b+\frac{a^2-b^2}{2}\left(-\mathrm{i} \theta_x-\theta_y \right)\right)\\
 (1-\epsilon)\left(a b+\frac{a^2-b^2}{2}\left(\mathrm{i} \theta_x-\theta_y \right)\right)&    b^2 - a b \theta_y 
    \end{pmatrix}\;
    \end{align}
where we have assumed that both $\theta_x$ and $\theta_y$ are
small. Suppose we have a total of $n$ copies. We choose to measure
$n_y$ copies in the $\sigma_z$ basis, and the remaining $n_x$ copies
in the $\sigma_y$ basis. When measuring in the $\sigma_z$ basis, the
two outcomes occur with probabilities
\begin{align}
  \label{eq:2}
  p(\pi_1|\theta_x,\theta_y) &= a(a+b\theta_y)\;,\\
  p(\pi_2|\theta_x,\theta_y) &= b(b-a\theta_y)\;,
\end{align}
which is independent of $\theta_x$. When $a=0$ or $a=1$, the outcomes
(up to first order) are also independent of $\theta_y$. When measuring in the $\sigma_y$ basis, the
two outcomes occur with probabilities
\begin{align}
  p(\pi_3|\theta_x,\theta_y) &= \frac{1}{2} + \frac{1-\epsilon}{2}\left(a^2-b^2\right)\theta_x\;,\\
  p(\pi_4|\theta_x,\theta_y) &= \frac{1}{2} - \frac{1-\epsilon}{2}\left(a^2-b^2\right)\theta_x\;.
\end{align}
The outcomes of these measurements are independent of $\theta_y$. From
these four outcomes, we construct the two independent unbiased estimators for
$\theta_x$ and $\theta_y$ as
\begin{align}
  \hat{\theta}_y &= \frac{n_1-n_2}{2 a\, b\, n_y}-\frac{a^2-b^2}{2 a\, b}\;,\\
  \hat{\theta}_x &= \frac{n_3-n_4}{n_x(1-\epsilon)\left(a^2-b^2\right)}\;,
\end{align}
where the different $n_j$ are the number of times we get the $j$-th outcome. Note that
$\hat\theta_y$ is not well defined when $a=1$ or $0$.

The variance of $\hat{\theta}_y$ is given by
\begin{align}
  \label{eq:4}
  \text{var}\left(\hat\theta_y\right) &=
                             \text{var}\left(\frac{n_1-n_2}{2a\,b\,n_y}\right)\\
 % &=
 %   \text{var}\left(\frac{2n_1-n_y}{2a\,b\,n_y}\right)\\
    %                       &=\frac{\text{var}(n_1)}{a^2 b^2 n_y^2}\\
                    %       &=\frac{n_y p_1 p_2}{ a^2 b^2 n_y^2}\\
  &=\frac{1}{n_y}\;.
\end{align}
We have used the fact that $n_1$ follows a binomial distribution and
has variance given by $\text{var}(n_1)=n_y p_1 p_2$ and that $p_1=a^2$
and $p_2=b^2$ when $\theta_y$ is small. Notice that the
variance of $\hat\theta_y$ does not depend on $a$ or $\epsilon$. Next, the variance of $\hat\theta_x$ is
\begin{align}
  \label{eq:5}
  \text{var}\left(\hat\theta_x\right) &= \text{var}\left(
                             \frac{n_3-n_4}{n_x(1-\epsilon)(a^2-b^2)}\right)\\
% &= \text{var}\left(
                             %\frac{2 n_3-n_x}{n_x(1-\epsilon)(a^2-b^2)}\right)\\
                          % &=\frac{4\text{var}(n_3)}{n_x^2(1-\epsilon)^2(a^2-b^2)^2}\\
                       % &=\frac{4n_x p_3 p_4}{n_x^2(1-\epsilon)^2(a^2-b^2)^2}\\
                        &=\frac{1}{n_x(1-\epsilon)^2(a^2-b^2)^2}\;.
\end{align}
We have used that fact that when $\theta_{x}$ is small, $p_{3}=p_{4}=\frac{1}{2}$. This is smallest when $a$ approaches $1$ or $0$ where
\begin{align}
  \label{eq:6}
  \text{var}\left(\hat\theta_x\right) = \frac{1}{n_x(1-\epsilon)^2}\;.
\end{align}
Putting everything together, we get
\begin{align}
  \label{eq:7}
\text{var}\left(\hat\theta_y\right)+\text{var}\left(\hat\theta_x\right)  = \frac{1}{n_y} + \frac{1}{n_x(1-\epsilon)^2}\;.
\end{align}
All that remains is to find $n_y$ and $n_x$ that minimises the above
expression subject to a fixed total number of probes, $n_y+n_x=n$. This optimal
fraction is given by
\begin{align}
  \label{eq:9}
  n_y=\frac{1-\epsilon}{2-\epsilon}n,\\
  n_x=\frac{1}{2-\epsilon}n\;,\\
\end{align}
which gives
\begin{align}
\text{var}\left(\hat\theta_y\right)+\text{var}\left(\hat\theta_x\right)  = \left(\frac{2-\epsilon}{1-\epsilon}\right)^2\frac{1}{n}\;.
\end{align}
This expression coincides with the Nagaoka bound given in Eq.~\eqref{nag12pd}.

\section{Measurement scheme which saturates the NHB for estimating three parameters in the phase damping channel
  with a two-qubit probe}
\label{phasedampapen}
The measurements presented in this appendix have already been presented in Ref.~\cite{conlon2021efficient}, however we include them here for completeness. We first describe the measurement strategy required to saturate the Nagaoka bound for estimating $\theta_x$ and $\theta_y$ simultaneously using a two qubit probe subject to the phase damping channel. We require the same POVMs as used in appendix \ref{app2q2cDC}. The probability for
each POVM outcome is
\begin{equation}
\begin{aligned}
  \begin{drcases}p_1\\
    p_2\end{drcases}&=\frac{1}{4}a(2-\epsilon)(a\pm \theta_x)\;,\\
  \begin{drcases}p_3\\
    p_4\end{drcases}&=\frac{1}{4}b(2-\epsilon)(b\pm \theta_y)\;,\\
  p_5&=1- \frac{1}{2}(2-\epsilon)\left(a^2+b^2\right)\;.
\end{aligned}
\end{equation}
The unbiased estimator coefficients for $\theta_{x}$ and $\theta_{y}$ are given by
\begin{equation}
  \begin{gathered}
\xi_{x,1} = -\xi_{x,2} =\frac{2}{(2-\epsilon)a} \;,\qquad\xi_{x,3}=\xi_{x,4}=\xi_{x,5}=0\;,\\
\xi_{y,3}=-\xi_{y,4}=\frac{2}{(2-\epsilon)b}\;,\qquad \xi_{y,1}=\xi_{y,2}=\xi_{y,5}=0\;.
\end{gathered}
\end{equation}
The fifth outcome $\Pi_5$ is only included to ensure the POVM elements sum to the identity, it does not improve the measurement precision. For a finite sample, to have a better estimate of $\theta_x$ and
$\theta_y$ it is beneficial to have both $a$ and $b$ large so
the outcomes $\Pi_1$ to $\Pi_4$ occur more often. However, in the
asymptotic limit, the variances in our estimate of $\theta_{x}$ and
$\theta_{y}$ are
\begin{equation}
  \begin{aligned}
    v_{x}&=\xi_{x,1}^2 \,p_1 + \xi_{x,2}^2\, p_2
    =\frac{4(p_1+p_2)}{(2-\epsilon)^2 a^2}=\frac{2}{2-\epsilon}\;,\\
    v_{y}&=\xi_{y,3}^2 \,p_3 + \xi_{y,4}^2\,
    p_4=\frac{4(p_3+p_4)}{(2-\epsilon)^2 b^2}=\frac{2}{2-\epsilon}\;,
  \end{aligned}
\end{equation}
which do not depend on $a$ or $b$. The sum $v_x+v_y=4/(2-\epsilon)$
saturates the Nagaoka bound as claimed.

We now consider estimating all three parameters $\theta_x$, $\theta_y$ and
$\theta_z$. One measurement strategy is to
use the same POVM outcomes for estimating
$\theta_x$ and $\theta_y$ but splitting $\Pi_5$ to get some
information on $\theta_z$. Ideally, we would like to set $a=b=0$ and use the following four
projectors
\begin{align}
  \label{nag_2pp}
  \Pi_1=\Pi_2&=\frac{1}{4}\begin{pmatrix}1&0&0&1\\
    0&0&0&0\\
    0&0&0&0\\
    1&0&0&1
  \end{pmatrix},\qquad
  \Pi_3=\Pi_4=\frac{1}{4}\begin{pmatrix}1&0&0&-1\\
    0&0&0&0\\
    0&0&0&0\\
    -1&0&0&1
  \end{pmatrix},\qquad
\Pi_5= \frac{1}{2}\begin{pmatrix}0&0&0&0\\
    0&1&\I&0\\
    0&-\I&1&0\\
    0&0&0&0
  \end{pmatrix}+
    \frac{1}{2}\begin{pmatrix}0&0&0&0\\
    0&1&-\I&0\\
    0&\I&1&0\\
    0&0&0&0
  \end{pmatrix}\;,
\end{align}
to obtain the most information on $\theta_z$ without affecting the estimate
of $\theta_x$ and $\theta_y$. But the problem is that at this
singular point, the first
four outcomes $\Pi_1$, $\Pi_2$, $\Pi_3$ and $\Pi_4$ do not give any
information on $\theta_x$ and $\theta_y$. To fix this, we need both $a$
and $b$ to be close to but not exactly zero. Writing
$\delta=(a^2+b^2)/2$, we can split $\Pi_5$ as
\begin{align}
  \Pi_{5}&=\begin{pmatrix}0&0&0&0\\
    0&1-\delta&-\delta&0\\
    0&-\delta&1-\delta&0\\
    0&0&0&0
  \end{pmatrix}\\
  &=\underbrace{\delta\begin{pmatrix}0&0&0&0\\
    0&1&-1&0\\
    0&-1&1&0\\
    0&0&0&0
  \end{pmatrix}}_{\Pi_5^{(3)}}+
\underbrace{          \frac{1-2\delta}{2}\begin{pmatrix}
             0&0&0&0\\
             0&1&-\I&0\\
             0&\I&1&0\\
             0&0&0&0
  \end{pmatrix}}_{\Pi_6^{(3)}}+
\underbrace{           \frac{1-2\delta}{2}\begin{pmatrix}0&0&0&0\\
    0&1&\I&0\\
    0&-\I&1&0\\
    0&0&0&0
  \end{pmatrix}}_{\Pi_7^{(3)}}\;,
\end{align}
which has outcome probabilities
\begin{equation}
\begin{aligned}
  p_5&=\delta \,\epsilon\;,\\
  \begin{drcases}p_6\\
    p_7\end{drcases}&=\frac{1}{2}(1-2\delta)\left(1\pm (1-\epsilon
    )\theta_z \right)\;.
\end{aligned}
\end{equation}
This together with
\begin{equation}
  \xi_{z,1}=\xi_{z,2}=\xi_{z,3}=\xi_{z,4}=\xi_{z,5}=0\;,\qquad
  \text{and}\qquad \xi_{z,6}=-\xi_{z,7}=\frac{1}{(1-\epsilon)(1-2\delta)}\;,
\end{equation}
give a variance for estimating $\theta_z$ as
$  v_z=\dfrac{1}{(1-\epsilon)^2(1-2\delta)} $
which approaches $v_z=\dfrac{1}{(1-\epsilon)^2}$ as $\delta$ tends to zero.

\bibliography{ref}

\begin{thebibliography}{159}
\providecommand{\natexlab}[1]{#1}
\providecommand{\url}[1]{\texttt{#1}}
\expandafter\ifx\csname urlstyle\endcsname\relax
  \providecommand{\doi}[1]{doi: #1}\else
  \providecommand{\doi}{doi: \begingroup \urlstyle{rm}\Url}\fi

\bibitem[Robertson(1929)]{robertson1929uncertainty}
Howard~Percy Robertson.
\newblock The uncertainty principle.
\newblock \emph{Physical Review}, 34\penalty0 (1):\penalty0 163, 1929.

\bibitem[Arthurs and Kelly(1965)]{arthurs1965bstj}
E~Arthurs and JL~Kelly.
\newblock Bstj briefs: On the simultaneous measurement of a pair of conjugate
  observables.
\newblock \emph{The Bell System Technical Journal}, 44\penalty0 (4):\penalty0
  725--729, 1965.

\bibitem[Heisenberg(1985)]{heisenberg1985quantentheoretische}
Werner Heisenberg.
\newblock {\"U}ber quantentheoretische umdeutung kinematischer und mechanischer
  beziehungen.
\newblock In \emph{Original Scientific Papers Wissenschaftliche
  Originalarbeiten}, pages 382--396. Springer, 1985.

\bibitem[Caves(1981)]{caves1981quantum}
Carlton~M Caves.
\newblock Quantum-mechanical noise in an interferometer.
\newblock \emph{Physical Review D}, 23\penalty0 (8):\penalty0 1693, 1981.

\bibitem[Barnett et~al.(2003)Barnett, Fabre, and
  Ma\^{i}tre]{barnett2003ultimate}
Stephen~M Barnett, Claude Fabre, and Agnes Ma\^{i}tre.
\newblock Ultimate quantum limits for resolution of beam displacements.
\newblock \emph{The European Physical Journal D-Atomic, Molecular, Optical and
  Plasma Physics}, 22\penalty0 (3):\penalty0 513--519, 2003.

\bibitem[Dorner et~al.(2009)Dorner, Demkowicz-Dobrza{\'n}ski, Smith, Lundeen,
  Wasilewski, Banaszek, and Walmsley]{dorner2009optimal}
U~Dorner, R~Demkowicz-Dobrza{\'n}ski, BJ~Smith, JS~Lundeen, W~Wasilewski,
  K~Banaszek, and IA~Walmsley.
\newblock Optimal quantum phase estimation.
\newblock \emph{Physical Review Letters}, 102\penalty0 (4):\penalty0 040403,
  2009.

\bibitem[Demkowicz-Dobrza{\'n}ski et~al.(2009)Demkowicz-Dobrza{\'n}ski, Dorner,
  Smith, Lundeen, Wasilewski, Banaszek, and Walmsley]{demkowicz2009quantum}
R~Demkowicz-Dobrza{\'n}ski, U~Dorner, BJ~Smith, JS~Lundeen, W~Wasilewski,
  K~Banaszek, and IA~Walmsley.
\newblock Quantum phase estimation with lossy interferometers.
\newblock \emph{Physical Review A}, 80\penalty0 (1):\penalty0 013825, 2009.

\bibitem[Zhuang et~al.(2018)Zhuang, Zhang, and Shapiro]{zhuang2018distributed}
Quntao Zhuang, Zheshen Zhang, and Jeffrey~H Shapiro.
\newblock Distributed quantum sensing using continuous-variable multipartite
  entanglement.
\newblock \emph{Physical Review A}, 97\penalty0 (3):\penalty0 032329, 2018.

\bibitem[Ge et~al.(2018)Ge, Jacobs, Eldredge, Gorshkov, and
  Foss-Feig]{ge2018distributed}
Wenchao Ge, Kurt Jacobs, Zachary Eldredge, Alexey~V Gorshkov, and Michael
  Foss-Feig.
\newblock Distributed quantum metrology with linear networks and separable
  inputs.
\newblock \emph{Physical review letters}, 121\penalty0 (4):\penalty0 043604,
  2018.

\bibitem[Conlon et~al.(2022{\natexlab{a}})Conlon, Michel, Guccione, McKenzie,
  Assad, and Lam]{conlon2022enhancing}
Lorc{\'a}n~O Conlon, Thibault Michel, Giovanni Guccione, Kirk McKenzie, Syed~M
  Assad, and Ping~Koy Lam.
\newblock Enhancing the precision limits of interferometric satellite geodesy
  missions.
\newblock \emph{npj Microgravity}, 8\penalty0 (1):\penalty0 21,
  2022{\natexlab{a}}.

\bibitem[Tsang et~al.(2016)Tsang, Nair, and Lu]{tsang2016quantum}
Mankei Tsang, Ranjith Nair, and Xiao-Ming Lu.
\newblock Quantum theory of superresolution for two incoherent optical point
  sources.
\newblock \emph{Physical Review X}, 6\penalty0 (3):\penalty0 031033, 2016.

\bibitem[Tsang(2019)]{tsang2019resolving}
Mankei Tsang.
\newblock Resolving starlight: a quantum perspective.
\newblock \emph{Contemporary Physics}, 60\penalty0 (4):\penalty0 279--298,
  2019.

\bibitem[Giovannetti et~al.(2001)Giovannetti, Lloyd, and
  Maccone]{giovannetti2001quantum}
Vittorio Giovannetti, Seth Lloyd, and Lorenzo Maccone.
\newblock Quantum-enhanced positioning and clock synchronization.
\newblock \emph{Nature}, 412\penalty0 (6845):\penalty0 417, 2001.

\bibitem[Lamine et~al.(2008)Lamine, Fabre, and Treps]{lamine2008quantum}
Brahim Lamine, Claude Fabre, and Nicolas Treps.
\newblock Quantum improvement of time transfer between remote clocks.
\newblock \emph{Physical Review Letters}, 101\penalty0 (12):\penalty0 123601,
  2008.

\bibitem[Brady et~al.(2022)Brady, Gao, Harnik, Liu, Zhang, and
  Zhuang]{brady2022entangled}
Anthony~J Brady, Christina Gao, Roni Harnik, Zhen Liu, Zheshen Zhang, and
  Quntao Zhuang.
\newblock Entangled sensor-networks for dark-matter searches.
\newblock \emph{PRX Quantum}, 3\penalty0 (3):\penalty0 030333, 2022.

\bibitem[Marchese et~al.(2023)Marchese, Belenchia, and
  Paternostro]{marchese2023optomechanics}
Marta~Maria Marchese, Alessio Belenchia, and Mauro Paternostro.
\newblock Optomechanics-based quantum estimation theory for collapse models.
\newblock \emph{Entropy}, 25\penalty0 (3):\penalty0 500, 2023.

\bibitem[Shi and Zhuang(2023)]{shi2023ultimate}
Haowei Shi and Quntao Zhuang.
\newblock Ultimate precision limit of noise sensing and dark matter search.
\newblock \emph{npj Quantum Information}, 9\penalty0 (1):\penalty0 27, 2023.

\bibitem[Higgins et~al.(2007)Higgins, Berry, Bartlett, Wiseman, and
  Pryde]{higgins2007entanglement}
Brendon~L Higgins, Dominic~W Berry, Stephen~D Bartlett, Howard~M Wiseman, and
  Geoff~J Pryde.
\newblock {Entanglement-free Heisenberg-limited phase estimation}.
\newblock \emph{Nature}, 450\penalty0 (7168):\penalty0 393--396, 2007.

\bibitem[Kacprowicz et~al.(2010)Kacprowicz, Demkowicz-Dobrza{\'n}ski,
  Wasilewski, Banaszek, and Walmsley]{kacprowicz2010experimental}
M~Kacprowicz, R~Demkowicz-Dobrza{\'n}ski, W~Wasilewski, K~Banaszek, and
  IA~Walmsley.
\newblock Experimental quantum-enhanced estimation of a lossy phase shift.
\newblock \emph{Nature Photonics}, 4\penalty0 (6):\penalty0 357, 2010.

\bibitem[Yonezawa et~al.(2012)Yonezawa, Nakane, Wheatley, Iwasawa, Takeda,
  Arao, Ohki, Tsumura, Berry, Ralph, et~al.]{yonezawa2012quantum}
Hidehiro Yonezawa, Daisuke Nakane, Trevor~A Wheatley, Kohjiro Iwasawa, Shuntaro
  Takeda, Hajime Arao, Kentaro Ohki, Koji Tsumura, Dominic~W Berry, Timothy~C
  Ralph, et~al.
\newblock Quantum-enhanced optical-phase tracking.
\newblock \emph{Science}, 337\penalty0 (6101):\penalty0 1514--1517, 2012.

\bibitem[Girolami et~al.(2014)Girolami, Souza, Giovannetti, Tufarelli,
  Filgueiras, Sarthour, Soares-Pinto, Oliveira, and
  Adesso]{girolami2014quantum}
Davide Girolami, Alexandre~M Souza, Vittorio Giovannetti, Tommaso Tufarelli,
  Jefferson~G Filgueiras, Roberto~S Sarthour, Diogo~O Soares-Pinto, Ivan~S
  Oliveira, and Gerardo Adesso.
\newblock Quantum discord determines the interferometric power of quantum
  states.
\newblock \emph{Physical Review Letters}, 112\penalty0 (21):\penalty0 210401,
  2014.

\bibitem[Strobel et~al.(2014)Strobel, Muessel, Linnemann, Zibold, Hume,
  Pezz{\`e}, Smerzi, and Oberthaler]{strobel2014fisher}
Helmut Strobel, Wolfgang Muessel, Daniel Linnemann, Tilman Zibold, David~B
  Hume, Luca Pezz{\`e}, Augusto Smerzi, and Markus~K Oberthaler.
\newblock {Fisher information and entanglement of non-Gaussian spin states}.
\newblock \emph{Science}, 345\penalty0 (6195):\penalty0 424--427, 2014.

\bibitem[Slussarenko et~al.(2017)Slussarenko, Weston, Chrzanowski, Shalm,
  Verma, Nam, and Pryde]{slussarenko2017unconditional}
Sergei Slussarenko, Morgan~M Weston, Helen~M Chrzanowski, Lynden~K Shalm,
  Varun~B Verma, Sae~Woo Nam, and Geoff~J Pryde.
\newblock Unconditional violation of the shot-noise limit in photonic quantum
  metrology.
\newblock \emph{Nature Photonics}, 11\penalty0 (11):\penalty0 700--703, 2017.

\bibitem[Zhang et~al.(2019)Zhang, Zheng, Liu, Zhao, Tang, Yonezawa, Zhang,
  Zhang, and Xiao]{zhang2019quantum}
Lidan Zhang, Kaimin Zheng, Fang Liu, Wei Zhao, Lei Tang, Hidehiro Yonezawa,
  Lijian Zhang, Yong Zhang, and Min Xiao.
\newblock Quantum-limited fiber-optic phase tracking beyond $\pi$ range.
\newblock \emph{Optics Express}, 27\penalty0 (3):\penalty0 2327--2334, 2019.

\bibitem[McCormick et~al.(2019)McCormick, Keller, Burd, Wineland, Wilson, and
  Leibfried]{mccormick2019quantum}
Katherine~C McCormick, Jonas Keller, Shaun~C Burd, David~J Wineland, Andrew~C
  Wilson, and Dietrich Leibfried.
\newblock Quantum-enhanced sensing of a single-ion mechanical oscillator.
\newblock \emph{Nature}, 572\penalty0 (7767):\penalty0 86--90, 2019.

\bibitem[Wang et~al.(2019)Wang, Wu, Ma, Cai, Hu, Mu, Xu, Chen, Wang, Song,
  et~al.]{wang2019heisenberg}
Weiting Wang, Yukai Wu, Yuwei Ma, Weizhou Cai, Ling Hu, Xianghao Mu, Yuan Xu,
  Zi-Jie Chen, Haiyan Wang, YP~Song, et~al.
\newblock {Heisenberg-limited single-mode quantum metrology in a
  superconducting circuit}.
\newblock \emph{Nature communications}, 10\penalty0 (1):\penalty0 1--6, 2019.

\bibitem[Aasi et~al.(2013)Aasi, Abadie, Abbott, Abbott, Abbott, Abernathy,
  Adams, Adams, Addesso, Adhikari, et~al.]{aasi2013enhanced}
Junaid Aasi, J~Abadie, BP~Abbott, Richard Abbott, TD~Abbott, MR~Abernathy, Carl
  Adams, Thomas Adams, Paolo Addesso, RX~Adhikari, et~al.
\newblock {Enhanced sensitivity of the LIGO gravitational wave detector by
  using squeezed states of light}.
\newblock \emph{Nature Photonics}, 7\penalty0 (8):\penalty0 613--619, 2013.

\bibitem[Guo et~al.(2020)Guo, Breum, Borregaard, Izumi, Larsen, Gehring,
  Christandl, Neergaard-Nielsen, and Andersen]{guo2020distributed}
Xueshi Guo, Casper~R Breum, Johannes Borregaard, Shuro Izumi, Mikkel~V Larsen,
  Tobias Gehring, Matthias Christandl, Jonas~S Neergaard-Nielsen, and Ulrik~L
  Andersen.
\newblock Distributed quantum sensing in a continuous-variable entangled
  network.
\newblock \emph{Nature Physics}, 16\penalty0 (3):\penalty0 281--284, 2020.

\bibitem[Liu et~al.(2021)Liu, Zhang, Li, Zhang, Yin, Fei, Li, Liu, Xu, Chen,
  et~al.]{liu2021distributed}
Li-Zheng Liu, Yu-Zhe Zhang, Zheng-Da Li, Rui Zhang, Xu-Fei Yin, Yue-Yang Fei,
  Li~Li, Nai-Le Liu, Feihu Xu, Yu-Ao Chen, et~al.
\newblock Distributed quantum phase estimation with entangled photons.
\newblock \emph{Nature Photonics}, 15\penalty0 (2):\penalty0 137--142, 2021.

\bibitem[Backes et~al.(2021)Backes, Palken, Kenany, Brubaker, Cahn, Droster,
  Hilton, Ghosh, Jackson, Lamoreaux, et~al.]{backes2021quantum}
Kelly~M Backes, Daniel~A Palken, S~Al Kenany, Benjamin~M Brubaker, SB~Cahn,
  A~Droster, Gene~C Hilton, Sumita Ghosh, H~Jackson, Steve~K Lamoreaux, et~al.
\newblock A quantum enhanced search for dark matter axions.
\newblock \emph{Nature}, 590\penalty0 (7845):\penalty0 238--242, 2021.

\bibitem[Casacio et~al.(2021)Casacio, Madsen, Terrasson, Waleed, Barnscheidt,
  Hage, Taylor, and Bowen]{casacio2021quantum}
Catxere~A Casacio, Lars~S Madsen, Alex Terrasson, Muhammad Waleed, Kai
  Barnscheidt, Boris Hage, Michael~A Taylor, and Warwick~P Bowen.
\newblock Quantum-enhanced nonlinear microscopy.
\newblock \emph{Nature}, 594\penalty0 (7862):\penalty0 201--206, 2021.

\bibitem[Marciniak et~al.(2022)Marciniak, Feldker, Pogorelov, Kaubruegger,
  Vasilyev, van Bijnen, Schindler, Zoller, Blatt, and
  Monz]{marciniak2022optimal}
Christian~D Marciniak, Thomas Feldker, Ivan Pogorelov, Raphael Kaubruegger,
  Denis~V Vasilyev, Rick van Bijnen, Philipp Schindler, Peter Zoller, Rainer
  Blatt, and Thomas Monz.
\newblock Optimal metrology with programmable quantum sensors.
\newblock \emph{Nature}, 603\penalty0 (7902):\penalty0 604--609, 2022.

\bibitem[Malia et~al.(2022)Malia, Wu, Mart{\'\i}nez-Rinc{\'o}n, and
  Kasevich]{malia2022distributed}
Benjamin~K Malia, Yunfan Wu, Juli{\'a}n Mart{\'\i}nez-Rinc{\'o}n, and Mark~A
  Kasevich.
\newblock Distributed quantum sensing with mode-entangled spin-squeezed atomic
  states.
\newblock \emph{Nature}, pages 1--5, 2022.

\bibitem[Nielsen et~al.(2023)Nielsen, Neergaard-Nielsen, Gehring, and
  Andersen]{PhysRevLett.130.123603}
Jens A.~H. Nielsen, Jonas~S. Neergaard-Nielsen, Tobias Gehring, and Ulrik~L.
  Andersen.
\newblock {Deterministic Quantum Phase Estimation beyond N00N States}.
\newblock \emph{Phys. Rev. Lett.}, 130:\penalty0 123603, Mar 2023.
\newblock \doi{10.1103/PhysRevLett.130.123603}.
\newblock URL \url{https://link.aps.org/doi/10.1103/PhysRevLett.130.123603}.

\bibitem[Baumgratz and Datta(2016)]{baumgratz2016quantum}
Tillmann Baumgratz and Animesh Datta.
\newblock Quantum enhanced estimation of a multidimensional field.
\newblock \emph{Physical review letters}, 116\penalty0 (3):\penalty0 030801,
  2016.

\bibitem[Hou et~al.(2020)Hou, Zhang, Xiang, Li, Guo, Chen, Liu, and
  Yuan]{hou2020minimal}
Zhibo Hou, Zhao Zhang, Guo-Yong Xiang, Chuan-Feng Li, Guang-Can Guo, Hongzhen
  Chen, Liqiang Liu, and Haidong Yuan.
\newblock Minimal tradeoff and ultimate precision limit of multiparameter
  quantum magnetometry under the parallel scheme.
\newblock \emph{Physical Review Letters}, 125\penalty0 (2):\penalty0 020501,
  2020.

\bibitem[Montenegro et~al.(2022)Montenegro, Jones, Bose, and
  Bayat]{montenegro2022sequential}
Victor Montenegro, Gareth~Si{\^o}n Jones, Sougato Bose, and Abolfazl Bayat.
\newblock Sequential measurements for quantum-enhanced magnetometry in spin
  chain probes.
\newblock \emph{Physical Review Letters}, 129\penalty0 (12):\penalty0 120503,
  2022.

\bibitem[Kaubruegger et~al.(2023)Kaubruegger, Shankar, Vasilyev, and
  Zoller]{kaubruegger2023optimal}
Raphael Kaubruegger, Athreya Shankar, Denis~V Vasilyev, and Peter Zoller.
\newblock Optimal and variational multi-parameter quantum metrology and vector
  field sensing.
\newblock \emph{arXiv preprint arXiv:2302.07785}, 2023.

\bibitem[Spagnolo et~al.(2012)Spagnolo, Aparo, Vitelli, Crespi, Ramponi,
  Osellame, Mataloni, and Sciarrino]{spagnolo2012quantum}
Nicol{\`o} Spagnolo, Lorenzo Aparo, Chiara Vitelli, Andrea Crespi, Roberta
  Ramponi, Roberto Osellame, Paolo Mataloni, and Fabio Sciarrino.
\newblock Quantum interferometry with three-dimensional geometry.
\newblock \emph{Scientific reports}, 2:\penalty0 862, 2012.

\bibitem[Humphreys et~al.(2013)Humphreys, Barbieri, Datta, and
  Walmsley]{humphreys2013quantum}
Peter~C Humphreys, Marco Barbieri, Animesh Datta, and Ian~A Walmsley.
\newblock Quantum enhanced multiple phase estimation.
\newblock \emph{Physical review letters}, 111\penalty0 (7):\penalty0 070403,
  2013.

\bibitem[Yue et~al.(2014)Yue, Zhang, and Fan]{yue2014quantum}
Jie-Dong Yue, Yu-Ran Zhang, and Heng Fan.
\newblock Quantum-enhanced metrology for multiple phase estimation with noise.
\newblock \emph{Scientific reports}, 4\penalty0 (1):\penalty0 5933, 2014.

\bibitem[Gagatsos et~al.(2016)Gagatsos, Branford, and
  Datta]{gagatsos2016gaussian}
Christos~N Gagatsos, Dominic Branford, and Animesh Datta.
\newblock Gaussian systems for quantum-enhanced multiple phase estimation.
\newblock \emph{Physical Review A}, 94\penalty0 (4):\penalty0 042342, 2016.

\bibitem[Ciampini et~al.(2016)Ciampini, Spagnolo, Vitelli, Pezz{\`e}, Smerzi,
  and Sciarrino]{ciampini2016quantum}
Mario~A Ciampini, Nicol{\`o} Spagnolo, Chiara Vitelli, Luca Pezz{\`e}, Augusto
  Smerzi, and Fabio Sciarrino.
\newblock Quantum-enhanced multiparameter estimation in multiarm
  interferometers.
\newblock \emph{Scientific reports}, 6\penalty0 (1):\penalty0 1--8, 2016.

\bibitem[Pezz{\`e} et~al.(2017)Pezz{\`e}, Ciampini, Spagnolo, Humphreys, Datta,
  Walmsley, Barbieri, Sciarrino, and Smerzi]{pezze2017optimal}
Luca Pezz{\`e}, Mario~A Ciampini, Nicol{\`o} Spagnolo, Peter~C Humphreys,
  Animesh Datta, Ian~A Walmsley, Marco Barbieri, Fabio Sciarrino, and Augusto
  Smerzi.
\newblock Optimal measurements for simultaneous quantum estimation of multiple
  phases.
\newblock \emph{Physical Review Letters}, 119\penalty0 (13):\penalty0 130504,
  2017.

\bibitem[Zhang and Chan(2017)]{zhang2017quantum}
Lu~Zhang and Kam Wai~Clifford Chan.
\newblock {Quantum multiparameter estimation with generalized balanced
  multimode NOON-like states}.
\newblock \emph{Physical Review A}, 95\penalty0 (3):\penalty0 032321, 2017.

\bibitem[Crowley et~al.(2014)Crowley, Datta, Barbieri, and
  Walmsley]{crowley2014tradeoff}
Philip~JD Crowley, Animesh Datta, Marco Barbieri, and Ian~A Walmsley.
\newblock Tradeoff in simultaneous quantum-limited phase and loss estimation in
  interferometry.
\newblock \emph{Physical Review A}, 89\penalty0 (2):\penalty0 023845, 2014.

\bibitem[Szczykulska et~al.(2017)Szczykulska, Baumgratz, and
  Datta]{szczykulska2017reaching}
Magdalena Szczykulska, Tillmann Baumgratz, and Animesh Datta.
\newblock Reaching for the quantum limits in the simultaneous estimation of
  phase and phase diffusion.
\newblock \emph{Quantum Science and Technology}, 2\penalty0 (4):\penalty0
  044004, 2017.

\bibitem[Cimini et~al.(2019)Cimini, Gianani, Ruggiero, Gasperi, Sbroscia,
  Roccia, Tofani, Bruni, Ricci, and Barbieri]{cimini2019quantum}
Valeria Cimini, Ilaria Gianani, Ludovica Ruggiero, Tecla Gasperi, Marco
  Sbroscia, Emanuele Roccia, Daniela Tofani, Fabio Bruni, Maria~Antonietta
  Ricci, and Marco Barbieri.
\newblock Quantum sensing for dynamical tracking of chemical processes.
\newblock \emph{Physical Review A}, 99\penalty0 (5):\penalty0 053817, 2019.

\bibitem[Chrostowski et~al.(2017)Chrostowski, Demkowicz-Dobrza{\'n}ski,
  Jarzyna, and Banaszek]{chrostowski2017super}
Andrzej Chrostowski, Rafa{\l} Demkowicz-Dobrza{\'n}ski, Marcin Jarzyna, and
  Konrad Banaszek.
\newblock On super-resolution imaging as a multiparameter estimation problem.
\newblock \emph{International Journal of Quantum Information}, 15\penalty0
  (08):\penalty0 1740005, 2017.

\bibitem[{\v{R}}eha{\v{c}}ek et~al.(2017){\v{R}}eha{\v{c}}ek, Hradil, Stoklasa,
  Pa{\'u}r, Grover, Krzic, and S{\'a}nchez-Soto]{vrehavcek2017multiparameter}
J~{\v{R}}eha{\v{c}}ek, Z~Hradil, B~Stoklasa, M~Pa{\'u}r, J~Grover, A~Krzic, and
  LL~S{\'a}nchez-Soto.
\newblock Multiparameter quantum metrology of incoherent point sources: towards
  realistic superresolution.
\newblock \emph{Physical Review A}, 96\penalty0 (6):\penalty0 062107, 2017.

\bibitem[Chiribella et~al.(2006)Chiribella, D'Ariano, and
  Sacchi]{chiribella2006joint}
Giulio Chiribella, GM~D'Ariano, and MF~Sacchi.
\newblock Joint estimation of real squeezing and displacement.
\newblock \emph{Journal of Physics A: Mathematical and General}, 39\penalty0
  (9):\penalty0 2127, 2006.

\bibitem[Monras and Illuminati(2011)]{monras2011measurement}
Alex Monras and Fabrizio Illuminati.
\newblock {Measurement of damping and temperature: Precision bounds in Gaussian
  dissipative channels}.
\newblock \emph{Physical Review A}, 83\penalty0 (1):\penalty0 012315, 2011.

\bibitem[Genoni et~al.(2013)Genoni, Paris, Adesso, Nha, Knight, and
  Kim]{Genoni2013}
MG~Genoni, MGA Paris, G~Adesso, H~Nha, PL~Knight, and MS~Kim.
\newblock Optimal estimation of joint parameters in phase space.
\newblock \emph{Phys. Rev. A}, 87\penalty0 (1):\penalty0 012107, 2013.
\newblock URL \url{http://pra.aps.org/abstract/PRA/v87/i1/e012107}.

\bibitem[Gao and Lee(2014)]{gao2014bounds}
Yang Gao and Hwang Lee.
\newblock {Bounds on quantum multiple-parameter estimation with Gaussian
  state}.
\newblock \emph{The European Physical Journal D}, 68\penalty0 (11):\penalty0
  1--7, 2014.

\bibitem[Bradshaw et~al.(2017)Bradshaw, Assad, and Lam]{bradshaw2017tight}
Mark Bradshaw, Syed~M Assad, and Ping~Koy Lam.
\newblock {A tight Cram{\'e}r--Rao bound for joint parameter estimation with a
  pure two-mode squeezed probe}.
\newblock \emph{Physics Letters A}, 381\penalty0 (32):\penalty0 2598--2607,
  2017.

\bibitem[Bradshaw et~al.(2018)Bradshaw, Lam, and Assad]{bradshaw2018ultimate}
Mark Bradshaw, Ping~Koy Lam, and Syed~M Assad.
\newblock {Ultimate precision of joint quadrature parameter estimation with a
  Gaussian probe}.
\newblock \emph{Physical Review A}, 97\penalty0 (1):\penalty0 012106, 2018.

\bibitem[Assad et~al.(2020)Assad, Li, Liu, Zhao, Zhao, Lam, Ou, and
  Li]{assad2020accessible}
Syed~M Assad, Jiamin Li, Yuhong Liu, Ningbo Zhao, Wen Zhao, Ping~Koy Lam,
  ZY~Ou, and Xiaoying Li.
\newblock Accessible precisions for estimating two conjugate parameters using
  {G}aussian probes.
\newblock \emph{Physical Review Research}, 2\penalty0 (2):\penalty0 023182,
  2020.

\bibitem[Park et~al.(2022)Park, Oh, Filip, and Marek]{park2022optimal}
Kimin Park, Changhun Oh, Radim Filip, and Petr Marek.
\newblock Optimal estimation of conjugate shifts in position and momentum by
  classically correlated probes and measurements.
\newblock \emph{Physical Review Applied}, 18\penalty0 (1):\penalty0 014060,
  2022.

\bibitem[Steinlechner et~al.(2013)Steinlechner, Bauchrowitz, Meinders,
  M{\"u}ller-Ebhardt, Danzmann, and Schnabel]{steinlechner2013quantum}
Sebastian Steinlechner, J{\"o}ran Bauchrowitz, Melanie Meinders, Helge
  M{\"u}ller-Ebhardt, Karsten Danzmann, and Roman Schnabel.
\newblock Quantum-dense metrology.
\newblock \emph{Nature Photonics}, 7\penalty0 (8):\penalty0 626--630, 2013.

\bibitem[Vidrighin et~al.(2014)Vidrighin, Donati, Genoni, Jin, Kolthammer, Kim,
  Datta, Barbieri, and Walmsley]{vidrighin2014joint}
Mihai~D Vidrighin, Gaia Donati, Marco~G Genoni, Xian-Min Jin, W~Steven
  Kolthammer, MS~Kim, Animesh Datta, Marco Barbieri, and Ian~A Walmsley.
\newblock Joint estimation of phase and phase diffusion for quantum metrology.
\newblock \emph{Nature communications}, 5\penalty0 (1):\penalty0 1--7, 2014.

\bibitem[Hou et~al.(2016)Hou, Zhu, Xiang, Li, and Guo]{hou2016achieving}
Zhibo Hou, Huangjun Zhu, Guo-Yong Xiang, Chuan-Feng Li, and Guang-Can Guo.
\newblock Achieving quantum precision limit in adaptive qubit state tomography.
\newblock \emph{npj Quantum Information}, 2\penalty0 (1):\penalty0 1--5, 2016.

\bibitem[Liu et~al.(2018)Liu, Li, Cui, Huo, Assad, Li, and Ou]{liu2018loss}
Yuhong Liu, Jiamin Li, Liang Cui, Nan Huo, Syed~M Assad, Xiaoying Li, and
  ZY~Ou.
\newblock Loss-tolerant quantum dense metrology with {SU}(1, 1) interferometer.
\newblock \emph{Optics Express}, 26\penalty0 (21):\penalty0 27705--27715, 2018.

\bibitem[Li et~al.(2023)Li, Conlon, Lam, and Assad]{li2023optimal}
Bacui Li, Lorcan~O Conlon, Ping~Koy Lam, and Syed~M Assad.
\newblock Optimal single qubit tomography: Realization of locally optimal
  measurements on a quantum computer.
\newblock \emph{arXiv preprint arXiv:2302.05140}, 2023.

\bibitem[Vaneph et~al.(2013)Vaneph, Tufarelli, and Genoni]{vaneph2013quantum}
Cyril Vaneph, Tommaso Tufarelli, and Marco~G Genoni.
\newblock Quantum estimation of a two-phase spin rotation.
\newblock \emph{Quantum Measurements and Quantum Metrology}, 1\penalty0
  (2013):\penalty0 12--20, 2013.

\bibitem[Suzuki(2015)]{suzuki2015parameter}
Jun Suzuki.
\newblock Parameter estimation of qubit states with unknown phase parameter.
\newblock \emph{International Journal of Quantum Information}, 13\penalty0
  (01):\penalty0 1450044, 2015.

\bibitem[Suzuki(2016)]{suzuki2016explicit}
Jun Suzuki.
\newblock {Explicit formula for the Holevo bound for two-parameter qubit-state
  estimation problem}.
\newblock \emph{Journal of Mathematical Physics}, 57\penalty0 (4):\penalty0
  042201, 2016.

\bibitem[Szczykulska et~al.(2016)Szczykulska, Baumgratz, and
  Datta]{szczykulska2016multi}
Magdalena Szczykulska, Tillmann Baumgratz, and Animesh Datta.
\newblock Multi-parameter quantum metrology.
\newblock \emph{Advances in Physics: X}, 1\penalty0 (4):\penalty0 621--639,
  2016.

\bibitem[Proctor et~al.(2018)Proctor, Knott, and
  Dunningham]{proctor2018multiparameter}
Timothy~J Proctor, Paul~A Knott, and Jacob~A Dunningham.
\newblock Multiparameter estimation in networked quantum sensors.
\newblock \emph{Physical review letters}, 120\penalty0 (8):\penalty0 080501,
  2018.

\bibitem[Gessner et~al.(2018)Gessner, Pezz{\`e}, and
  Smerzi]{gessner2018sensitivity}
Manuel Gessner, Luca Pezz{\`e}, and Augusto Smerzi.
\newblock Sensitivity bounds for multiparameter quantum metrology.
\newblock \emph{Physical review letters}, 121\penalty0 (13):\penalty0 130503,
  2018.

\bibitem[Tsang et~al.(2020{\natexlab{a}})Tsang, Albarelli, and
  Datta]{tsang2019quantum}
Mankei Tsang, Francesco Albarelli, and Animesh Datta.
\newblock Quantum semiparametric estimation.
\newblock \emph{Physical Review X}, 10:\penalty0 031023, Jul
  2020{\natexlab{a}}.

\bibitem[Carollo et~al.(2019)Carollo, Spagnolo, Dubkov, and
  Valenti]{carollo2019quantumness}
Angelo Carollo, Bernardo Spagnolo, Alexander~A Dubkov, and Davide Valenti.
\newblock On quantumness in multi-parameter quantum estimation.
\newblock \emph{Journal of Statistical Mechanics: Theory and Experiment},
  2019\penalty0 (9):\penalty0 094010, 2019.

\bibitem[Tsang et~al.(2020{\natexlab{b}})Tsang, Albarelli, and
  Datta]{tsang2020quantum}
Mankei Tsang, Francesco Albarelli, and Animesh Datta.
\newblock Quantum semiparametric estimation.
\newblock \emph{Physical Review X}, 10\penalty0 (3):\penalty0 031023,
  2020{\natexlab{b}}.

\bibitem[Demkowicz-Dobrza{\'n}ski et~al.(2020)Demkowicz-Dobrza{\'n}ski,
  G{\'o}recki, and Gu{\c{t}}{\u{a}}]{demkowicz2020multi}
Rafa{\l} Demkowicz-Dobrza{\'n}ski, Wojciech G{\'o}recki, and
  M{\u{a}}d{\u{a}}lin Gu{\c{t}}{\u{a}}.
\newblock Multi-parameter estimation beyond quantum {F}isher information.
\newblock \emph{Journal of Physics A: Mathematical and Theoretical}, \text{in
  press}, 2020.
\newblock \doi{https://doi.org/10.1088/1751- 8121/ab8ef3}.

\bibitem[Razavian et~al.(2020)Razavian, Paris, and
  Genoni]{razavian2020quantumness}
Sholeh Razavian, Matteo~GA Paris, and Marco~G Genoni.
\newblock On the quantumness of multiparameter estimation problems for qubit
  systems.
\newblock \emph{Entropy}, 22\penalty0 (11):\penalty0 1197, 2020.

\bibitem[Gessner et~al.(2020)Gessner, Smerzi, and
  Pezz{\`e}]{gessner2020multiparameter}
Manuel Gessner, Augusto Smerzi, and Luca Pezz{\`e}.
\newblock Multiparameter squeezing for optimal quantum enhancements in sensor
  networks.
\newblock \emph{Nature communications}, 11\penalty0 (1):\penalty0 1--9, 2020.

\bibitem[Lu and Wang(2021)]{lu2021incorporating}
Xiao-Ming Lu and Xiaoguang Wang.
\newblock Incorporating heisenberg’s uncertainty principle into quantum
  multiparameter estimation.
\newblock \emph{Physical Review Letters}, 126\penalty0 (12):\penalty0 120503,
  2021.

\bibitem[Gebhart et~al.(2021)Gebhart, Smerzi, and
  Pezz{\`e}]{gebhart2021bayesian}
Valentin Gebhart, Augusto Smerzi, and Luca Pezz{\`e}.
\newblock Bayesian quantum multiphase estimation algorithm.
\newblock \emph{Physical Review Applied}, 16\penalty0 (1):\penalty0 014035,
  2021.

\bibitem[Albarelli and Demkowicz-Dobrza{\'n}ski(2022)]{albarelli2022probe}
Francesco Albarelli and Rafa{\l} Demkowicz-Dobrza{\'n}ski.
\newblock Probe incompatibility in multiparameter noisy quantum metrology.
\newblock \emph{Physical Review X}, 12\penalty0 (1):\penalty0 011039, 2022.

\bibitem[Huang et~al.(2021)Huang, Lupo, and Kok]{huang2021quantum}
Zixin Huang, Cosmo Lupo, and Pieter Kok.
\newblock Quantum-limited estimation of range and velocity.
\newblock \emph{PRX Quantum}, 2\penalty0 (3):\penalty0 030303, 2021.

\bibitem[Gianani et~al.(2021)Gianani, Albarelli, Verna, Cimini,
  Demkowicz-Dobrzanski, and Barbieri]{gianani2021kramers}
Ilaria Gianani, Francesco Albarelli, Adriano Verna, Valeria Cimini, Rafal
  Demkowicz-Dobrzanski, and Marco Barbieri.
\newblock {Kramers--Kronig relations and precision limits in quantum phase
  estimation}.
\newblock \emph{Optica}, 8\penalty0 (12):\penalty0 1642--1645, 2021.

\bibitem[Hanamura et~al.(2021)Hanamura, Asavanant, Fukui, Konno, and
  Furusawa]{hanamura2021estimation}
Fumiya Hanamura, Warit Asavanant, Kosuke Fukui, Shunya Konno, and Akira
  Furusawa.
\newblock {Estimation of Gaussian random displacement using non-Gaussian
  states}.
\newblock \emph{Physical Review A}, 104\penalty0 (6):\penalty0 062601, 2021.

\bibitem[Di~Fresco et~al.(2022)Di~Fresco, Spagnolo, Valenti, and
  Carollo]{di2022multiparameter}
Giovanni Di~Fresco, Bernardo Spagnolo, Davide Valenti, and Angelo Carollo.
\newblock Multiparameter quantum critical metrology.
\newblock \emph{SciPost Physics}, 13\penalty0 (4):\penalty0 077, 2022.

\bibitem[Hosseiny et~al.(2022)Hosseiny, Jahromi, Radgohar, and
  Amniat-Talab]{hosseiny2022estimating}
Seyed~Mohammad Hosseiny, Hossein~Rangani Jahromi, Roya Radgohar, and Mahdi
  Amniat-Talab.
\newblock Estimating energy levels of a three-level atom in single and
  multi-parameter metrological schemes.
\newblock \emph{Physica Scripta}, 97\penalty0 (12):\penalty0 125402, 2022.

\bibitem[Fadel et~al.(2022)Fadel, Yadin, Mao, Byrnes, and
  Gessner]{fadel2022multiparameter}
Matteo Fadel, Benjamin Yadin, Yuping Mao, Tim Byrnes, and Manuel Gessner.
\newblock Multiparameter quantum metrology and mode entanglement with spatially
  split nonclassical spin states.
\newblock \emph{arXiv preprint arXiv:2201.11081}, 2022.

\bibitem[Len(2022)]{len2022multiparameter}
Yink~Loong Len.
\newblock Multiparameter estimation for qubit states with collective
  measurements: a case study.
\newblock \emph{New Journal of Physics}, 24\penalty0 (3):\penalty0 033037,
  2022.

\bibitem[Xie et~al.(2022)Xie, Xu, and Wang]{xie2022quantum}
Dong Xie, Chunling Xu, and An~Min Wang.
\newblock Quantum thermometry in diffraction-limited systems.
\newblock \emph{Physical Review A}, 106\penalty0 (5):\penalty0 052407, 2022.

\bibitem[Liu et~al.(2019)Liu, Yuan, Lu, and Wang]{liu2019quantum}
Jing Liu, Haidong Yuan, Xiao-Ming Lu, and Xiaoguang Wang.
\newblock {Quantum Fisher information matrix and multiparameter estimation}.
\newblock \emph{Journal of Physics A: Mathematical and Theoretical},
  53\penalty0 (2):\penalty0 023001, 2019.

\bibitem[Albarelli et~al.(2020)Albarelli, Barbieri, Genoni, and
  Gianani]{albarelli2020perspective}
Francesco Albarelli, Marco Barbieri, Marco~G Genoni, and Ilaria Gianani.
\newblock A perspective on multiparameter quantum metrology: from theoretical
  tools to applications in quantum imaging.
\newblock \emph{Physics Letters A}, 384\penalty0 (12):\penalty0 126311, 2020.

\bibitem[Sidhu and Kok(2020)]{sidhu2020geometric}
Jasminder~S Sidhu and Pieter Kok.
\newblock Geometric perspective on quantum parameter estimation.
\newblock \emph{AVS Quantum Science}, 2\penalty0 (1):\penalty0 014701, 2020.

\bibitem[Polino et~al.(2020)Polino, Valeri, Spagnolo, and
  Sciarrino]{polino2020photonic}
Emanuele Polino, Mauro Valeri, Nicol{\`o} Spagnolo, and Fabio Sciarrino.
\newblock Photonic quantum metrology.
\newblock \emph{AVS Quantum Science}, 2\penalty0 (2):\penalty0 024703, 2020.

\bibitem[Bennett and Shor(1998)]{bennett1998quantum}
Charles~H. Bennett and Peter~W. Shor.
\newblock Quantum information theory.
\newblock \emph{IEEE transactions on information theory}, 44\penalty0
  (6):\penalty0 2724--2742, 1998.

\bibitem[Giovannetti et~al.(2006)Giovannetti, Lloyd, and
  Maccone]{giovannetti2006quantum}
Vittorio Giovannetti, Seth Lloyd, and Lorenzo Maccone.
\newblock Quantum metrology.
\newblock \emph{Physical review letters}, 96\penalty0 (1):\penalty0 010401,
  2006.

\bibitem[Matsumoto(2002)]{Matsumoto2002}
K.~Matsumoto.
\newblock {A new approach to the {C}ram{\'e}r-{Rao}-type bound of the
  pure-state model}.
\newblock \emph{J. Phys. A: Math. Gen.}, 35\penalty0 (13):\penalty0 3111--3123,
  Mar 2002.

\bibitem[Helstrom(1967)]{helstrom1967minimum}
Carl~W Helstrom.
\newblock Minimum mean-squared error of estimates in quantum statistics.
\newblock \emph{Physics Letters A}, 25\penalty0 (2):\penalty0 101--102, 1967.

\bibitem[Helstrom(1968)]{helstrom1968minimum}
Carl~W Helstrom.
\newblock The minimum variance of estimates in quantum signal detection.
\newblock \emph{IEEE Transactions on Information Theory}, 14\penalty0
  (2):\penalty0 234--242, 1968.

\bibitem[Yuen and Lax(1973)]{yuen1973}
H~Yuen and Melvin Lax.
\newblock Multiple-parameter quantum estimation and measurement of
  nonselfadjoint observables.
\newblock \emph{IEEE Transactions on Information Theory}, 19\penalty0
  (6):\penalty0 740--750, 1973.

\bibitem[Holevo(2011)]{holevo2011probabilistic}
Alexander~S Holevo.
\newblock \emph{Probabilistic and statistical aspects of quantum theory},
  volume~1.
\newblock Springer Science \& Business Media, 2011.

\bibitem[Holevo(1973)]{holevo1973statistical}
Alexander~S Holevo.
\newblock Statistical decision theory for quantum systems.
\newblock \emph{Journal of multivariate analysis}, 3\penalty0 (4):\penalty0
  337--394, 1973.

\bibitem[Kahn and Gu{\c{t}}{\u{a}}(2009)]{kahn2009local}
Jonas Kahn and M{\u{a}}d{\u{a}}lin Gu{\c{t}}{\u{a}}.
\newblock Local asymptotic normality for finite dimensional quantum systems.
\newblock \emph{Communications in Mathematical Physics}, 289\penalty0
  (2):\penalty0 597--652, 2009.

\bibitem[Yamagata et~al.(2013)Yamagata, Fujiwara, Gill,
  et~al.]{yamagata2013quantum}
Koichi Yamagata, Akio Fujiwara, Richard~D Gill, et~al.
\newblock Quantum local asymptotic normality based on a new quantum likelihood
  ratio.
\newblock \emph{Annals of Statistics}, 41\penalty0 (4):\penalty0 2197--2217,
  2013.

\bibitem[Yang et~al.(2019)Yang, Chiribella, and Hayashi]{yang2019attaining}
Yuxiang Yang, Giulio Chiribella, and Masahito Hayashi.
\newblock Attaining the ultimate precision limit in quantum state estimation.
\newblock \emph{Communications in Mathematical Physics}, 368\penalty0
  (1):\penalty0 223--293, 2019.

\bibitem[Yu et~al.(2022)Yu, Liu, Yang, Gong, Cao, Zhang, Liu, Heyl, Ozawa,
  Goldman, et~al.]{yu2022quantum}
Min Yu, Yu~Liu, Pengcheng Yang, Musang Gong, Qingyun Cao, Shaoliang Zhang,
  Haibin Liu, Markus Heyl, Tomoki Ozawa, Nathan Goldman, et~al.
\newblock {Quantum Fisher information measurement and verification of the
  quantum Cram{\'e}r--Rao bound in a solid-state qubit}.
\newblock \emph{npj Quantum Information}, 8\penalty0 (1):\penalty0 56, 2022.

\bibitem[Li et~al.(2022)Li, Chen, and Cappellaro]{li2022geometric}
Changhao Li, Mo~Chen, and Paola Cappellaro.
\newblock {A geometric perspective: experimental evaluation of the quantum
  Cramer-Rao bound}.
\newblock \emph{arXiv preprint arXiv:2204.13777}, 2022.

\bibitem[Conlon et~al.(2022{\natexlab{b}})Conlon, Suzuki, Lam, and
  Assad]{conlon2022gap}
Lorc{\'a}n~O Conlon, Jun Suzuki, Ping~Koy Lam, and Syed~M Assad.
\newblock The gap persistence theorem for quantum multiparameter estimation.
\newblock \emph{arXiv preprint arXiv:2208.07386}, 2022{\natexlab{b}}.

\bibitem[Nagaoka(2005{\natexlab{a}})]{nagaoka2005new}
Hiroshi Nagaoka.
\newblock A new approach to {C}ram{\'e}r--{R}ao bounds for quantum state
  estimation.
\newblock In \emph{Asymptotic Theory Of Quantum Statistical Inference: Selected
  Papers}, pages 100--112. World Scientific, 2005{\natexlab{a}}.

\bibitem[Hayashi(1997)]{hayashi1997linear}
Masahito Hayashi.
\newblock {A linear programming approach to attainable Cramer-Rao type bounds}.
\newblock In \emph{Asymptotic Theory Of Quantum Statistical Inference: Selected
  Papers}, pages 150--161. 1997.

\bibitem[Conlon et~al.(2021)Conlon, Suzuki, Lam, and
  Assad]{conlon2021efficient}
Lorc{\'a}n~O Conlon, Jun Suzuki, Ping~Koy Lam, and Syed~M Assad.
\newblock Efficient computation of the nagaoka--hayashi bound for
  multiparameter estimation with separable measurements.
\newblock \emph{npj Quantum Information}, 7\penalty0 (1):\penalty0 110, 2021.

\bibitem[Nagaoka(2005{\natexlab{b}})]{nagaoka2005generalization}
Hiroshi Nagaoka.
\newblock A generalization of the simultaneous diagonalization of {H}ermitian
  matrices and its relation to quantum estimation theory.
\newblock In \emph{Asymptotic Theory Of Quantum Statistical Inference: Selected
  Papers}, pages 133--149. World Scientific, 2005{\natexlab{b}}.

\bibitem[Hayashi and Ouyang(2022)]{hayashi2022tight}
Masahito Hayashi and Yingkai Ouyang.
\newblock {Tight Cram$\backslash$'$\{$e$\}$ r-Rao type bounds for
  multiparameter quantum metrology through conic programming}.
\newblock \emph{arXiv preprint arXiv:2209.05218}, 2022.

\bibitem[Roccia et~al.(2017)Roccia, Gianani, Mancino, Sbroscia, Somma, Genoni,
  and Barbieri]{roccia2017entangling}
Emanuele Roccia, Ilaria Gianani, Luca Mancino, Marco Sbroscia, Fabrizia Somma,
  Marco~G Genoni, and Marco Barbieri.
\newblock Entangling measurements for multiparameter estimation with two
  qubits.
\newblock \emph{Quantum Science and Technology}, 3\penalty0 (1):\penalty0
  01LT01, 2017.

\bibitem[Hou et~al.(2018)Hou, Tang, Shang, Zhu, Li, Yuan, Wu, Xiang, Li, and
  Guo]{hou2018deterministic}
Zhibo Hou, Jun-Feng Tang, Jiangwei Shang, Huangjun Zhu, Jian Li, Yuan Yuan,
  Kang-Da Wu, Guo-Yong Xiang, Chuan-Feng Li, and Guang-Can Guo.
\newblock Deterministic realization of collective measurements via photonic
  quantum walks.
\newblock \emph{Nature Communications}, 9\penalty0 (1):\penalty0 1--7, 2018.

\bibitem[Parniak et~al.(2018)Parniak, Bor{\'o}wka, Boroszko, Wasilewski,
  Banaszek, and Demkowicz-Dobrza{\'n}ski]{parniak2018beating}
Micha{\l} Parniak, Sebastian Bor{\'o}wka, Kajetan Boroszko, Wojciech
  Wasilewski, Konrad Banaszek, and Rafa{\l} Demkowicz-Dobrza{\'n}ski.
\newblock {Beating the Rayleigh limit using two-photon interference}.
\newblock \emph{Physical review letters}, 121\penalty0 (25):\penalty0 250503,
  2018.

\bibitem[Wu et~al.(2019)Wu, Yuan, Xiang, Li, Guo, and
  Perarnau-Llobet]{wu2019experimentally}
Kang-Da Wu, Yuan Yuan, Guo-Yong Xiang, Chuan-Feng Li, Guang-Can Guo, and
  Mart{\'\i} Perarnau-Llobet.
\newblock Experimentally reducing the quantum measurement back action in work
  distributions by a collective measurement.
\newblock \emph{Science Advances}, 5\penalty0 (3):\penalty0 eaav4944, 2019.

\bibitem[Wu et~al.(2020)Wu, B{\"a}umer, Tang, Hovhannisyan, Perarnau-Llobet,
  Xiang, Li, and Guo]{wu2020minimizing}
Kang-Da Wu, Elisa B{\"a}umer, Jun-Feng Tang, Karen~V Hovhannisyan, Mart{\'\i}
  Perarnau-Llobet, Guo-Yong Xiang, Chuan-Feng Li, and Guang-Can Guo.
\newblock Minimizing backaction through entangled measurements.
\newblock \emph{Physical Review Letters}, 125\penalty0 (21):\penalty0 210401,
  2020.

\bibitem[Yuan et~al.(2020)Yuan, Hou, Tang, Streltsov, Xiang, Li, and
  Guo]{yuan2020direct}
Yuan Yuan, Zhibo Hou, Jun-Feng Tang, Alexander Streltsov, Guo-Yong Xiang,
  Chuan-Feng Li, and Guang-Can Guo.
\newblock Direct estimation of quantum coherence by collective measurements.
\newblock \emph{npj Quantum Information}, 6\penalty0 (1):\penalty0 1--5, 2020.

\bibitem[Conlon et~al.(2023{\natexlab{a}})Conlon, Vogl, Marciniak, Pogorelov,
  Yung, Eilenberger, Berry, Santana, Blatt, Monz,
  et~al.]{conlon2023approaching}
Lorc{\'a}n~O Conlon, Tobias Vogl, Christian~D Marciniak, Ivan Pogorelov,
  Simon~K Yung, Falk Eilenberger, Dominic~W Berry, Fabiana~S Santana, Rainer
  Blatt, Thomas Monz, et~al.
\newblock Approaching optimal entangling collective measurements on quantum
  computing platforms.
\newblock \emph{Nature Physics}, 19:\penalty0 351–357, 2023{\natexlab{a}}.

\bibitem[Conlon et~al.(2023{\natexlab{b}})Conlon, Eilenberger, Lam, and
  Assad]{conlon2023discriminating}
Lorcan~O Conlon, Falk Eilenberger, Ping~Koy Lam, and Syed~M Assad.
\newblock Discriminating qubit states with entangling collective measurements.
\newblock \emph{arXiv preprint arXiv:2302.08882}, 2023{\natexlab{b}}.

\bibitem[Mart{\'\i}nez et~al.(2023)Mart{\'\i}nez, G{\'o}mez, Cari{\~n}e,
  Pereira, Delgado, Walborn, Tavakoli, and Lima]{martinez2023certification}
Daniel Mart{\'\i}nez, Esteban~S G{\'o}mez, Jaime Cari{\~n}e, Luciano Pereira,
  Aldo Delgado, Stephen~P Walborn, Armin Tavakoli, and Gustavo Lima.
\newblock Certification of a non-projective qudit measurement using multiport
  beamsplitters.
\newblock \emph{Nature Physics}, 19\penalty0 (2):\penalty0 190--195, 2023.

\bibitem[Albarelli et~al.(2019)Albarelli, Friel, and
  Datta]{albarelli2019evaluating}
Francesco Albarelli, Jamie~F Friel, and Animesh Datta.
\newblock {Evaluating the Holevo Cram{\'e}r-Rao Bound for Multiparameter
  Quantum Metrology}.
\newblock \emph{Physical Review Letters}, 123\penalty0 (20):\penalty0 200503,
  2019.

\bibitem[Sidhu et~al.(2021)Sidhu, Ouyang, Campbell, and Kok]{sidhu2021tight}
Jasminder~S. Sidhu, Yingkai Ouyang, Earl~T. Campbell, and Pieter Kok.
\newblock Tight bounds on the simultaneous estimation of incompatible
  parameters.
\newblock \emph{Phys. Rev. X}, 11:\penalty0 011028, Feb 2021.
\newblock \doi{10.1103/PhysRevX.11.011028}.

\bibitem[Genoni et~al.(2011)Genoni, Olivares, and Paris]{genoni2011optical}
Marco~G Genoni, Stefano Olivares, and Matteo~GA Paris.
\newblock Optical phase estimation in the presence of phase diffusion.
\newblock \emph{Physical review letters}, 106\penalty0 (15):\penalty0 153603,
  2011.

\bibitem[Datta et~al.(2011)Datta, Zhang, Thomas-Peter, Dorner, Smith, and
  Walmsley]{datta2011quantum}
Animesh Datta, Lijian Zhang, Nicholas Thomas-Peter, Uwe Dorner, Brian~J Smith,
  and Ian~A Walmsley.
\newblock Quantum metrology with imperfect states and detectors.
\newblock \emph{Physical Review A}, 83\penalty0 (6):\penalty0 063836, 2011.

\bibitem[Ragy et~al.(2016)Ragy, Jarzyna, and
  Demkowicz-Dobrza{\'n}ski]{ragy2016compatibility}
Sammy Ragy, Marcin Jarzyna, and Rafa{\l} Demkowicz-Dobrza{\'n}ski.
\newblock Compatibility in multiparameter quantum metrology.
\newblock \emph{Physical Review A}, 94\penalty0 (5):\penalty0 052108, 2016.

\bibitem[Kull et~al.(2020)Kull, Gu{\'e}rin, and
  Verstraete]{kull2020uncertainty}
Ilya Kull, Philippe~Allard Gu{\'e}rin, and Frank Verstraete.
\newblock Uncertainty and trade-offs in quantum multiparameter estimation.
\newblock \emph{Journal of Physics A: Mathematical and Theoretical}, 2020.

\bibitem[Ballester(2005)]{ballester2005optimal}
Manuel~A Ballester.
\newblock {Optimal estimation of SU (d) using exact and approximate 2-designs}.
\newblock \emph{arXiv preprint quant-ph/0507073}, 2005.

\bibitem[Imai and Fujiwara(2007)]{imai2007geometry}
Hiroshi Imai and Akio Fujiwara.
\newblock {Geometry of optimal estimation scheme for SU (D) channels}.
\newblock \emph{Journal of Physics A: Mathematical and Theoretical},
  40\penalty0 (16):\penalty0 4391, 2007.

\bibitem[Napolitano et~al.(2011)Napolitano, Koschorreck, Dubost, Behbood,
  Sewell, and Mitchell]{napolitano2011interaction}
Mario Napolitano, Marco Koschorreck, Brice Dubost, Naeimeh Behbood, RJ~Sewell,
  and Morgan~W Mitchell.
\newblock {Interaction-based quantum metrology showing scaling beyond the
  Heisenberg limit}.
\newblock \emph{Nature}, 471\penalty0 (7339):\penalty0 486--489, 2011.

\bibitem[Hayashi et~al.(2022)Hayashi, Liu, and Yuan]{hayashi2022global}
Masahito Hayashi, Zi-Wen Liu, and Haidong Yuan.
\newblock {Global Heisenberg scaling in noisy and practical phase estimation}.
\newblock \emph{Quantum Science and Technology}, 7\penalty0 (2):\penalty0
  025030, 2022.

\bibitem[Pinel et~al.(2013)Pinel, Jian, Treps, Fabre, and
  Braun]{pinel2013quantum}
Olivier Pinel, Pu~Jian, Nicolas Treps, Claude Fabre, and Daniel Braun.
\newblock {Quantum parameter estimation using general single-mode Gaussian
  states}.
\newblock \emph{Physical Review A}, 88\penalty0 (4):\penalty0 040102, 2013.

\bibitem[Holevo(1976)]{holevo1976noncommutative}
AS~Holevo.
\newblock {Noncommutative analogues of the Cram{\'e}r-Rao inequality in the
  quantum measurement theory}.
\newblock In \emph{Proceedings of the Third Japan---USSR Symposium on
  Probability Theory}, pages 194--222. Springer, 1976.

\bibitem[Gill and Massar(2000)]{gill2005state}
Richard~D Gill and Serge Massar.
\newblock State estimation for large ensembles.
\newblock \emph{Physical Review A}, 61\penalty0 (4):\penalty0 042312, 2000.

\bibitem[Suzuki(2019)]{suzuki2019information}
Jun Suzuki.
\newblock Information geometrical characterization of quantum statistical
  models in quantum estimation theory.
\newblock \emph{Entropy}, 21\penalty0 (7):\penalty0 703, 2019.

\bibitem[Kraus(1983)]{kraus1983states}
Karl Kraus.
\newblock \emph{States, effects and operations: fundamental notions of quantum
  theory}.
\newblock Springer, 1983.

\bibitem[Serafini(2017)]{serafini2017quantum}
Alessio Serafini.
\newblock \emph{Quantum Continuous Variables: A Primer of Theoretical Methods}.
\newblock CRC Press, 2017.

\bibitem[Suzuki(2020)]{suzuki2020nuisance}
Jun Suzuki.
\newblock Nuisance parameter problem in quantum estimation theory: {T}radeoff
  relation and qubit examples.
\newblock \emph{Journal of Physics A: Mathematical and Theoretical},
  53\penalty0 (26):\penalty0 264001, 2020.

\bibitem[Suzuki et~al.(2020)Suzuki, Yang, and Hayashi]{suzuki2020quantum}
Jun Suzuki, Yuxiang Yang, and Masahito Hayashi.
\newblock Quantum state estimation with nuisance parameters.
\newblock \emph{Journal of Physics A: Mathematical and Theoretical}, in press,
  2020.
\newblock \doi{https://doi.org/10.1088/1751-8121/ab8b78}.

\bibitem[Ozaydin(2014)]{ozaydin2014phase}
Fatih Ozaydin.
\newblock {Phase damping destroys quantum Fisher information of W states}.
\newblock \emph{Physics Letters A}, 378\penalty0 (43):\penalty0 3161--3164,
  2014.

\bibitem[Ma et~al.(2011)Ma, Huang, Wang, and Sun]{ma2011quantum}
Jian Ma, Yi-xiao Huang, Xiaoguang Wang, and CP~Sun.
\newblock {Quantum Fisher information of the Greenberger-Horne-Zeilinger state
  in decoherence channels}.
\newblock \emph{Physical Review A}, 84\penalty0 (2):\penalty0 022302, 2011.

\bibitem[Huelga et~al.(1997)Huelga, Macchiavello, Pellizzari, Ekert, Plenio,
  and Cirac]{huelga1997improvement}
Susanna~F Huelga, Chiara Macchiavello, Thomas Pellizzari, Artur~K Ekert,
  Martin~B Plenio, and J~Ignacio Cirac.
\newblock Improvement of frequency standards with quantum entanglement.
\newblock \emph{Physical Review Letters}, 79\penalty0 (20):\penalty0 3865,
  1997.

\bibitem[Myatt et~al.(2000)Myatt, King, Turchette, Sackett, Kielpinski, Itano,
  Monroe, and Wineland]{myatt2000decoherence}
Christopher~J Myatt, Brian~E King, Quentin~A Turchette, Cass~A Sackett, David
  Kielpinski, Wayne~M Itano, CWDJ Monroe, and David~J Wineland.
\newblock Decoherence of quantum superpositions through coupling to engineered
  reservoirs.
\newblock \emph{Nature}, 403\penalty0 (6767):\penalty0 269--273, 2000.

\bibitem[Turchette et~al.(2000)Turchette, Myatt, King, Sackett, Kielpinski,
  Itano, Monroe, and Wineland]{turchette2000decoherence}
QA~Turchette, CJ~Myatt, BE~King, CA~Sackett, David Kielpinski, WM~Itano,
  Ch~Monroe, and DJ~Wineland.
\newblock Decoherence and decay of motional quantum states of a trapped atom
  coupled to engineered reservoirs.
\newblock \emph{Physical Review A}, 62\penalty0 (5):\penalty0 053807, 2000.

\bibitem[Nielsen and Chuang(2002)]{nielsen2002quantum}
Michael~A Nielsen and Isaac Chuang.
\newblock Quantum computation and quantum information, 2002.

\bibitem[Preskill(2015)]{preskillnotes}
John Preskill.
\newblock Quantum information: Ch3.
\newblock 2015.

\bibitem[Fujiwara(2001)]{fujiwara2001estimation}
Akio Fujiwara.
\newblock {Estimation of SU (2) operation and dense coding: An information
  geometric approach}.
\newblock \emph{Physical Review A}, 65\penalty0 (1):\penalty0 012316, 2001.

\bibitem[Demkowicz-Dobrza{\'n}ski et~al.(2012)Demkowicz-Dobrza{\'n}ski,
  Ko{\l}ody{\'n}ski, and Gu{\c{t}}{\u{a}}]{demkowicz2012elusive}
Rafa{\l} Demkowicz-Dobrza{\'n}ski, Jan Ko{\l}ody{\'n}ski, and
  M{\u{a}}d{\u{a}}lin Gu{\c{t}}{\u{a}}.
\newblock {The elusive Heisenberg limit in quantum-enhanced metrology}.
\newblock \emph{Nature communications}, 3\penalty0 (1):\penalty0 1--8, 2012.

\bibitem[Hayashi and Matsumoto(2008)]{hayashi2008asymptotic}
Masahito Hayashi and Keiji Matsumoto.
\newblock Asymptotic performance of optimal state estimation in qubit system.
\newblock \emph{Journal of Mathematical Physics}, 49\penalty0 (10):\penalty0
  102101, 2008.

\bibitem[Friel et~al.(2020)Friel, Palittapongarnpim, Albarelli, and
  Datta]{friel2020attainability}
Jamie Friel, Pantita Palittapongarnpim, Francesco Albarelli, and Animesh Datta.
\newblock {Attainability of the Holevo-Cram\'{e}r-Rao bound for two-qubit 3D
  magnetometry}.
\newblock \emph{arXiv preprint arXiv:2008.01502}, 2020.

\bibitem[Hayashi(2016)]{hayashi2016quantum}
Masahito Hayashi.
\newblock \emph{Quantum information theory}.
\newblock Springer, 2016.

\bibitem[{\v{S}}afr{\'a}nek(2017)]{vsafranek2017discontinuities}
Dominik {\v{S}}afr{\'a}nek.
\newblock {Discontinuities of the quantum Fisher information and the Bures
  metric}.
\newblock \emph{Physical Review A}, 95\penalty0 (5):\penalty0 052320, 2017.

\bibitem[{\v{S}}afr{\'a}nek(2018)]{vsafranek2018simple}
Dominik {\v{S}}afr{\'a}nek.
\newblock {Simple expression for the quantum Fisher information matrix}.
\newblock \emph{Physical Review A}, 97\penalty0 (4):\penalty0 042322, 2018.

\bibitem[Seveso et~al.(2019)Seveso, Albarelli, Genoni, and
  Paris]{seveso2019discontinuity}
Luigi Seveso, Francesco Albarelli, Marco~G Genoni, and Matteo~GA Paris.
\newblock {On the discontinuity of the quantum Fisher information for quantum
  statistical models with parameter dependent rank}.
\newblock \emph{Journal of Physics A: Mathematical and Theoretical},
  53\penalty0 (2):\penalty0 02LT01, 2019.

\bibitem[Rezakhani et~al.(2019)Rezakhani, Hassani, and
  Alipour]{rezakhani2019continuity}
AT~Rezakhani, M~Hassani, and S~Alipour.
\newblock {Continuity of the quantum Fisher information}.
\newblock \emph{Physical Review A}, 100\penalty0 (3):\penalty0 032317, 2019.

\bibitem[Goldberg et~al.(2021)Goldberg, Romero, Sanz, and
  S{\'a}nchez-Soto]{goldberg2021taming}
Aaron~Z Goldberg, Jos{\'e}~L Romero, {\'A}ngel~S Sanz, and Luis~L
  S{\'a}nchez-Soto.
\newblock {Taming singularities of the quantum Fisher information}.
\newblock \emph{International Journal of Quantum Information}, page 2140004,
  2021.

\bibitem[Ye and Lu(2022)]{ye2022quantum}
Yating Ye and Xiao-Ming Lu.
\newblock {Quantum Cram{\'e}r-Rao bound for quantum statistical models with
  parameter-dependent rank}.
\newblock \emph{Physical Review A}, 106\penalty0 (2):\penalty0 022429, 2022.

\bibitem[Suzuki(2023)]{suzuki2023bayesian}
Jun Suzuki.
\newblock Bayesian nagaoka-hayashi bound for multiparameter quantum-state
  estimation problem.
\newblock \emph{arXiv preprint arXiv:2302.14223}, 2023.

\bibitem[Rubio and Dunningham(2020)]{rubio2020bayesian}
Jes{\'u}s Rubio and Jacob Dunningham.
\newblock Bayesian multiparameter quantum metrology with limited data.
\newblock \emph{Physical Review A}, 101\penalty0 (3):\penalty0 032114, 2020.

\bibitem[Thearle et~al.(2016)Thearle, Assad, and Symul]{thearle2016estimation}
Oliver Thearle, Syed~M Assad, and Thomas Symul.
\newblock Estimation of output-channel noise for continuous-variable quantum
  key distribution.
\newblock \emph{Physical Review A}, 93\penalty0 (4):\penalty0 042343, 2016.

\bibitem[Wang et~al.(2017)Wang, Yin, Wang, Chen, Guo, and
  Han]{wang2017measurement}
Chao Wang, Zhen-Qiang Yin, Shuang Wang, Wei Chen, Guang-Can Guo, and Zheng-Fu
  Han.
\newblock Measurement-device-independent quantum key distribution robust
  against environmental disturbances.
\newblock \emph{Optica}, 4\penalty0 (9):\penalty0 1016--1023, 2017.

\bibitem[Wang et~al.(2022)Wang, Yin, He, Chen, Wang, Ye, Zhou, Fan-Yuan, Wang,
  Chen, et~al.]{wang2022twin}
Shuang Wang, Zhen-Qiang Yin, De-Yong He, Wei Chen, Rui-Qiang Wang, Peng Ye, Yao
  Zhou, Guan-Jie Fan-Yuan, Fang-Xiang Wang, Wei Chen, et~al.
\newblock Twin-field quantum key distribution over 830-km fibre.
\newblock \emph{Nature photonics}, 16\penalty0 (2):\penalty0 154--161, 2022.

\end{thebibliography}
\bibliographystyle{unsrtnat}

\end{document}